\documentclass[12pt]{article}
\usepackage{epsfig}
\usepackage{bbm,latexsym}
\usepackage{amssymb,amsmath}
\usepackage{slashed}
\newcommand{\bone}{\mathbbm{1}}
\newcommand{\dd}{\mathrm{d}}
\newtheorem{theorem}{Theorem}
\newtheorem{lemma}{Lemma}
\newcommand{\mt}{\mathcal{T}}
\newcommand{\diag}{\mbox{diag}}
\newcommand{\tr}{\mathrm{Tr}}
\newcommand{\vecp}[1]{\vec{#1}^{\raisebox{-1pt}{$\scriptstyle\,\prime$}}}
\allowdisplaybreaks
\begin{document}
\title{
\normalsize \hfill UWThPh-2021-14 \\[10mm]
\LARGE Notes on basis-independent computations \\ 
with the Dirac algebra}

\author{
W.~Grimus\thanks{E-mail: walter.grimus@univie.ac.at}\;
\addtocounter{footnote}{1}
\\[5mm]
\small University of Vienna, Faculty of Physics \\
\small Boltzmanngasse 5, A--1090 Vienna, Austria
}

\date{February 25, 2022}

\maketitle

\begin{abstract}
In these notes we 
first
review Pauli's proof of his `fundamental theorem' that 
states the equivalence of any two sets of 
Dirac matrices $\{ \gamma^\mu \}$. 
Due to this theorem 
not only all physical results in the context of the Dirac equation have to be 
independent of the basis chosen for the Dirac matrices, but it should 
also be possible to obtain the results without resorting to a specific 
basis in the course of the computation. Indeed, we demonstrate this in the 
case of the behaviour of Dirac spinors under Lorentz transformations, 
the quantization of the Dirac field, the expectation value of the spin 
operator and several other topics.
In particular, we emphasize the totally different physics and 
mathematics background of the matrix $\beta$, used in the definition 
of the conjugate Dirac spinor, and $\gamma^0$. 
Finally, we compare the basis-independent manipulations with those 
performed in the Weyl basis of Dirac matrices.
The present notes provide a self-contained introduction to the Dirac 
theory by solely exploiting the simplicity of the Dirac algebra and the 
power of Pauli's Theorem.
\end{abstract}

\newpage

\section*{Notation}
\paragraph{Units:}
We use natural units $\hbar = c = 1$.
\paragraph{Metric tensor and Minkowski space:}
The metric tensor is
\[
\left( g^{\mu\nu} \right) = \left( g_{\mu\nu} \right) = 
\left( \begin{array}{rrrr}
1 & 0 & 0 & 0 \\
0 & -1 & 0 & 0 \\
0 & 0 & -1 & 0 \\
0 & 0 & 0 & -1 
\end{array} \right).
\]
Greek indices run over $0,1,2,3$, while Roman indices refer only to the spatial 
indices $1,2,3$. Four-vectors are not specially marked, but three-vectors 
always have an arrow above the letter. Indices of four-vectors are lowered 
or raised with the metric tensor. Let 
\[
v = (v^\mu) = \left( \begin{array}{c} v^0 \\ \vec v \end{array} \right)
\quad \mbox{and} \quad 
w = (w^\nu) = \left( \begin{array}{c} w^0 \\ \vec w \end{array} \right)
\]
be four-vectors. Then the Minkowski scalar-product is given by
\[
v \cdot w = g_{\mu\nu} v^\mu w^\nu = v_\mu w^\mu = v^\mu w_\mu = 
v^0 w^0 - \vec v \cdot \vec w.
\]
For $v = w$ we write $v \cdot v \equiv v^2$.
Note that scalar products of both three and four-vectors are indicated by 
a dot. Here and in the following the Einstein sum convention is understood.
\paragraph{Unit matrices:}
An $n \times n$ unit matrix is denoted by $\bone_n$.
For $n=2$ we skip the index and just write $\bone$.
\paragraph{Pauli matrices:} 
In the context of the Weyl basis the symbols $\sigma^\mu$ and 
$\bar\sigma^\mu$ are very useful.
The notation $\sigma^\mu$ includes the 
unit matrix and the Pauli matrices, \textit{i.e.}
\[
\sigma^0 = \left( \begin{array}{cc} 1 & 0 \\ 0 & 1 \end{array} \right),
\quad
\sigma^1 = \left( \begin{array}{cc} 0 & 1 \\ 1 & 0 \end{array} \right),
\quad
\sigma^2 = \left( \begin{array}{cc} 0 & -i \\ i & 0 \end{array} \right),
\quad
\sigma^3 = \left( \begin{array}{cc} 1 & 0 \\ 0 & -1 \end{array} \right).
\]
For the `three-vector' consisting of the proper Pauli matrices we write 
$\vec \sigma$. Thus we have a `four-vector' 
$(\sigma^\mu) = (\bone, \vec\sigma)$, while $\bar\sigma^\mu$ is defined by 
$(\bar\sigma^\mu) = (\bone, -\vec\sigma)$. 
Lowering the indices of these symbols is done in usual way, 
\textit{i.e.}\ $\sigma_\mu = g_{\mu\nu} \sigma^\nu$ and
$\bar\sigma_\mu = g_{\mu\nu} \bar\sigma^\nu$. 
Consequently, if $v$ is a four-vector, then 
\[
v \cdot \sigma = v^0 \bone - \vec v \cdot \vec\sigma
\quad \mbox{and} \quad 
v \cdot \bar\sigma = v^0 \bone + \vec v \cdot \vec\sigma.
\]

\newpage

\section{Introduction}
The Dirac equation
\begin{equation}\label{dirac-eq}
\left( i \gamma^\mu \partial_\mu - m \right) \psi(x) = 0
\end{equation}
is a relativistic wave equation for fermions put forward by 
Dirac in~1928~\cite{dirac1,dirac2} (see~\cite{pais} for the 
historical background).  
The abstract quantities $\gamma^\mu$ ($\mu = 0,1,2,3$) are defined by 
postulating that 
\begin{eqnarray}
\lefteqn{
\left( i \gamma^\nu \partial_\nu + m \right)
\left( i \gamma^\mu \partial_\mu - m \right) \psi(x) 
} \nonumber \\ && = 
-\left( \gamma^\nu \gamma^\mu \partial_\nu \partial_\mu + m^2 \right) \psi(x) 
\nonumber \\  && =
-\left( \frac{1}{2} \left( \gamma^\mu \gamma^\nu + \gamma^\nu \gamma^\mu \right)
\partial_\mu \partial_\nu + m^2 \right) \psi(x) = 0,
\label{intermediate}
\end{eqnarray}
leads to the Klein--Gordon equation for each component of 
$\psi$~\cite{dirac1,dirac2}. In this equation 
the algebraic identity $(a+b)(a-b) = a^2 - b^2$ with $b=m$ 
and the mass $m$ being a `$c$-number' 
have been used. Equation~(\ref{intermediate}) agrees with the 
Klein--Gordon equation
\begin{equation}
\left(\, g^{\mu\nu}\partial_\mu \partial_\nu + m^2\, \right) \psi(x) = 0, 
\end{equation}
provided
\begin{equation}\label{anticom_4}
\gamma^\mu \gamma^\nu + \gamma^\nu \gamma^\mu = 2 g^{\mu\nu} \bone_4.
\end{equation}
These relations among the quantities $\gamma^\mu$ define the Dirac algebra. 
The $4 \times 4$ unit matrix $\bone_4$ indicates that the 
$\gamma^\mu$ are thought to be realized as $4 \times 4$ matrices, 
the so-called Dirac or gamma matrices. The representation of the Dirac 
algebra by $4 \times 4$ matrices is indeed the unique possibility 
for $N=4$ space-time dimensions featuring four quantities $\gamma^\mu$, 
provided the representation is irreducible.
This will be demonstrated in section~\ref{pauli}.

The purpose of the manuscript is twofold. Firstly, in section~\ref{pauli} we 
allow for $N$ space-time dimensions and $d \times d$ Dirac matrices, and 
review Pauli's proof~\cite{pauli} of the theorem named after him, 
by adapting it to arbitrary $N$. In the course of this we obtain 
a relation between $N$ and $d$ that establishes $d=4$ for $N=4$.
Secondly, we revert to $N=4$ and discuss 
different topics in the following 
eight
sections with emphasis on 
basis-independent computations, \textit{i.e.}\ by exploiting solely 
Pauli's Theorem and 
equation~(\ref{anticom_4}) for the computations without 
ever taking refuge to a special realization of the Dirac matrices. 
These topics are 
\begin{itemize}
\item Dirac equation and Lorentz invariance,
\item conjugate spinors and Lorentz invariance,
\item plane-wave solutions of the Dirac equation,
\item charge conjugation, 
\item quantization of the Dirac field,
\item time reversal,
\item expectation value of the spin operator,
\item ultrarelativistic particles and helicity.
\end{itemize}
However, at the very end, in section~\ref{weyl basis} 
we choose the Weyl basis for the Dirac matrices and repeat the computations 
performed earlier. The reason is that in this basis the behaviour of the 
Dirac spinors under Lorentz transformations is particularly 
transparent. 
Moreover, 
we find it instructive to compare the computations in the Weyl basis 
with the basis-independent treatment that is the focus of our notes.
Most physics aspects of
the topics presented here can also be found in textbooks, for instance 
in~\cite{jauch,bjorken,bailin,itzykson,peskin,horejsi}. The book by 
Jauch and Rohrlich~\cite{jauch} provides a 
rather
thorough basis-independent 
discussion of the Dirac matrices, including a proof of Pauli's Theorem, 
but the notation needs getting used to. The connection between the Dirac 
algebra and the massive spin-1/2 representation of the Poincar\'e group 
is elucidated in the book by Bailin~\cite{bailin}. Often 
the so-called `standard representation' is utilized in textbooks for 
specific computations, see for instance~\cite{bjorken,itzykson,horejsi}, 
but in~\cite{peskin} the Weyl representation of the Dirac matrices 
is preferred. As for basis-independent computations, we also refer the reader 
to~\cite{pal}; there is some overlap, but in general 
the focus of the present manuscript is different from that of~\cite{pal}. 
Another difference is that in~\cite{pal} the popular hermiticity properties 
of the Dirac matrices are postulated whereas we keep them 
completely general throughout, except in section~\ref{weyl basis}.

For reading this manuscript, some basic knowledge of linear algebra and 
group theory, familiarity with Schur's Lemma and 
some introductory knowledge of (quantum) field theory are assumed. 
Otherwise, the material should be 
self-contained.
In particular, concerning the group $SL(2,\mathbbm{C})$ and the Lorentz 
group, the reader can find all necessary material in appendix~\ref{sl2c}.
We use the phrase `irrep' 
as an abbreviation for `irreducible representation'.

\section{Pauli's Theorem}
\label{pauli}
In this section we review Pauli's proof of his equivalence 
theorem, named `fundamental theorem' in his paper~\cite{pauli} 
(see also~\cite{jauch}). 
However, it is an amusing fact that his proof and related 
considerations for an $N$-dimensional 
space-time are not more complicated than for $N=4$, provided $N$ is 
\emph{even}. On the contrary, we actually believe that 
it is more transparent and instructive to discuss, in this section, 
the Dirac algebra for arbitrary but even $N$ ($N \geq 2$).
The case of $N$ odd is slightly more involved and we adapt Pauli's 
proof to this situation after the discussion of $N$ even.
We conclude this section by discussing some implications of 
Pauli's Theorem. 
An alternative proof of Pauli's Theorem for arbitrary $N$, 
based on the theory of finite 
groups, can be found in appendix~\ref{a-proof}. 

\subsection{Preliminaries}
In $N$ space-time dimensions there are $N$ Dirac matrices. 
The anticommutation relations read
\begin{equation}\label{anticom_N}
\gamma^\mu \gamma^\nu + \gamma^\nu \gamma^\mu = 2 g^{\mu\nu} \bone_d
\quad \mbox{with} \quad
\mu,\nu = 0,1, \ldots, N-1
\quad \mbox{and} \quad 
\left( g^{\mu\nu} \right) = \diag \big( 1, 
\underbrace{-1, \ldots, -1}_{N-1\,\mathrm{times}} \big).
\end{equation}
We search for irreps of the Dirac algebra with 
$d \times d$ Dirac matrices. This amounts to obtaining a relation between 
$N$ and $d$.

In the following discussion we utilize the 
$d \times d$ matrices $G^r$ defined as  
\begin{equation}\label{G}
\bone_d, \; \gamma^\mu, \; \gamma^\mu \gamma^\nu \,(\mu < \nu), \dots,
\; \gamma^{\mu_1} \gamma^{\mu_2} \cdots \gamma^{\mu_p} \,
(\mu_1 < \mu_2 < \cdots < \mu_p), \ldots, \;
\gamma^0 \gamma^1 \cdots \gamma^{N-1}.
\end{equation}
There are 
\begin{equation}
\sum_{p=0}^N \genfrac(){0pt}{0}{N}{p} = 2^N
\end{equation}
such matrices. Any $G^r$ is thus a product of $p$ different Dirac matrices 
with $0 \leq p \leq N$. The case $p=0$ refers to the unit matrix $\bone_d$.

Note that equation~(\ref{anticom_N}) implies that the Dirac matrices 
are non-singular because
\begin{equation}\label{sign}
\left( \gamma^\mu \right)^{-1} = \epsilon(\mu) \gamma^\mu
\quad \mbox{with} \quad \epsilon(\mu) = 
\left\{ \begin{array}{rcl}
1 & \mbox{for} & \mu = 0, \\
-1 & \mbox{for} & \mu = 1,\ldots,N-1.
\end{array} \right.
\end{equation}

\subsection{Proof of Pauli's Theorem for $N$ even}
We remark that in the following theorem the Dirac algebra is not 
necessarily irreducible.
\begin{theorem}[\boldmath$N$\unboldmath\ even]\label{Neven}
Let $N$ be even and the $\gamma^\mu$ be 
a $d$-dimensional representation of the Dirac algebra. Then, 
$\tr\,G^r = 0$ for any $G^r \neq \bone_d$ and, 
moreover,
$\tr \left( (G^r)^{-1} G^s \right) = d\,\delta_{rs}$. Therefore, the matrices 
$G^r$ are linearly independent.
\end{theorem}
\textbf{Proof:} 
Suppose we have $p$ indices $\mu_k$ such that 
$0 \leq \mu_1 < \mu_2 < \ldots \mu_p \leq N-1$ with $1 \leq p \leq N$. 
Let $p$ be even. Then 
\[
\gamma^{\mu_1} 
\left( \gamma^{\mu_1} \gamma^{\mu_2} \cdots \gamma^{\mu_p} \right) 
\left( \gamma^{\mu_1} \right)^{-1} = 
-\left( \gamma^{\mu_1} \gamma^{\mu_2} \cdots \gamma^{\mu_p} \right)
\]
according to equation~(\ref{anticom_N}).
Therefore, the trace of this product is zero. If $p$ is odd, 
taking into account that $N$ is even, 
there is always an index $\nu$ different from all 
$\mu_1, \mu_2, \ldots, \mu_p$. Then 
\[
\gamma^\nu 
\left( \gamma^{\mu_1} \gamma^{\mu_2} \cdots \gamma^{\mu_p} \right) 
\left( \gamma^\nu \right)^{-1} = 
- \left( \gamma^{\mu_1} \gamma^{\mu_2} \cdots \gamma^{\mu_p} \right)
\]
and we conclude again that the trace must be 
zero. The second part of the theorem follows from the first one because, 
apart from a sign, $(G^r)^{-1}$ is just a product of different 
Dirac matrices and, applying equation~(\ref{anticom_N}) a sufficient 
number of times, the same is true for $(G^r)^{-1} G^s$. The unit matrix is 
obtained if and only if $G^r = G^s$, \textit{i.e.}\ $r=s$.
In order to prove linear independence of the matrices $G^s$, we 
assume that there is a relation $\sum_s c_s G^s = 0$ with 
complex coefficients $c_s$. Multiplying this 
relation by $(G^r)^{-1}$ and taking the trace, we find $c_r = 0$. Since $r$ 
was arbitrary, we conclude $c_s = 0$ $\forall \, s$ and the $G^s$ are 
linearly independent. Q.E.D.

The following theorem is the generalization of Pauli's Theorem 
for $N$ space-time dimensions when $N$ is even. 
\begin{theorem}[\boldmath$N$\unboldmath\ even]\label{irrep}
An irrep of the Dirac algebra referring to an 
$N$-dimensional space-time with $N$ even is unique up to similarity 
transformations and has dimension $d = 2^{N/2}$.
\end{theorem}
\textbf{Proof:} 
First we focus on the uniqueness.
We assume that we have a $d$-dimensional and a $d'$-dimensional 
irrep of the Dirac matrices. With these we build 
matrices $G^r$ and ${G'}^r$, as defined in the 
beginning of the present section. We define the matrix 
\begin{equation}\label{S}
S = \sum_s \left(  G^s \right)^{-1} F {G'}^s
\end{equation}
such that \emph{all} $2^N$ indices $s$ occur in this sum.
The $d \times d'$ matrix $F$ is arbitrary for the time being. Note that 
\begin{equation}\label{srt}
G^s G^r = \epsilon_{sr} G^t, \quad
{G'}^s {G'}^r = \epsilon_{sr} {G'}^t
\quad \mbox{with} \quad 
\epsilon_{sr}^2 = 1,
\end{equation}
where $t$ and $\epsilon_{sr}$ are determined by equation~(\ref{anticom_N}). 
The key observation in equation~(\ref{srt}) is that, 
if we fix $r$ and vary $s$ over all $2^N$ possibilities, then 
also $t$ varies over all $2^N$ possibilities, although in a different order.
Therefore, we obtain 
\begin{equation}\label{GSGS}
\left( G^r \right)^{-1} S\, {G'}^r = S 
\quad \mbox{or} \quad 
S\, {G'}^r = G^r S
\; \forall\,r.
\end{equation}
Taking into account that we deal with irreps, Schur's Lemma 
tells us that $S$ is either zero or invertible.
Note that $S$ depends on the matrix $F$, which is arbitrary.
Let us first assume that $S$ is zero for all $F$. In particular, we can 
choose an $F$ such that $F_{kl} = 1$ but all other elements are zero. In this 
way we deduce from equation~(\ref{GSGS})
\begin{equation}
\left( (G^r)^{-1} \right)_{jk} \left( {G'}^r \right)_{lm} = 0 \; \forall\;
j,k = 1,\ldots,d 
\quad \mbox{and} \quad
\forall \; l,m = 1,\ldots,d'.
\end{equation}
However, from this equation we are lead to conclude 
that the set of the ${G'}^r$ 
(or that of the $G^r$) is linearly dependent. This is a contradiction, 
\textit{cf.}\ Theorem~\ref{Neven}. Therefore, 
the assumption that $S = 0$ for all $F$ is wrong. There is thus an $F$ such 
that $S$ is invertible. For this $S$ we have 
${G'}^r = S^{-1} G^r S\;\forall r$ and the two irreps are equivalent. 
In particular, $d = d'$. This concludes the uniqueness part of the theorem.

In order to obtain a relation between $d$ and $N$, 
we set ${G'}^r = G^r$ in the definition of $S$. Then, in analogy to  
equation~(\ref{GSGS}), we obtain
\begin{equation}
S\, G^r = G^r S \; \forall\,r.
\end{equation}
Since we assume an irrep of the Dirac algebra, 
according to Schur's Lemma we have $S = \lambda \bone_d$ 
or 
\begin{equation}\label{GSGlambda}
\sum_s \left(  G^s \right)^{-1} F G^s = \lambda \bone_d,
\end{equation}
where $\lambda$ is some complex number depending on $F$. 
Taking the trace of this relation 
and taking into account that in $S$ there are $2^N$ summands, we find 
\begin{equation}\label{lambda}
\lambda = \frac{2^N}{d}\, \tr F.
\end{equation}
Now we specify again to an $F$ with $F_{kl} = 1$ and zero otherwise. 
Inserting this $F$ into equation~(\ref{GSGlambda}) and using 
$\tr F = \delta_{kl}$, we obtain 
\begin{equation}\label{sggs}
\sum_s \left( (G^s)^{-1} \right)_{jk} \left( G^s \right)_{lm} = 
\frac{2^N}{d}\, \delta_{kl}\, \delta_{jm}.
\end{equation}
In order to determine the dimension $d$, we sum, in this equation, over 
$j=k=1,\ldots,d$ and $l=m=1,\ldots,d$. 
Then the left-hand side of equation~(\ref{sggs}) yields 
$\sum_s \tr \left( G^s \right)^{-1} \tr\,G^s$. 
Using $\tr\,G^r = 0$ except for $G^r = \bone_d$ according to 
Theorem~\ref{Neven}, we finally arrive at
\begin{equation}
d^2 = \frac{2^N}{d} \times d
\quad \mbox{or} \quad d = 2^{N/2}
\end{equation}
for $N$ even. Q.E.D.

The following theorem is a byproduct of the previous theorems. 
\begin{theorem}[\boldmath$N$\unboldmath\ even]\label{basis}
For the $d = 2^{N/2}$-dimensional irrep of the Dirac 
matrices referring to an $N$-dimensional space-time, the matrices 
$\{ G^r | r = 1, \ldots, 2^N \}$ 
form a basis in the vector space of matrices on $\mathbbm{C}^d$.
\end{theorem}
\textbf{Proof:} 
The dimension of the vector space of matrices on $\mathbbm{C}^d$ is 
$d^2 = 2^N$, which is identical with the number of matrices $G^r$. 
Since the $G^r$ are linearly independent, the theorem is proven. Q.E.D.

\addtocounter{theorem}{-3}
\subsection{The case of $N$ odd}
\label{caseNodd}
There is a crucial difference between $N$ even and $N$ odd because, if  
$N$ is odd, all $\gamma^\mu$ commute with 
\begin{equation}\label{A}
A \equiv \gamma^0 \gamma^1 \cdots \gamma^{N-1}. 
\end{equation}
In other words, 
for $p=N$ there is no such index $\nu$ that has been used in the proof of 
Theorem~\ref{Neven} ($N$ even) and in general the trace of $A$ will be 
different from zero. Moreover, in an irrep of the Dirac algebra, 
$A$ will be proportional to the unit matrix and $\tr\,A$ will be definitely 
nonzero.
Therefore, we have to modify Theorem~\ref{Neven} in the following way.
\begin{theorem}[\boldmath$N$\unboldmath\ odd]\label{Nodd}
Let $N$ be odd and let the $\gamma^\mu$ be 
a $d$-dimensional representation of the Dirac algebra. Then, 
$\tr\,G^r = 0$ for all $G^r$ different from both $\bone_d$ and 
$\gamma^0 \gamma^1 \cdots \gamma^{N-1}$.
Moreover, writing $N = 2K+1$, then 
$\tr \left( (G^r)^{-1} G^s \right) = d\,\delta_{rs}$ if both $G^r$ and $G^s$ 
contain $K$ or less Dirac matrices. Therefore, the set of those matrices $G^r$ 
that contain products of $K$ or less Dirac matrices is linearly independent.
In addition, if the representation is irreducible, then 
\[
\gamma^0 \gamma^1 \cdots \gamma^{N-1} = \omega \bone_d
\quad \mbox{with} \quad 
\omega^4 = 1.
\]
\end{theorem}
The latter statement is trivial because $\left( G^r \right)^2 = \pm \bone_d$ 
for all $r$. 

It is nevertheless interesting to determine $\omega$ for $A$. 
By using equation~(\ref{anticom_N}), we compute
\begin{equation}\label{A2}
A^2 = (-1)^q \bone_d \quad \mbox{with} \quad 
q = N(N-1)/2 + N-1
\quad \Rightarrow \quad \omega^2 = (-1)^q = (-1)^K.
\end{equation}
We thus obtain
\begin{equation}\label{omega}
\omega = \left\{
\begin{array}{ccccl}
\pm 1 &\mathrm{for}& K=0,2,4,6,\ldots &\mathrm{or}& N = 1,5,9,13,\ldots,
\\
\pm i &\mathrm{for}& K=1,3,5,7,\ldots &\mathrm{or}& N = 3,7,11,15,\ldots.
\end{array} \right.
\end{equation}
For completeness we have included the trivial case $N=1$.

The relation $A = \omega \bone_d$ for irreps procures that matrices 
$G^r$ that are products of more than $K = (N-1)/2$ Dirac matrices 
can be reduced to matrices $G^r$ that are products of $K = (N-1)/2$ or
less Dirac matrices. Therefore, only $2^{N-1}$ matrices 
are linearly independent as stated in the theorem, 
conveniently chosen to be those with 
$0 \leq p \leq K$. In particular, this reduction has to be 
applied to $G^t$ on the right-hand side of equation~(\ref{srt}), whenever 
$G^t$ is a product of more than $K$ different Dirac matrices. Such a 
reduction produces an imaginary $\epsilon_{rs}$ if $N = 3,7,11,\ldots$. 

Equation~(\ref{omega}) suggests to define $\omega_+ = +1$ or $+i$ 
and $\omega_- = -1$ or $-i$, and likewise $A_\pm = \omega_\pm \bone_d$ 
for irreps. We emphasize that an irrep where $A$ is represented by $A_+$ is 
necessarily inequivalent to an irrep where $A$ is represented by $A_-$ 
since $A_-$ cannot be obtained by a similarity transformation from $A_+$. 
Having two realizations of $A$, the above mentioned reduction can be done 
in two ways, either with $A_+$ or with $A_-$.  
In particular, this choice is important 
for $G^t$ on the right-hand side of equation~(\ref{srt}) because this equation 
is used in the proof of Theorem~\ref{irrep} when equation~(\ref{GSGS}) 
is derived. Actually, it is straightforward to check that the proof of 
Theorem~\ref{irrep} can be adapted to $N$ odd provided in $S$ we use only 
the $2^{N-1}$ Dirac matrices with $0 \leq p \leq K$. In this way, 
equation~(\ref{lambda}) is modified to 
$\lambda = \frac{2^{N-1}}{d}\, \tr F$ and the dimension of the irreducible 
representation is $d = 2^{(N-1)/2}$. We are thus led to the following theorem.
\begin{theorem}[\boldmath$N$\unboldmath\ odd]\label{irrep-odd}
In an $N$-dimensi\-o\-nal space-time with $N$ odd 
there are exactly two inequivalent irreps of the Dirac matrices. 
Both have dimension $d = 2^{(N-1)/2}$. 
If the set $\{ \gamma^\mu \}$ represents one such irrep, 
then the other one is given by the set $\{ -\gamma^\mu \}$. 
\end{theorem}

Finally, we can reformulate Theorem~\ref{basis} for $N$ odd.
\begin{theorem}[\boldmath$N$\unboldmath\ odd]\label{basis-odd}
For the two $d = 2^{(N-1)/2}$-dimensional irreps of the Dirac 
matrices referring to an $N$-dimensional space-time with $N$ odd, the 
$2^{N-1}$ matrices $G^r$ that have $0 \leq p \leq (N-1)/2$
form a basis in the vector space of matrices on $\mathbbm{C}^d$.
\end{theorem}

\subsection{On the similarity transformation between equivalent irreps}
If we have two equivalent irreps of Dirac matrices, then Schur's Lemma 
has a strong impact on the similarity transformation that connects the 
two irreps.
\begin{theorem}\label{similarity}
Let $\gamma^\mu$ and ${\gamma'}^\mu$ be irreducible and equivalent 
representations of the Dirac algebra referring to an 
$N$-dimensional space-time, 
and let $S$ be the matrix that effects the similarity transformation,
\textit{i.e.}\  
$S^{-1} \gamma^\mu S = {\gamma'}^\mu$. Then $S$ is unique up to a 
multiplicative complex constant.
\end{theorem}
\textbf{Proof:}
Suppose there are two matrices $S_1$ and $S_2$ with 
\[
S_1^{-1} \gamma^\mu S_1 = S_2^{-1} \gamma^\mu S_2 = {\gamma'}^\mu.
\]
Consequently, 
\[
\left( S_1 S_2^{-1} \right)^{-1} \gamma^\mu \left( S_1 S_2^{-1} \right) = 
\gamma^\mu 
\quad \Rightarrow \quad
S_1 S_2^{-1} = c \bone_d
\]
according to Schur's Lemma,
since we assumed an irrep of the Dirac algebra.
Q.E.D. 

Note that up to now the metric tensor $g^{\mu\nu}$ played no 
role in our discussion, except for $N$ odd where it had an impact on $\omega$, 
equation~(\ref{omega}). Actually, Theorems~\ref{Neven}, \ref{irrep} 
and~\ref{basis} ($N$ even and odd) hold for any choice of signs in 
$g^{\mu\nu}$, not only for those of equation~(\ref{anticom_N}).   
Moreover, up to now we have not assumed any  
hermiticity properties of the Dirac matrices. In general, such an assumption 
is not necessary, but it is quite popular in textbooks. However, 
as soon as we want to impose hermiticity or antihermiticity on the 
Dirac matrices, we have to take into account the signs $g^{\mu\mu}$ 
for the following reason. Suppose a quadratic matrix $M$ fulfills 
$M^2 = \epsilon \bone_d$ and $M^\dagger = \eta M$ with 
$\epsilon^2 = \eta^2 = 1$. Then 
\begin{equation}
0 \leq M^\dagger M = \eta M^2 = \eta \epsilon \bone_d 
\quad \Rightarrow \quad \eta = \epsilon.
\end{equation}
Therefore, if $g^{\mu\mu} = 1$ or $\left( \gamma^\mu \right)^2 = \bone_d$, 
then $\gamma^\mu$ must be assumed hermitian, 
if $g^{\mu\mu} = -1$ or $\left( \gamma^\mu \right)^2 = -\bone_d$, 
then $\gamma^\mu$ must be assumed antihermitian. 
Consequently, with the metric tensor given 
in equation~(\ref{anticom_N}), the only consistent choice 
of hermiticity properties is given by 
\begin{equation}\label{special}
\left( \gamma^0 \right)^\dagger = \gamma^0,
\quad
\left( \gamma^j \right)^\dagger = -\gamma^j \;(j = 1, \ldots, N-1).
\end{equation}
In the case of $N=4$, the Weyl basis discussed in section~\ref{weyl basis} 
is an example of a set of Dirac matrices with these hermiticity properties.

For Dirac matrices obeying equation~(\ref{special}), 
Theorem~\ref{similarity} is modified in the following way.
\begin{theorem}\label{unitary}
Let $\gamma^\mu$ and ${\gamma'}^\mu$ be both irreps 
of the Dirac algebra referring to an $N$-dimensio\-nal space-time. 
Let us furthermore assume that both sets of Dirac matrices obey 
equation~(\ref{special}) and that $S^{-1} \gamma^\mu S = {\gamma'}^\mu$. 
Then $S = c S'$ where $S'$ is unitary and $c > 0$.
\end{theorem}
\textbf{Proof:}
Taking the hermitian conjugate of the similarity transformation, we obtain
\[
S^\dagger \gamma^\mu \left( S^\dagger \right)^{-1} = {\gamma'}^\mu.
\]
With $S_1 = S$ and $S_2 = \left( S^\dagger \right)^{-1}$, application of 
Theorem~\ref{similarity} delivers 
\[
S S^\dagger = a \bone_d.
\]
Taking into account that $S S^\dagger$ is a positive matrix, we find $a > 0$.
Therefore, $S' = S/\sqrt{a}$ and $c = \sqrt{a}$. Q.E.D. \\[2mm]
It is thus reasonable to perform the similarity transformation with 
a unitary matrix $S$ in situations where Theorem~\ref{unitary} is applicable.
Such an $S$ is then unique up to an overall phase factor, which can be 
chosen according to a suitable convention.

\subsection{Chiral projectors}
\label{chiral pro}
We first define the \emph{chiral matrix}
\begin{equation}\label{gammachir}
\gamma_\mathrm{chir} = i^{N/2-1} \gamma^0 \gamma^1 \cdots \gamma^{N-1}.
\end{equation}
Note that, if we have an irrep of the Dirac algebra, 
this matrix makes sense only for $N$ even 
because $\gamma_\mathrm{chir} \propto \bone_d$ for $N$ odd. Moreover, 
an essential property for the application of the chiral matrix is 
\begin{equation}\label{essential}
\gamma^\mu \gamma_\mathrm{chir} = -\gamma_\mathrm{chir} \gamma^\mu
\;\; \forall\, \mu,
\end{equation}
which again only holds for $N$ even. Therefore, 
from now on we confine ourselves to $N = 2K$ with $K \in \mathbbm{N}$. 
In the remainder of this subsection the set of 
Dirac matrices is not necessarily irreducible.

The chiral matrix has the following property.
\begin{theorem}\label{chiral2}
\[
\left( \gamma_\mathrm{chir} \right)^2 = \bone_d.
\]
\end{theorem}
\textbf{Proof:}
With equations~(\ref{A}) and~(\ref{A2}) the square of the chiral matrix 
is given by
\[
\left( \gamma_\mathrm{chir} \right)^2 = \left( i^{N/2-1} A \right)^2 = 
(-1)^{N/2-1+q} \bone_d
\]
with $q$ defined in equation~(\ref{A2}). Inserting $N=2K$, 
the exponent of $-1$ is 
\[
N/2 - 1 + q = N/2-1 + N(N-1)/2 + N-1 = 2(K^2 + K - 1).
\]
This is an even number and the theorem is proven. Q.E.D. \\[2mm]
Now we define the chiral projectors
\begin{equation}\label{chiral projectors}
\gamma_+ = \frac{1}{2} \left( \bone_d + \gamma_\mathrm{chir} \right),
\quad
\gamma_- = \frac{1}{2} \left( \bone_d - \gamma_\mathrm{chir} \right).
\end{equation}
Though the $\gamma_\pm$ are in general not hermitian, they have all other
properties required of complementary projectors:
\begin{equation}
\gamma_+^2 = \gamma_+, \quad
\gamma_-^2 = \gamma_-, \quad
\gamma_+ + \gamma_- = \bone_d, \quad
\gamma_+\gamma_- = \gamma_-\gamma_+ = 0.
\end{equation}
Therefore, according to Lemma~\ref{lemma} of appendix~\ref{selfinverse}, 
the representation space 
$\mathbbm{C}^d$ can be decomposed as
\begin{equation}\label{V+-}
\mathbbm{C}^d = \mathcal{V}_+ \oplus \mathcal{V}_-
\quad \mbox{with} \quad
\mathcal{V}_+ = \gamma_+ \mathbbm{C}^d, \;\;
\mathcal{V}_- = \gamma_- \mathbbm{C}^d.
\end{equation}
Because of Theorem~\ref{chiral2} and $\tr\,\gamma_\mathrm{chir} = 0$, 
the multiplicity of the eigenvalue $+1$ of $\gamma_\mathrm{chir}$ 
equals that of the eigenvalue $-1$. Therefore, 
\begin{equation}\label{dimV+-}
\dim \mathcal{V}_+ = \dim \mathcal{V}_- = 2^{N/2-1}.
\end{equation}

In the following special case the chiral projectors are even hermitian.
\begin{theorem}
If the Dirac matrices obey equation~(\ref{special}), then
\begin{equation}\label{chiral-hermitian}
\left( \gamma_\mathrm{chir} \right)^\dagger = \gamma_\mathrm{chir}.
\end{equation}
\end{theorem}
\textbf{Proof:}
We form the hermitian conjugate 
\[
\left( \gamma_\mathrm{chir} \right)^\dagger = \left( -i \right)^{N/2-1}
\gamma^{N-1} \cdots \gamma^2 \gamma^1 \left( -1 \right)^{N-1}.
\]
The factor $\left( -1 \right)^{N-1}$ takes into account that $N-1$ Dirac
matrices are antihermitian. The number of transpositions for 
restoring the order of the Dirac matrices with ascending indices is 
$N(N-1)/2$. Adding all powers of $-1$, we end up with the same even number 
$N/2 - 1 + q$ as in the proof of Theorem~\ref{chiral2}. Q.E.D.

\subsection{An application of Pauli's Theorem}
\label{application}
As an example for the usefulness of Pauli's Theorem, we prove 
the existence of a $C$-type transformation, where $C$ denotes 
charge conjugation. Such a transformation is defined via the 
similarity transformation
\begin{equation}
C^{-1} \gamma^\mu C = \epsilon \left( \gamma^\mu \right)^T
\quad \mbox{with} \quad \epsilon^2 = 1.
\end{equation}
Denoting the anticommutator by $\{ .,. \}$, the essential observation is 
\begin{equation}
\{ \gamma^\mu, \gamma^\nu \} = 2 g^{\mu\nu} \bone_d
\quad \Leftrightarrow \quad
\{ \epsilon \left( \gamma^\mu \right)^T, \epsilon \left( \gamma^\nu \right)^T 
\} = 2 g^{\mu\nu} \bone_d.
\end{equation}
Therefore, in the case of $N$ even, the uniqueness of the irrep 
of the Dirac algebra, Theorem~\ref{irrep} ($N$ even), provides the existence 
of a matrix $C$ for both signs $\epsilon = \pm 1$. That both signs are 
possible is trivial for $N$ even, because the very same uniqueness implies 
the existence of a similarity transformation that effects 
$\gamma^\mu \to -\gamma^\mu$ $\forall\, \mu$.

The latter similarity transformation does not exist for $N$ odd since it 
would induce $A \to -A$, \textit{cf.}\ equation~(\ref{A}), which is 
impossible because $A \propto \bone_d$. However, precisely this relation
allows us to find the permitted sign via
\begin{equation}
A = C^{-1} A C = \epsilon^N \left( \gamma^0 \right)^T 
\left( \gamma^1 \right)^T \cdots \left( \gamma^{N-1} \right)^T = 
\epsilon^N (-1)^{N(N-1)/2} A.
\end{equation}
Taking into account that $N$ is odd, we find 
\begin{equation}
\epsilon^N (-1)^{N(N-1)/2} = \epsilon (-1)^{(N-1)/2} = 1
\quad \Rightarrow \quad 
\epsilon = (-1)^{(N-1)/2}.
\end{equation}
We conclude this section by listing $\epsilon$ for $N = 1$ to $7$:
\begin{equation}
\begin{array}{c|cccccccc}
N &        1 & 2   & 3 & 4   & 5 & 6   & 7 \\ \hline
\epsilon & + & \pm & - & \pm & + & \pm & -
\end{array}
\end{equation}
For $N=4$, the standard definition of the matrix $C$ has $\epsilon = -1$.

\section{Dirac equation and Lorentz invariance}
\label{deli}
From now on we revert to $N = 4$ space-time dimensions, 
in which case the irrep of the Dirac algebra is unique and 
the Dirac matrices are $4 \times 4$ because the dimension of the 
irrep is $d = 2^{N/2} = 4$---\textit{cf.}\ Theorem~\ref{irrep} ($N$ even).
For the sake of clarity we reformulate 
the version of Pauli's Theorem we will apply henceforth.
\begin{theorem}[$N=4$]\label{pauli4}
For any two sets $\{ \gamma^\mu \}$ and 
$\{ {\gamma'}^\mu \}$ of four $4 \times 4$ matrices satisfying 
equation~(\ref{anticom_4}) there is \emph{always} a similarity 
transformation linking the two sets. In other words there exists a matrix 
$S$ such that $S^{-1} \gamma^\mu S = {\gamma'}^\mu$. 
\end{theorem}

After this introduction, we intend to discuss the property 
of solutions of the Dirac equation, equation~(\ref{dirac-eq}),
under Lorentz transformations. For the basics of Lorentz transformations and 
the group $SL(2,\mathbbm{C})$ we refer the reader to appendix~\ref{sl2c}. 
Assuming that we have a solution $\psi(x)$ in one inertial frame, 
we make the ansatz
\begin{equation}\label{Spsi}
\psi'(x') = \mathcal{S} \psi(L^{-1}x')
\end{equation}
with the Lorentz transformation $L \in \mathbbm{L}$, $x = L^{-1} x'$, and 
a yet unknown $4 \times 4$ matrix $\mathcal{S}$ for the solution in 
another inertial frame. Form invariance of the Dirac equation under 
transformations between inertial frames then leads to
\begin{equation}
\left( i\gamma^\mu \partial'_\mu - m \right) \psi'(x') = \mathcal{S} \left(
i\, \mathcal{S}^{-1} \gamma^\mu \mathcal{S} 
\left( L^{-1} \right)^\nu_{\hphantom{\nu}\mu} \partial_\nu - m \right)
\psi(x) = 0,
\end{equation}
whence the condition 
\begin{equation}\label{Scond}
\mathcal{S}^{-1} \gamma^\mu \mathcal{S} = L^\mu_{\hphantom{\mu}\lambda} 
\gamma^\lambda
\end{equation}
follows. At this point we have to ask the crucial question whether 
such an $S$ exists. The answer is affirmative because, 
defining ${\gamma'}^\mu = L^\mu_{\hphantom{\mu}\lambda} \gamma^\lambda$ and 
making use of equations~(\ref{anticom_4}) and~(\ref{LT}),
we find the anticommutator 
$\{ {\gamma'}^\mu, {\gamma'}^\nu \} = 2 g^{\mu\nu} \bone_4$~\cite{jauch,bailin}. 
Therefore, Theorem~\ref{pauli4} ensures the existence of $\mathcal{S}$ for 
every $L$. Moreover, from Theorem~\ref{similarity} we 
know that $\mathcal{S}$ is unique up to a multiplicative constant. 

Obviously, the collection of matrices $\mathcal{S}$ forms a group. 
Since the Lorentz group is a Lie group, we assume this for the group 
generated by the matrices $\mathcal{S}$ as well. 
The most important Lie group property is 
that elements in a neighbourhood of the unit element can be written as 
exponentials. As for a Lorentz transformation $L$, 
this amounts to\footnote{It is interesting to note that 
the exponential function in this equation 
is surjective, \textit{i.e.}\ every element 
of the proper orthochronous Lorentz group $\mathbbm{L}^\uparrow_+$ can be 
written as an exponential, not only those in a neighbourhood of the unit 
element---see for instance~\cite{moskowitz}.}
\begin{equation}\label{Linfinitesimal}
L^\mu_{\hphantom{\mu}\nu} = \left( e^{\omega} \right)^\mu_{\hphantom{\mu}\nu} = 
\delta^\mu_{\hphantom{\mu}\nu} + \omega^\mu_{\hphantom{\mu}\nu} + \cdots,
\end{equation}\label{omega-anti}
where the group parameters $\omega^\mu_{\hphantom{\mu}\nu}$ fulfill
\begin{equation}\label{omega-antisym}
\omega_{\mu\nu} = -\omega_{\nu\mu}
\quad \mbox{with} \quad 
\omega_{\mu\nu} = g_{\mu\lambda}\, \omega^\lambda_{\hphantom{\lambda}\nu}.
\end{equation}
This follows straightforwardly from the definition of a Lorentz 
transformation, equation~(\ref{LgLg}). Therefore, for the $\mathcal{S}$ 
related to $L$ via equation~(\ref{Scond}), we make the ansatz
\begin{equation}\label{Sexp}
\mathcal{S}(\omega) = \exp \left( -\frac{i}{4}\, 
\omega_{\alpha\beta}\, \sigma^{\alpha\beta} \right)
\end{equation}
with group generators $\sigma^{\alpha\beta} = -\sigma^{\beta\alpha}$. 
According to the assumed Lie-group property, this ansatz is expected 
to be valid in a neighbourhood of the unit element.
Expanding equation~(\ref{Scond}) in $\omega$, the first order gives
\begin{equation}
\frac{i}{4}\, \omega_{\alpha\beta}\, 
[ \sigma^{\alpha\beta}, \gamma^\mu] = \omega^\mu_{\hphantom{\mu}\nu} \gamma^\nu 
\end{equation}
and, accordingly,
\begin{equation}\label{sigmacond}
\frac{i}{2}\, [ \sigma^{\alpha\beta}, \gamma^\mu] = 
g^{\mu\alpha} \gamma^\beta - g^{\mu\beta} \gamma^\alpha.
\end{equation}
The group generators $\sigma^{\alpha\beta}$ can be expanded in terms of 
products of Dirac matrices, \textit{cf.} Theorem~\ref{basis} ($N$ even). 
We explicitly exclude the unit matrix from this expansion because $\bone_4$ 
does not contribute to the right-hand side of equation~(\ref{sigmacond}).
To proceed further, we use the relation 
\begin{equation}\label{cac}
[AB,C] = A \{B,C\} - \{A,C\} B,
\end{equation}
which formulates the commutator by anticommutators. 
With this equation it is obvious that the generators are 
products of two Dirac matrices. Indeed, having dropped the unit matrix in 
the expansion of the generators, we find a unique solution of 
equation~(\ref{sigmacond}) given by
\begin{equation}
\sigma^{\alpha\beta} = i\gamma^\alpha \gamma^\beta = 
\frac{i}{2}\, [\gamma^\alpha, \gamma^\beta] \quad (\alpha \neq \beta).
\end{equation}
In addition, using the relation $\det \exp M = \exp \tr M$ for an arbitrary 
quadratic matrix $M$, we conclude 
\begin{equation}
\tr\, \sigma^{\alpha\beta} = 0 \quad  \Rightarrow \quad 
\det \mathcal{S}(\omega) = 1.
\end{equation}

Let us explore the structure of the group generated by the 
matrices of the form of equation~(\ref{Sexp}). In this context we need 
the chiral projectors of equation~(\ref{chiral projectors}).
In four space-time dimensions, the chiral matrix $\gamma_\mathrm{chir}$ of 
equation~(\ref{gammachir}) is denoted by $\gamma_5$ and given by
\begin{equation}
\gamma_5 = i \gamma^0 \gamma^1\gamma^2 \gamma^3.
\end{equation}
Specializing the findings of section~\ref{chiral pro} to $N=4$, the chiral 
projectors decompose $\mathbbm{C}^4$, the space on which the Dirac matrices 
act, as
\begin{equation}\label{V+V-}
\mathbbm{C}^4 = \mathcal{V}_+ \oplus \mathcal{V}_- 
\quad \mbox{with} \quad 
\mathcal{V}_\pm = \gamma_\pm \mathbbm{C}^4 
\quad \mbox{and} \quad 
\dim \mathcal{V}_\pm = 2.
\end{equation}
Since the $\sigma^{\alpha\beta}$ commute with $\gamma_5$, \textit{cf.}\ 
equation~(\ref{essential}), we are allowed to write
\begin{equation}\label{sigma+-}
\sigma^{\alpha\beta} = \gamma_+ \sigma^{\alpha\beta} \gamma_+ + 
\gamma_- \sigma^{\alpha\beta} \gamma_-.
\end{equation}
These `chiral' generators have the properties
\begin{equation}
\left( \gamma_+ \sigma^{\alpha\beta} \gamma_+ \right) 
\left( \gamma_- \sigma^{\alpha\beta} \gamma_- \right) = 
\left( \gamma_- \sigma^{\alpha\beta} \gamma_- \right) 
\left( \gamma_+ \sigma^{\alpha\beta} \gamma_+ \right) = 0
\end{equation}
and, consequently,
\begin{equation}
[ \gamma_+ \sigma^{\alpha\beta} \gamma_+, 
\gamma_- \sigma^{\alpha\beta} \gamma_- ] = 0.
\end{equation}
It is thus suggestive to define 
\begin{equation}\label{S+S-}
\mathcal{S}_+(\omega) = \exp \left( -\frac{i}{4}\, 
\omega_{\alpha\beta}\, \gamma_+ \sigma^{\alpha\beta} \gamma_+ \right)
\quad \mbox{and} \quad
\mathcal{S}_-(\omega) = \exp \left( -\frac{i}{4}\, 
\omega_{\alpha\beta}\, \gamma_- \sigma^{\alpha\beta} \gamma_- \right).
\end{equation}
In this way, we can write $\mathcal{S}(\omega)$ as the product
\begin{equation}\label{Sexp1}
\mathcal{S}(\omega) = \mathcal{S}_+(\omega) \mathcal{S}_-(\omega) =
\mathcal{S}_-(\omega) \mathcal{S}_+(\omega)
\end{equation}
such that 
\begin{equation}
\mathcal{S}_+(\omega) \mathcal{V}_+ = \mathcal{V}_+,
\quad
\left. \mathcal{S}_+(\omega) \right|_{\mathcal{V}_-} = 
\left. \mathrm{id} \right|_{\mathcal{V}_-}
\quad \mbox{and} \quad 
\mathcal{S}_-(\omega) \mathcal{V}_- = \mathcal{V}_-,
\quad
\left. \mathcal{S}_-(\omega) \right|_{\mathcal{V}_+} = 
\left. \mathrm{id} \right|_{\mathcal{V}_+}.
\end{equation}
Therefore, the matrices $\mathcal{S}(\omega)$ generate a 
reducible representation 
of the group we search for. Let us consider 
$\mathcal{S}_+(\omega)$ and $\mathcal{S}_-(\omega)$ separately. 
The respective groups act on two-dimensional spaces and, since
\begin{equation}
\tr \left( \gamma_- \sigma^{\alpha\beta} \gamma_- \right) = 
\tr \left( \gamma_+ \sigma^{\alpha\beta} \gamma_+ \right) = 0,
\end{equation}
they have six traceless generators. Therefore, 
\begin{equation}
\det \mathcal{S}_+(\omega) = \det \mathcal{S}_-(\omega) = 1.
\end{equation}
Note that the general linear group $GL(2, \mathbbm{C})$ of 
complex $2 \times 2$ matrices has eight generators, 
while its subgroup $SL(2, \mathbbm{C})$, consisting of all elements 
that have determinant~1, has six generators. Therefore, 
the matrices $\mathcal{S}_+(\omega)$ and $\mathcal{S}_-(\omega)$ generate 
each an irrep of the group 
$SL(2, \mathbbm{C})$.\footnote{As a side note, we mention that 
in $SL(2, \mathbbm{C})$ the exponential function reaches 
`almost' all elements, only a subset of measure zero cannot be obtained 
in this way. However, elements in this subset can be represented 
as a product of two exponentials~\cite{moskowitz}.\label{side note}}

Having thus found the group, it remains to investigate the relationship 
between $\mathcal{S}_+(\omega)$ and $\mathcal{S}_-(\omega)$. For this 
purpose we have to introduce the matrix $\beta$ defined via
\begin{equation}\label{beta}
\beta \gamma^\mu \beta^{-1} = \left( \gamma^\mu \right)^\dagger.
\end{equation}
The existence of $\beta$ is ensured by Theorem~\ref{pauli4}. 
It is easy to check that
\begin{equation}\label{beta5}
\beta^{-1} \gamma_5^\dagger \beta = -\gamma_5
\end{equation}
and 
\begin{equation}\label{bsigma}
\beta^{-1} \left( \sigma^{\alpha\beta} \right)^\dagger \beta = 
\sigma^{\alpha\beta}.
\end{equation}

Equations~(\ref{beta5}) and~(\ref{bsigma}) allow us to relate the 
generators of $\mathcal{S}_+(\omega)$ to those of $\mathcal{S}_-(\omega)$ by
\begin{equation}
\beta^{-1} \left( \gamma_+ \sigma^{\alpha\beta} \gamma_+ \right)^\dagger \beta = 
 \left( \gamma_- \sigma^{\alpha\beta} \gamma_- \right).
\end{equation}
Consequently,
\begin{equation}\label{+-}
\beta^{-1} \left( \mathcal{S}_+^{-1}(\omega) \right)^\dagger \beta = 
\mathcal{S}_-(\omega).
\end{equation}
In other words, the representation of $SL(2, \mathbbm{C})$ generated by 
the matrices $\mathcal{S}(\omega)$ can be conceived as the direct sum of 
two inequivalent irreps, namely 
the defining irrep and its complex-conjugate 
contragredient representation acting on $\mathcal{V}_-$ and 
$\mathcal{V}_+$, respectively, the two-dimensional subspaces of 
$\mathbbm{C}^4$ defined in equation~(\ref{V+V-}). 
For further material on $SL(2,\mathbbm{C})$ we refer the reader to 
appendix~\ref{sl2c}.

With very little effort one can make the connection from equation~(\ref{S+S-}) 
to $SL(2,\mathbbm{C})$ more transparent. The relations 
\begin{equation}
\sigma^{01} = i \sigma^{23} \gamma_5, \quad
\sigma^{02} = i \sigma^{31} \gamma_5, \quad
\sigma^{03} = i \sigma^{12} \gamma_5
\end{equation}
allow us to replace the $\sigma^{0j}$ ($j=1,2,3$) in the exponent of 
$\mathcal{S}(\omega)$ by $\sigma^{kl}$ ($k,l = 1,2,3$ and $k \neq l$).
Then, using $\gamma_5 \gamma_- = -\gamma_-$ and 
$\gamma_5 \gamma_+ = \gamma_+$,
we obtain
\begin{subequations}
\begin{eqnarray}
\frac{1}{2}\, \omega_{\alpha\beta}\, \gamma_- \sigma^{\alpha\beta} \gamma_- 
&=& 
\sum_{j=1}^3 z^{(-)}_j \gamma_- T_j \gamma_-,
\\
\frac{1}{2}\, \omega_{\alpha\beta}\, \gamma_+ \sigma^{\alpha\beta} \gamma_+ 
&=& 
\sum_{j=1}^3 z^{(+)}_j \gamma_+ T_j \gamma_+
\end{eqnarray}
\end{subequations}
with
\begin{subequations}
\begin{eqnarray}
&&
z^{(-)}_1 = \omega_{23} - i \omega_{01}, \quad
z^{(-)}_2 = \omega_{31} - i \omega_{02}, \quad
z^{(-)}_3 = \omega_{12} - i \omega_{03}, 
\\
&&
z^{(+)}_1 = \omega_{23} + i \omega_{01}, \quad
z^{(+)}_2 = \omega_{31} + i \omega_{02}, \quad
z^{(+)}_3 = \omega_{12} + i \omega_{03}
\end{eqnarray}
\end{subequations}
and
\begin{equation}
T_1 = \sigma^{23}, \quad
T_2 = \sigma^{31}, \quad
T_3 = \sigma^{12}.
\end{equation}
It is straightforward to check that these matrices behave like Pauli 
matrices:
\begin{equation}
T_j T_k = \delta_{jk} \bone_4 + i \sum_{l=1}^3 \varepsilon_{jkl} T_l.
\end{equation}
In this equation, $\varepsilon_{jkl}$ is the Levi--Civita 
symbol ($\varepsilon_{123} = 1$). Taking into account that the chiral 
projectors $\gamma_\pm$ commute with the $T_j$, 
we finally arrive at the form 
\begin{subequations}\label{S+-z}
\begin{eqnarray}
&&
\mathcal{S}_-(\omega) = \exp \left( -\frac{i}{2} 
\sum_{j=1}^3 z^{(-)}_j \gamma_- T_j \gamma_- \right) = 
\gamma_+ + \gamma_- \exp \left( -\frac{i}{2} 
\sum_{j=1}^3 z^{(-)}_j T_j \right) \gamma_-, \hphantom{X}
\\ &&
\mathcal{S}_+(\omega) = \exp \left( -\frac{i}{2} 
\sum_{j=1}^3 z^{(+)}_j \gamma_+ T_j \gamma_+ \right) = 
\gamma_- + \gamma_+ \exp \left( -\frac{i}{2} 
\sum_{j=1}^3 z^{(+)}_j T_j \right) \gamma_+
\end{eqnarray}
\end{subequations}
of the transformation matrices of equation~(\ref{S+S-}). 
This form of $\mathcal{S}_\pm(\omega)$ corresponds to the familiar 
parameterization of $SL(2,\mathbbm{C})$ matrices in the literature.
We emphasize that in the manipulations leading to equation~(\ref{S+-z}) 
we have exclusively used the anticommutation relations of 
equation~(\ref{anticom_4}).

At last, we examine the ambiguity in $\mathcal{S}$ for a given $L$. 
This ambiguity arises because equation~(\ref{Scond}) 
determines $\mathcal{S}$ only up to a multiplicative constant 
$c$---\textit{cf.}\ Theorem~\ref{similarity}.
By assuming traceless generators $\sigma^{\alpha\beta}$, we have enforced 
$\det \mathcal{S} = 1$, which imposes $c^4 = 1$. In addition, the Dirac 
algebra allows us to make the decomposition 
$\mathcal{S} = \left. \mathcal{S}_- \right|_{\mathcal{V}_-}
\oplus \left. \mathcal{S}_+ \right|_{\mathcal{V}_+}$ 
with the chiral generators of equation~(\ref{sigma+-}). 
Since the chiral generators are also traceless, 
the freedom in $c$ is reduced to $c = \pm 1$ according to 
$\det \left(\left. \mathcal{S}_- \right|_{\mathcal{V}_-}\right) = 
\det \left(\left. \mathcal{S}_+ \right|_{\mathcal{V}_+}\right) = 1$.
In other words, in the framework of 
$SL(2, \mathbbm{C})$, for every Lorentz transformation $L$ 
there are exactly two matrices $\mathcal{S}$ that solve 
equation~(\ref{Scond}) and the two solutions differ only by a sign.

\section{Conjugate spinors and Lorentz invariance}
\label{conjugate spinors}
The matrix $\beta$, defined in equation~(\ref{beta}),  
is also required for the definition of the conjugate Dirac spinor
\begin{equation}\label{conjugate spinor}
\bar\psi = \psi^\dagger \beta.
\end{equation}
Conjugate spinors are important for the formation of fermion bilinears with 
definite transformation properties under Lorentz transformations.
For the time being we leave out a possible $x$-dependence of spinors 
because it is not relevant in the present context. 
The transformation property of the conjugate spinor 
under Lorentz transformations or, more precisely, under 
$SL(2,\mathbbm{C})$ derives from that of the spinor, equation~(\ref{Spsi}):
\begin{equation}\label{Scjsp}
\bar\psi \to \psi^\dagger \mathcal{S}^\dagger(\omega) \beta = 
\bar\psi \mathcal{S}^{-1}(\omega),
\end{equation}
where in the second step we have applied 
\begin{equation}
\beta^{-1} \mathcal{S}^\dagger(\omega) \beta = \mathcal{S}^{-1}(\omega),
\end{equation}
which follows from equation~(\ref{bsigma}).

Suppose we have two Dirac spinors $\psi_1$ and $\psi_2$. Then 
making use of equations~(\ref{Scond}) and~(\ref{Scjsp}), we 
find that the bilinears
\begin{equation}\label{bilinears}
\bar\psi_1 \psi_2, \quad 
\bar\psi_1 \gamma^\mu \psi_2, \quad 
\bar\psi_1 \sigma^{\mu\nu} \psi_2
\end{equation}
transform as scalar, vector, and antisymmetric tensor, respectively, 
under Lorentz transformations. The corresponding symmetric tensor is 
proportional to $g^{\mu\nu}$ according to equation~(\ref{anticom_4}) and 
is thus invariant under Lorentz transformations. 
One can also define pseudoscalars, pseudovectors and pseudotensors as
\begin{equation}\label{pseudobilinears}
\bar\psi_1 \gamma_5 \psi_2, \quad 
\bar\psi_1 \gamma^\mu \gamma_5 \psi_2, \quad 
\bar\psi_1 \sigma^{\mu\nu} \gamma_5 \psi_2,
\end{equation}
respectively. Since $[\gamma_5, \mathcal{S}(\omega)] = 0$, 
these bilinears transform under Lorentz transformations just 
as the bilinears of equation~(\ref{bilinears}).

However, under parity, bilinears and `pseudo-bilinears' behave differently.
Denoting parity as a member of the Lorentz group by $P$, it is given by
\begin{equation}
P = \diag \left( 1,-1,-1,-1 \right).
\end{equation}
Formally, the matrix $P$ is identical with the metric tensor, but its totally 
different physical content makes the usage of an extra notation meaningful. 
Because of $\det P = -1$ and $P^0_{\hphantom{0}0} =1$, parity is an element 
of the orthochronous Lorentz group, \textit{i.e.}\ 
$P \in \mathbbm{L}^\uparrow$ but $P \not\in \mathbbm{L}_+$. In order to obtain 
the counterpart of parity in the space of Dirac spinors we have to employ 
equation~(\ref{Scond}). Denoting this matrix by $\mathcal{S}_P$,  
it is obviously given by
\begin{equation}
\mathcal{S}_P = \gamma^0
\end{equation}
and a Dirac spinor and its conjugate spinor transform as
\begin{equation}
\psi \to \gamma^0 \psi \quad \mbox{and} \quad 
\bar \psi \to \psi^\dagger \left( \gamma^0 \right)^\dagger \beta = 
\bar \psi \beta^{-1} \left( \gamma^0 \right)^\dagger \beta = 
\bar \psi \gamma^0,
\end{equation}
respectively. 
Actually, according to Theorem~\ref{similarity}, $\mathcal{S}_P$ is 
determined only up to a multiplicative constant $c$. However, with 
the requirement that parity is a selfinverse operation on Dirac spinors, 
one obtains $c = \pm 1$.

With the sign function defined in equation~(\ref{sign}), 
the bilinears of equation~(\ref{bilinears}) behave as 
\begin{equation}
\bar\psi_1 \psi_2 \to \bar\psi_1 \psi_2, \quad 
\bar\psi_1 \gamma^\mu \psi_2 \to 
\epsilon(\mu) \bar\psi_1 \gamma^\mu \psi_2, \quad 
\bar\psi_1 \sigma^{\mu\nu} \psi_2 \to 
\epsilon(\mu) \epsilon(\nu) \bar\psi_1 \sigma^{\mu\nu} \psi_2
\end{equation}
under parity. However, since $\gamma_5$ anticommutes with $\gamma^0$, in 
the transformation of the `pseudo-bilinears' one gets 
an additional minus sign:
\begin{eqnarray}
&& 
\bar\psi_1 \gamma_5 \psi_2 \to -\bar\psi_1 \gamma_5 \psi_2, \quad 
\bar\psi_1 \gamma^\mu \gamma_5 \psi_2 \to 
-\epsilon(\mu) \bar\psi_1 \gamma^\mu \gamma_5 \psi_2, 
\nonumber \\ &&
\bar\psi_1 \sigma^{\mu\nu} \gamma_5 \psi_2 \to 
-\epsilon(\mu) \epsilon(\nu) \bar\psi_1 \sigma^{\mu\nu} \gamma_5 \psi_2.
\end{eqnarray}

In order to formulate fermionic Lagrangians, one needs conjugate 
spinors too, but in addition a property of $\beta$ we have not yet 
elaborated on. Taking the hermitian 
conjugate of equation~(\ref{beta}) and shifting the $\beta$ matrices to 
the right-hand side, we obtain
\begin{equation}
\left( \gamma^\mu \right)^\dagger = 
\beta^\dagger \gamma^\mu \left( \beta^\dagger \right)^{-1}. 
\end{equation}
Therefore, according to Theorem~\ref{similarity},  
$\beta^\dagger = b \beta$ with $b \in \mathbbm{C}$. The hermitian conjugate 
of this relation leads to $bb^* = |b|^2 = 1$. 
Writing $b = e^{2i\sigma}$, we obtain
\begin{equation}\label{beta1}
\left( e^{i\sigma} \beta \right)^\dagger = e^{i\sigma} \beta.
\end{equation}

It is a fundamental property of the action that it is real. Therefore, 
the Lagrangian (density) $\mathcal{L}$ has to be real as well or, 
in the context of particle physics, hermitian---possibly 
up to partial integrations if a term in $\mathcal{L}$ contains derivatives. 
Let us consider the simplest term in a fermionic Lagrangian, namely a 
mass term $\bar\psi \psi$ where $\psi$ is a Dirac spinor.
Then, 
\begin{equation}
\left( e^{i\sigma} \bar\psi \psi \right)^\dagger = 
e^{-i\sigma} \psi^\dagger \beta^\dagger \psi = 
e^{i\sigma} \psi^\dagger \beta \psi = 
e^{i\sigma} \bar\psi \psi.
\end{equation}
This suggests to absorb $e^{i\sigma}$ into $\beta$ and 
to use without loss of generality the convention
\begin{equation}\label{beta-conv}
\beta^\dagger = \beta.
\end{equation}
In this way one gets rid of the awkward phase factor $e^{i\sigma}$ in the 
Lagrangian.

If the Dirac matrices obey equation~(\ref{special}), then, 
using the sign function $\epsilon(\mu)$ defined in equation~(\ref{sign}),
we find
\begin{equation}\label{beta=gamma0}
\left( \gamma^\mu \right)^\dagger = \epsilon(\mu) \gamma^\mu
\quad \Rightarrow \quad \beta = \gamma^0.
\end{equation}
Most textbooks, but not~\cite{jauch,bailin}, 
take advantage of this identification. 
However, since this identification already requires a special 
basis of the Dirac matrices and, moreover, the physics behind $\beta$ and 
$\gamma^0$ is different, we keep these matrices apart in 
the present manuscript.

\section{Plane-wave solutions of the Dirac equation}
\label{plane-wave}
Plane-wave solutions of the Dirac equation are solutions of the form 
$\psi(x) = u\, e^{-i p \cdot x}$ and $\psi(x) = v\, e^{i p \cdot x}$, 
where $u$ and $v$ are $x$-independent Dirac spinors and $p$ is the 4-momentum. 
Plugging these ans\"atze into the Dirac equation~(\ref{dirac-eq})
leads to 
\begin{subequations}\label{plane}
\begin{eqnarray}
\left( \slashed{p} - m \right) u &=& 0, \label{u} \\
\left( \slashed{p} + m \right) v &=& 0. \label{v}
\end{eqnarray}
\end{subequations}
Since $d=4$, there are four linearly independent plane-wave solutions.
Clearly, $p$ alone is not sufficient to characterize all four solutions.
It turns out that a second 4-vector is needed, the so-called spin
vector $s$, see for instance~\cite{bjorken}. 
The two 4-vectors are characterized by\footnote{We confine ourselves 
to massive fermions.}
\begin{equation}\label{ps}
p^2 = m^2 \neq 0, \quad s^2 = -1, \quad s \cdot p = 0.
\end{equation}
The meaning of the conditions for $s$ will shortly become clear. 
Equation~(\ref{plane}) suggests to define the projectors
\begin{equation}
\Lambda_+(p) = \frac{1}{2m} \left( \slashed{p} + m \right), \quad
\Lambda_-(p) = \frac{1}{2m} \left( -\slashed{p} + m \right),
\end{equation}
such that 
\begin{equation}
\Lambda_+(p)\, u = u, \quad \Lambda_+(p)\, v = 0, \quad 
\Lambda_-(p)\, u = 0, \quad \Lambda_-(p)\, v = v.
\end{equation}
The matrices $\Lambda_\pm(p)$ are projectors in the sense of 
Lemma~\ref{lemma} of appendix~\ref{selfinverse}
because $\left( \slashed{p}/m \right)^2 = \bone_4$. They fulfill 
\begin{equation}
\left( \Lambda_+(p) \right) ^2 = \Lambda_+(p), \quad
\left( \Lambda_-(p) \right) ^2 = \Lambda_-(p), \quad
\Lambda_+(p) \Lambda_-(p) = \Lambda_-(p) \Lambda_+(p) = 0,
\end{equation}
and 
\begin{equation}
\Lambda_+(p) + \Lambda_-(p) = \bone_4,
\end{equation}
but they will not be hermitian in general. A second set of projectors, 
associated with the spin vector, is given by
\begin{equation}
\Sigma_+(s) = \frac{1}{2} \left( \bone_4 + \gamma_5 \slashed{s} \right), \quad
\Sigma_-(s) = \frac{1}{2} \left( \bone_4 - \gamma_5 \slashed{s} \right).
\end{equation}
They fulfill
\begin{equation}
\Sigma_+(s) + \Sigma_-(s) = \bone_4
\end{equation}
and 
\begin{equation}
\left( \Sigma_+(s) \right)^2 = \Sigma_+(s), \quad
\left( \Sigma_-(s) \right)^2 = \Sigma_-(s), \quad
\Sigma_+(s) \Sigma_-(s) = \Sigma_-(s) \Sigma_+(s) = 0
\end{equation}
due to 
\begin{equation}
\left( \gamma_5 \slashed{s} \right)^2 = -s^2 \bone_4 = \bone_4.
\end{equation}
This explains the condition $s^2 = -1$ and guarantees that the $\Sigma_\pm(s)$ 
are projectors in the sense of Lemma~\ref{lemma}. 
Moreover, because of $s \cdot p = 0$, 
\begin{equation}
\{ \slashed{p}, \slashed{s} \} = 2 (s \cdot p) \bone_4= 0
\quad \Rightarrow \quad
\slashed{p} \left( \gamma_5 \slashed{s} \right) = 
\left( \gamma_5 \slashed{s} \right) \slashed{p} 
\end{equation}
and 
\begin{equation}\label{comm}
[ \Lambda_+(p), \Sigma_+(s) ] = 
[ \Lambda_-(p), \Sigma_+(s) ] = 
[ \Lambda_+(p), \Sigma_-(s) ] = 
[ \Lambda_-(p), \Sigma_-(s) ] = 0.
\end{equation}

Since the four projectors commute with each other, Lemma~\ref{lemma} 
ensures that we can find a \emph{common} basis of eigenvectors. We denote 
them by $u(p,\pm s)$ and $v(p,\pm s)$. In the table below we list 
their eigenvalues with respect to the projectors:
\begin{equation}\label{evp}
\begin{array}{c|cccc}
& \Lambda_+(p) & \Lambda_-(p) & \Sigma_+(s) & \Sigma_-(s) \\ \hline
u(p,+s)  & 1 & 0 & 1 & 0 \\
u(p,-s) & 1 & 0 & 0 & 1 \\
v(p,+s)  & 0 & 1 & 1 & 0 \\
v(p,-s) & 0 & 1 & 0 & 1
\end{array}
\end{equation}
We have thus characterized, for a given 4-momentum $p$ and a spin vector $s$, 
the four linearly 
independent plane-wave solutions in a completely basis-independent way, 
\textit{i.e.}\ without an explicit realization of the Dirac matrices.

Finally, we discuss the action of the projectors 
on the conjugate spinors. From 
the definition of the matrix $\beta$, equation~(\ref{beta}), 
we immediately obtain 
\begin{equation}
\Lambda_\pm^\dagger(p) \beta = \beta \Lambda_\pm(p).
\end{equation}
Moreover, using equation~(\ref{beta5}), we derive 
\begin{equation}
\left( \gamma_5 \slashed{s} \right)^\dagger \beta = 
\beta \beta^{-1} \slashed{s}^\dagger \gamma_5^\dagger \beta =
-\beta \slashed{s} \gamma_5 =
\beta \gamma_5 \slashed{s}.
\end{equation}
Therefore, 
\begin{equation}
\Sigma_\pm^\dagger(s) \beta = \beta\, \Sigma_\pm(s).
\end{equation}
As a consequence, the action of the projectors on the conjugate spinors 
are again given by the table of equation~(\ref{evp}), \textit{i.e.}
\begin{equation}
\bar u(p,s) \Lambda_+(p) = 
\bar u(p,s) \Sigma_+(p) = \bar u(p,s), \quad
\bar u(p,s) \Lambda_-(p) = 
\bar u(p,s) \Sigma_-(p) = 0, \;\mbox{etc.}
\end{equation}

\section{Charge conjugation}
The \emph{charge-conjugation matrix} $C$ is defined by 
\begin{equation}\label{C}
C^{-1} \gamma^\mu C = -\left( \gamma^\mu \right)^T
\end{equation}
in particle physics, see also section~\ref{application}.
The existence of $C$ is guaranteed by Theorem~\ref{pauli4} because the matrices 
$-\left( \gamma^\mu \right)^T$ fulfill the anticommutation relations of 
equation~(\ref{anticom_4}) if the $\gamma^\mu$ do so.
By transposition of equation~(\ref{C}) we obtain
\begin{equation}
\left( C^T \right)^{-1} \gamma^\mu\, C^T = -\left( \gamma^\mu \right)^T.
\end{equation}
Consequently, according to Theorem~\ref{similarity}, we have the relation
$C^T = a C$ or, after transposition, $C = a C^T$. Therefore, $a^2 = 1$ and 
$C$ is either symmetric or antisymmetric:
\begin{equation}\label{C1}
C^T = \epsilon C \quad \mbox{with} \quad \epsilon^2 = 1.
\end{equation}
This equation and equations~(\ref{anticom_4}) and (\ref{C}) allow us 
to make the list~\cite{jauch}
\begin{subequations}\label{symmetry}
\begin{eqnarray}
C^T &=& \hphantom{-}\epsilon C, \\
\left( \gamma^\mu C \right)^T &=& -\epsilon \gamma^\mu C, \\
\left( \gamma^\mu \gamma^\nu C \right)^T &=& 
-\epsilon \gamma^\mu \gamma^\nu C
\;\; (\mu < \nu), \\
\left( \gamma^\mu \gamma^\nu \gamma^\lambda C \right)^T &=& 
\hphantom{-}\epsilon \gamma^\mu \gamma^\nu \gamma^\lambda C
\;\; (\mu < \nu < \lambda), \\
\left( \gamma^0 \gamma^1 \gamma^2 \gamma^3 C \right)^T &=& 
\hphantom{-}\epsilon \gamma^0 \gamma^1 \gamma^2 \gamma^3 C.
\end{eqnarray}
\end{subequations}
These matrices are all linearly independent, \textit{cf.} 
Theorem~\ref{Neven} ($N$ even). In equation~(\ref{symmetry}) 
10 matrices obtain $-\epsilon$ and 6 matrices 
obtain $+\epsilon$ upon transposition. Since in the set of $4 \times 4$ 
matrices there are 10 linearly independent symmetric matrices
and 6 linearly independent antisymmetric matrices, 
the only consistent sign in equation~(\ref{C1}) is $\epsilon = -1$ or
\begin{equation}\label{Canti}
C^T = -C.
\end{equation}

There is a relation between the matrices $\beta$ and $C$. Because when 
applying equation~(\ref{C}) and, subsequently, equation~(\ref{beta}), 
one obtains
\begin{equation}\label{Cbg}
\left( C \beta^T \right)^{-1} \gamma^\mu \left( C \beta^T \right) = 
-\left( \gamma^\mu \right)^*
\quad \Rightarrow \quad
\left( \left( C \beta^T \right) \left( C \beta^T \right)^* \right)^{-1} 
\gamma^\mu \left( C \beta^T \right) \left( C \beta^T \right)^* = \gamma^\mu.
\end{equation}
Therefore $\left( C \beta^T \right) \left( C \beta^T \right)^*$ 
commutes with all $\gamma^\mu$. Hence, 
\begin{equation}\label{cbd}
\left( C \beta^T \right) \left( C \beta^T \right)^* = 
\left( C \beta^T \right)^* \left( C \beta^T \right) = 
d\, \bone_4
\quad \mbox{with} \quad d \in \mathbbm{R}.
\end{equation}

It is an interesting observation that $d$ is basis-independent~\cite{jauch}. 
This can be derived in the following way.
When we perform a similarity transformation from one irrep 
of the Dirac algebra to another, the matrices $\beta$ and 
$C$ transform as well:
\begin{equation}\label{sbc}
{\gamma'}^\mu = S^{-1} \gamma^\mu S \quad \Rightarrow \quad
C' = S^{-1} C (S^{-1})^T, \quad \beta' = S^\dagger \beta S.
\end{equation}
Using these relations we obtain
\begin{equation}
C' {\beta'}^T = S^{-1} \left( C \beta^T \right) S^*,
\end{equation}
whence we conclude 
\begin{equation}
\left( C' {\beta'}^T \right) \left( C' {\beta'}^T \right)^* = 
S^{-1} \left( C \beta^T \right) \left( C \beta^T \right)^* S = 
S^{-1} d\, \bone_4 S = d\, \bone_4.
\end{equation}
This proves the above statement. 
Note that $\beta$ and $C$ are only fixed up to multiplicative 
constants---\textit{cf.}\ Theorem~\ref{similarity}---whose 
phases drop out in
$\left( C \beta^T \right) \left( C \beta^T \right)^*$. Therefore, 
it is the sign of $d$ which is independent of 
any conventions for $\beta$ and $C$ and any 
basis of the Dirac matrices. 

One can obtain the sign of $d$ by using a special basis of the Dirac 
matrices~\cite{jauch}. Assuming hermiticity properties, 
equation~(\ref{special}), is already special enough. 
Because then Theorem~\ref{unitary} implies that, apart from multiplicative 
real constant, $C$ is unitary and $\beta$ can be identified with 
$\gamma^0$---\textit{cf.} equation~(\ref{beta=gamma0}).
With this input we compute
\begin{equation}
\left( C \beta^T \right) \left( C \beta^T \right)^* = 
C \left( \gamma^0 \right)^T C^* \left( \gamma^0 \right)^\dagger = 
-C \left( \gamma^0 \right)^T C^\dagger \gamma^0 = 
\left( \gamma^0 \right)^2 = \bone_4.
\end{equation}
Therefore, we conclude that $d > 0$. 

Since $\beta$ can be identified with $\gamma^0$ in any basis 
obeying equation~(\ref{special}), there must be some 
general relation linking the two matrices. To investigate this we consider 
the product $\beta \gamma^0$. Note that with the convention 
$\beta^\dagger = \beta$ this matrix is hermitian:
\begin{equation}
\left( \beta \gamma^0 \right)^\dagger = \beta \gamma^0.
\end{equation}
Under similarity transformations, \textit{cf.}\ equation~(\ref{sbc}), 
$\beta \gamma^0$ behaves as
\begin{equation}
\beta' {\gamma'}^0 = S^\dagger \left( \beta \gamma^0 \right) S.
\end{equation}
Let us again take refuge to equation~(\ref{special}). From  
$\beta = \gamma^0$ we infer $\beta \gamma^0 = \bone_4$ and 
$\beta' {\gamma'}^0 = S^\dagger S$. Taking into account that with the convention 
$\beta^\dagger = \beta$ we still can choose the overall sign of $\beta$, we 
conclude that, in general, $\beta \gamma^0$ is either positive or negative 
definite.

Can one prove $d > 0$ in a completely basis-independent way? The answer is 
affirmative if we prove first in a basis-independent way that 
$\beta \gamma^0$ is a definite matrix~\cite{sexl}.
\begin{theorem}
With the phase convention $\beta^\dagger = \beta$ the matrix 
$\beta \gamma^0$ is either positive or negative definite.
\end{theorem}
\textbf{Proof:} Using the sign function $\epsilon(\mu)$ defined in 
equation~(\ref{sign}), one readily obtains
\begin{equation}\label{beta-gamma0}
\left( \beta \gamma^0 \right) \gamma^\mu \left( \beta \gamma^0 \right)^{-1} = 
\epsilon(\mu) \left( \gamma^\mu \right)^\dagger \equiv 
{\gamma'}^\mu.
\end{equation}
We have thus two sets of Dirac matrices, $\{ \gamma^\mu \}$ and 
$\{ {\gamma'}^\mu \}$. On the one hand, 
equation~(\ref{beta-gamma0}) states explicitly the similarity 
transformation connecting the two sets. On the other hand, 
equation~(\ref{S}), which is essential for the proof of 
Theorem~\ref{irrep}, constitutes a general formula, in terms of the 
matrices $\gamma^\mu$ and ${\gamma'}^\mu$,
for obtaining the matrix $S$ of the similarity transformation  
${\gamma'}^\mu = S^{-1} \gamma^\mu S$, provided we can find a matrix $F$ 
such that $S$ is invertible. Explicitly, this $S$ reads 
\begin{eqnarray}
S &=& F + \sum_\mu \left( \gamma^\mu \right)^{-1} F {\gamma'}^\mu + 
\sum_{\mu < \nu} \left( \gamma^\mu \gamma^\nu \right)^{-1} 
F \left( {\gamma'}^\mu {\gamma'}^\nu \right) 
\nonumber \\ && +
\sum_{\mu < \nu < \lambda} \left( \gamma^\mu \gamma^\nu \gamma^\lambda \right)^{-1} 
F \left( {\gamma'}^\mu {\gamma'}^\nu {\gamma'}^\lambda \right) + 
\left( \gamma^0 \gamma^1 \gamma^2 \gamma^3 \right)^{-1} 
F \left( {\gamma'}^0 {\gamma'}^1 {\gamma'}^2 {\gamma'}^3 \right).
\end{eqnarray}
Taking advantage of equation~(\ref{sign}), we obtain
\begin{equation}
S = F + \sum_\mu \gamma^\mu F \left( \gamma^\mu \right)^\dagger + 
\sum_{\mu < \nu} \gamma^\nu \gamma^\mu 
F \left( \gamma^\mu \right)^\dagger \left( \gamma^\nu \right)^\dagger + \cdots 
\end{equation}
and finally 
\begin{eqnarray}
S &=& F + \sum_\mu \gamma^\mu F \left( \gamma^\mu \right)^\dagger + 
\sum_{\mu < \nu} \left( \gamma^\nu \gamma^\mu \right) 
F \left( \gamma^\nu \gamma^\mu \right)^\dagger 
\nonumber \\ && +
\sum_{\mu < \nu < \lambda} \left( \gamma^\lambda \gamma^\nu \gamma^\mu \right)
F \left( \gamma^\lambda \gamma^\nu \gamma^\mu \right)^\dagger + 
\left( \gamma^3 \gamma^2 \gamma^1 \gamma^0 \right) 
F \left( \gamma^3 \gamma^2 \gamma^1 \gamma^0 \right)^\dagger. 
\end{eqnarray}
We still have to freedom to choose $F$. Obviously, taking $F = \bone_4$,
the thus obtained $S$ is positive definite and, therefore, invertible.
According to Theorem~\ref{similarity}, there is a constant $c$ such that 
$S = c \left( \beta \gamma^0 \right)^{-1}$. Since both matrices are 
hermitian, $c$ must be real and $\beta \gamma^0 = cS^{-1}$
is a definite matrix. Q.E.D.
\begin{theorem}
The constant $d$ defined in equation~(\ref{cbd}) is positive.
\end{theorem}
\textbf{Proof:}
We denote by $\langle x | y \rangle = x^\dagger y$ the usual scalar product on 
$\mathbbm{C}^4$. Let $x$ be an arbitrary but non-zero vector in 
$\mathbbm{C}^4$. Then we consider~\cite{sexl}
\begin{eqnarray}
d\,\langle x | \beta \gamma^0 x \rangle &=& 
\langle x | \beta \gamma^0 \left( C \beta^T \right) 
\left( C \beta^T \right)^* x \rangle 
\nonumber \\ &=& -
\langle x | \beta C \left( \gamma^0 \right)^T \beta^T 
\left( C \beta^T \right)^* x \rangle 
\nonumber \\ &=& -
\langle C^\dagger \beta^\dagger x | \left( \beta \gamma^0 \right)^T 
\left( C \beta^T \right)^* x \rangle 
\nonumber \\ &=& 
\langle \left( C \beta^T \right)^* x | \left( \beta \gamma^0 \right)^* 
\left( C \beta^T \right)^* x \rangle 
\nonumber \\ &=& 
\left( \langle \left( C \beta^T \right) x^* | \left( \beta \gamma^0 \right) 
\left( C \beta^T \right) x^* \rangle \right)^* .
\end{eqnarray}
We have thus obtained 
\begin{equation}
d\,\langle x | \beta \gamma^0 x \rangle =
\left( \langle y | \beta \gamma^0 y \rangle \right)^*
\quad \mbox{with} \quad y = \left( C \beta^T \right) x^*.
\end{equation}
Due to the fact that $\beta \gamma^0$ is hermitian and definite,  
the two scalar products in this equation are real and have the same sign, 
which entails $d > 0$. Q.E.D.

Now we come to the actual topic of this section.
The operation of charge conjugation on a 4-spinor is defined by 
\begin{equation}\label{cc}
\psi^c = C \bar\psi^T = C \beta^T \psi^*.
\end{equation}
The asterisk indicates complex conjugation.\footnote{If $\psi$ is 
a fermion field-operator, then the asterisk denotes the hermitian conjugate 
of each of its four components.} 
Applying charge conjugation a second time, we obtain
\begin{equation}
\left( \psi^c \right)^c = C \beta^T \left( C \beta^T \psi^* \right)^* = 
\left( C \beta^T \right) \left( C \beta^T \right)^* \psi.
\end{equation}
Above we have just proven that the matrix 
$\left( C \beta^T \right) \left( C \beta^T \right)^*$ is 
proportional to the unit matrix with a positive constant $d$.
Therefore, $\sqrt{d}$ is real. Adopting 
the convention that $C/\sqrt{d}$ defines a new matrix $C$,  
we find 
\begin{equation}\label{cbcb}
\left( C \beta^T \right) \left( C \beta^T \right)^* = \bone_4
\quad \Rightarrow \quad 
\left( \psi^c \right)^c = \psi.
\end{equation}
In this way, $d > 0$ enables us to define 
charge conjugation as a selfinverse operation on Dirac spinors.
Henceforth we will stick to this definition.

To discuss the behaviour of the projectors defined in 
section~\ref{plane-wave} under charge conjugation, we need the relation
\begin{equation}\label{C5}
C^{-1} \gamma_5 C = \gamma_5^T,
\end{equation}
which is obtained by  by a straightforward computation.
Under a similarity transformation with $C \beta^T$ the 
projectors $\Lambda_\pm(p)$ and $\Sigma_\pm(s)$ have the following 
properties:
\begin{subequations}\label{ccP}
\begin{eqnarray}
\left( C \beta^T \right)^{-1} \Lambda_+(p) \left( C \beta^T \right) = 
\Lambda_-^*(p), 
&&
\left( C \beta^T \right)^{-1} \Lambda_-(p) \left( C \beta^T \right) = 
\Lambda_+^*(p),
\\
\left( C \beta^T \right)^{-1} \Sigma_+(s) \left( C \beta^T \right) = 
\Sigma_+^*(s), 
&&
\left( C \beta^T \right)^{-1} \Sigma_-(s) \left( C \beta^T \right) = 
\Sigma_-^*(s).
\end{eqnarray}
\end{subequations}
\textbf{Proof:} It suffices to consider $\slashed{p}$ and 
$\left( \gamma_5 \slashed{s} \right)$. In the first case we 
simply apply equation~(\ref{Cbg}) and obtain
\[
\left( C \beta^T \right)^{-1} \slashed{p} \left( C \beta^T \right) = 
-\slashed{p}^*.
\]
In the second case, after 
combining equations~(\ref{beta5}) and~(\ref{C5}) leading to
\[
\left( C \beta^T \right)^{-1} \gamma_5 \left( C \beta^T \right) = 
-\gamma_5^*,
\]
we readily obtain
\[
\left( C \beta^T \right)^{-1} \left( \gamma_5 \slashed{s} \right) 
\left( C \beta^T \right) = 
\left( \gamma_5 \slashed{s} \right)^*.
\]
Q.E.D.

Equation~(\ref{ccP}) tells us that the operation of charge conjugation 
defined in equation~(\ref{cc}) transforms $u$-spinors into $v$-spinors 
and vice versa, while the spin vector $s$ remains the same:
\begin{subequations}
\begin{eqnarray}
&& 
\Lambda_-(p) u^c(p,s) = \Sigma_+(s) u^c(p,s) = u^c(p,s), \\
&&
\Lambda_+(p) v^c(p,s) = \Sigma_+(s) v^c(p,s) = v^c(p,s), \\ 
&&
\Lambda_+(p) u^c(p,s) = \Sigma_-(s) u^c(p,s) = 0, \\
&&
\Lambda_-(p) v^c(p,s) = \Sigma_-(s) v^c(p,s) = 0.
\end{eqnarray}
\end{subequations}

We conclude this section with a discussion of the behaviour of the 
fermion bilinears that were introduced in section~\ref{conjugate spinors} 
under charge conjugation. The result is 
\begin{subequations}\label{Cbil}
\begin{alignat}{6}
&\bar\psi_1 \psi_2 & 
&\to & 
&\bar\psi_2 \psi_1, & 
\qquad
&\bar\psi_1 \gamma_5 \psi_2 &
&\to &
&\bar\psi_2 \gamma_5 \psi_1, 
\\
&\bar\psi_1 \gamma^\mu \psi_2 &
&\to & \;-
&\bar\psi_2 \gamma^\mu \psi_1, &
&\bar\psi_1 \gamma^\mu \gamma_5 \psi_2 &
&\to &
&\bar\psi_2 \gamma^\mu \gamma_5 \psi_1, 
\label{VA} \\
&\bar\psi_1 \sigma^{\mu\nu} \psi_2 &
&\to & -
&\bar\psi_2 \sigma^{\mu\nu} \psi_1, &
&\bar\psi_1 \sigma^{\mu\nu} \gamma_5 \psi_2 &
&\to & \;-
&\bar\psi_2 \sigma^{\mu\nu} \gamma_5 \psi_1.
\end{alignat}
\end{subequations}
This can straightforwardly be derived by using the two relations
\begin{equation}\label{Cb1}
\left( C \beta^T \right)^\dagger \beta C = -\bone_4
\end{equation}
and
\begin{equation}
\psi_1^T \Gamma^T \bar\psi_2^T = -\bar\psi_2 \Gamma \psi_1,
\end{equation}
where $\Gamma$ is a matrix in Dirac space.
The first relation is obtained by transposing 
the first formula in equation~(\ref{cbcb}). The second one takes into 
account the Grassmann nature of fermion field variables, \textit{i.e.}\ 
$\left( \psi_1 \right)_a \left( \bar\psi_2 \right)_b = -
\left( \bar\psi_2 \right)_b \left( \psi_1 \right)_a$, where $a$ and $b$ 
are Dirac indices. As an example of a proof, 
we pick out one relation of equation~(\ref{Cbil}), 
namely the axial-vector current in equation~(\ref{VA}): 
\begin{eqnarray}
\bar\psi_1 \gamma^\mu \gamma_5 \psi_2  & \to &
\psi_1^T \left( C \beta^T \right)^\dagger \beta \gamma^\mu \gamma_5 
\left( C \beta^T \right) \psi_2^* 
\nonumber \\ && = 
\psi_1^T \left(\left( C \beta^T \right)^\dagger \beta C \right)
\left( C^{-1} \gamma^\mu \gamma_5 C \right) \beta^T \psi_2^* 
\nonumber \\ && = 
\psi_1^T \left( \gamma^\mu \right)^T \gamma_5 ^T \beta^T \psi_2^* = 
\psi_1^T \left( \gamma_5 \gamma^\mu \right)^T \bar\psi_2^T
\nonumber \\ && = 
-\bar\psi_2 \gamma_5 \gamma^\mu \psi_1 = 
\bar\psi_2 \gamma^\mu \gamma_5 \psi_1.
\end{eqnarray}
In the third line we have used equations~(\ref{C}) and~(\ref{C5}).
As a bonus, we realize by dropping $\gamma_5$ in this computation 
that the vector current gets a minus sign under charge conjugation.

\section{Quantization of the Dirac field}
The Lagrangian of the free Dirac field is given by
\begin{equation}\label{lagrangian}
\mathcal{L} = \bar\psi\,i \gamma^\mu \partial_\mu \psi - m\, \bar\psi \psi.
\end{equation}
Clearly, the derivative with respect to $\bar\psi_a$ leads to the Dirac 
equation---\textit{cf.}\ equation~(\ref{dirac-eq}). While there is 
no conjugate momentum for $\bar\psi_a$, the conjugate momentum for 
$\psi_a$ is obtained by
\begin{equation}
\frac{\partial \mathcal{L}}{\partial \dot \psi_a} = 
i \left( \bar\psi \gamma^0 \right)_a.
\end{equation}
This leads to the canonical-quantization conditions 
\begin{subequations}\label{quant}
\begin{eqnarray}
&&
\{ \psi_a(x), \psi_b(y) \}_{x^0 = y^0} =  
\{ \psi^\dagger_a(x), \psi^\dagger_b(y) \}_{x^0 = y^0} = 0,
\label{quant-a} \\[2mm]
&&
\{ \psi_a(x), \left( \bar\psi(y) \gamma^0 \right)_b \}_{x^0 = y^0} = 
\delta_{ab}\, \delta(\vec x - \vec y\,).
\label{quant-b}
\end{eqnarray}
\end{subequations}
The Dirac field $\psi(x)$ can be expanded in terms of the 4-spinors of the 
plane-wave solutions discussed in section~\ref{plane-wave}, which in turn 
leads to an expansion in terms of annihilation operators $b(p,s)$ 
and creation operators $d^\dagger(p,s)$:
\begin{equation}\label{psi}
\psi(x) = \int 
\frac{\dd^3 p}{\sqrt{(2\pi)^3 2E_p}} \sum_{\epsilon = \pm 1} 
\left( u(p,\epsilon s)\, b(p,\epsilon s)\, e^{-ip \cdot x} + 
v(p,\epsilon s)\, d^\dagger(p,\epsilon s)\, e^{ip \cdot x} \right).
\end{equation}
The factor $\left( (2\pi)^3 2E_p \right)^{-1/2}$ in the integral is a 
very useful convention as will be seen in the following.
As for the spin vector $s$, we assume that there is some prescription that
assigns an $s$ fulfilling equation~(\ref{ps}) to every 4-momentum 
$p$.\footnote{For instance, equation~(\ref{s}) with a fixed unit vector 
$\hat s$ could be regarded as such a prescription.}

In the following we will derive the anticommutation relations of 
the creation and annihilation operators from those of equation~(\ref{quant}). 
Moreover, we will stipulate a suitable normalization for them. Such a 
normalization condition can be chosen freely because those operators 
appear in the products 
$u(p,\epsilon s)\, b(p,\epsilon s)$ and 
$v(p,\epsilon s)\, d^\dagger(p,\epsilon s)$. 
However, with fixed normalization conditions for $b(p,\epsilon s)$ and 
$d(p,\epsilon s)$, those of $u(p,\epsilon s)$ and $v(p,\epsilon s)$ are 
determined as well.

In order to execute this plan, we have to project out of $\psi(x)$ the 
annihilation and creation operators. In this context we need the relations 
\begin{equation}\label{L0L0}
\Lambda_+(p) \gamma^0 \Lambda_-(\tilde p) = 
\Lambda_-(p) \gamma^0 \Lambda_+(\tilde p) = 0
\quad \mbox{with} \quad
\tilde p = \left( \begin{array}{r}  p^0 \\ -\vec p 
\end{array} \right)
\end{equation}
and
\begin{equation}\label{L0LL}
\Lambda_+(p) \gamma^0 \Lambda_+(p) = \frac{E_p}{m}\, \Lambda_+(p),
\quad
\Lambda_-(p) \gamma^0 \Lambda_-(p) = -\frac{E_p}{m}\, \Lambda_-(p).
\end{equation}
\textbf{Proof:}
Equation~(\ref{L0L0}) follows trivially from the anticommutation relations 
of the Dirac matrices and 
\[
\Lambda_+(\tilde p) \Lambda_-(\tilde p) = 
\Lambda_-(\tilde p) \Lambda_+(\tilde p) = 0.
\]
The second relation of equation~(\ref{L0LL}) is obtained from the first 
one by $m \to -m$. It remains to consider
\begin{eqnarray*}
\Lambda_+(p) \gamma^0 \Lambda_+(p) &=& 
\gamma^0 \Lambda_+(\tilde p) \Lambda_+(p) \\
&=& \frac{1}{4m^2}\, \gamma^0 
\left( E_p \gamma^0 + \vec p \cdot \vec \gamma + m \right)
\left( E_p \gamma^0 - \vec p \cdot \vec \gamma + m \right). 
\end{eqnarray*}
Expanding the parentheses and simplification leads to
\[
\frac{1}{2m^2}
\left( E_p^2 \gamma^0 + m E_p - E_p\, \vec p \cdot \vec \gamma \right) = 
\frac{E_p}{m}\, \Lambda_+(p).
\]
Q.E.D.

To proceed further, we define 
\begin{subequations}\label{BDdaggerdef}
\begin{eqnarray}
B(p,s) &=& 
\frac{1}{\sqrt{(2\pi)^3 2E_p}} \int \dd^3x\, \bar u(p,s) \gamma ^0 \psi(x)
e^{i p \cdot x},
\label{Bdef} \\
D^\dagger (p,s) &=& 
\frac{1}{\sqrt{(2\pi)^3 2E_p}} \int \dd^3x\, \bar v(p,s) \gamma ^0 \psi(x)
e^{-i p \cdot x}.
\label{Ddaggerdef}
\end{eqnarray}
\end{subequations}
Performing the integrals, we obtain 
\begin{subequations}\label{BDdagger}
\begin{eqnarray}
B(p,s) &=& \sum_{\epsilon = \pm 1} 
\frac{1}{2E_p} \bar u(p,s) \gamma^0 u(p, \epsilon s)
\, b(p, \epsilon s), 
\label{B} \\
D^\dagger (p,s) &=& \sum_{\epsilon = \pm 1} 
\frac{1}{2E_p} \bar v(p,s) \gamma^0 v(p, \epsilon s)
\, d^\dagger(p, \epsilon s).
\label{Ddagger}
\end{eqnarray}
\end{subequations}
Because of 
\begin{eqnarray}
\bar u(p,s) \gamma^0 v(p',s')\, \delta (\vec p + \vecp{p}) &=&
\bar u(p,s) \gamma^0 v(\tilde p, \tilde s)\, \delta (\vec p + \vecp{p}) 
\nonumber \\ &=& 
\bar u(p,s) \Lambda_+ (p) \gamma^0 \Lambda_- (\tilde p) 
v(\tilde p, \tilde s)\, \delta (\vec p + \vecp{p}) \,=\, 0
\label{=0}
\end{eqnarray}
according to equation~(\ref{L0L0}), where $\tilde s$ a the spin vector 
associated with $\tilde p$, the $v$-term of $\psi$ does not 
contribute to $B(p,s)$. The same applies to the $u$-term and $D(p,s)$. 
To treat the term in the sum of equation~(\ref{B}), 
we make use of the first relation of equation~(\ref{L0LL}) and obtain 
\begin{equation}
\frac{1}{2E_p} \bar u(p,s) \gamma^0 u(p, \epsilon s) =
\frac{1}{2E_p} \bar u(p,s) \Lambda_+(p) \gamma^0 \Lambda_+(p) u(p, \epsilon s) =
\frac{1}{2m} \bar u(p,s) u(p,s) \delta_{1 \epsilon}.
\end{equation}
With the second relation of equation~(\ref{L0LL}) we perform the 
analogous steps in equation~(\ref{Ddagger}).
Finally, we arrive at the intermediate result
\begin{subequations}\label{uub-vvd}
\begin{eqnarray}
B(p,s) &=& 
\hphantom{-}\frac{1}{2m}\, \bar u(p,s) u(p,s)\, b(p,s),
\label{uub} \\
D^\dagger (p,s) &=& 
-\frac{1}{2m}\, \bar v(p,s) v(p,s)\, d^\dagger(p,s).
\label{vvd}
\end{eqnarray}
\end{subequations}

Now we invoke the quantization conditions. 
Equation~(\ref{quant-a}), which has anticommutators that are zero, 
immediately translates into zero anticommutators among the operators 
of equation~(\ref{BDdaggerdef}) and, because of equation~(\ref{uub-vvd}), 
into those of the annihilation and creation operators:
\begin{equation}\label{bd=01}
\{ b(p,s), b(p',s') \} = \{ d^\dagger(p,s), d^\dagger(p',s') \} = 
\{ b(p,s), d^\dagger(p',s') \} = 0.
\end{equation}
Moving on to equation~(\ref{quant-b}), we first consider 
$\{ B(p,s), D(p',s') \}$. Since we need $D(p',s')$, 
we have to take the hermitian 
conjugate of the integrand of equation~(\ref{Ddaggerdef}) 
leading to the expression
\begin{equation}\label{auxiliary}
\left( \bar v(p',s') \gamma^0 \psi(y) \right)^\dagger = 
\psi^\dagger(y) \left( \gamma^0 \right)^\dagger \beta^\dagger v(p',s') =
\bar\psi(y) \beta^{-1} \left( \gamma^0 \right)^\dagger \beta 
v(p',s') = \bar\psi(y) \gamma^0 v(p',s').
\end{equation}
Note that we have used the convention $\beta^\dagger = \beta$.
Then the evaluation of the anticommutator is 
straightforward:\footnote{Note that $B(p,s)$ and $D(p',s')$ are 
time-independent. Therefore, in the following computation we are allowed 
to choose equal times in the anticommutator of the field operators.}
\begin{eqnarray}
\lefteqn{ \{ B(p,s), D(p',s') \} = 
\frac{1}{(2\pi)^3 \sqrt{2E_p\, 2E_{p'}}} }
\nonumber \\ && \times 
\int \dd^3x \left( \bar u(p,s) \gamma ^0 \right)_a e^{i p \cdot x} 
\int \dd^3y\, v_b(p',s') e^{i p' \cdot y}
\{ \psi_a(x), \left( \bar\psi(y)\gamma ^0 \right)_b \}_{x^0 = y^0}
\nonumber \\
&& = \frac{1}{\sqrt{2E_p\, 2E_{p'}}} \bar u(p,s) \gamma^0 v(p',s')
\delta (\vec p + \vecp{p} )\, e^{i(E_p + E_{p'}) x^0}
\nonumber \\ &&
= \frac{1}{2E_p} \bar u(p,s) \gamma^0 v(\tilde p,\tilde s) 
\delta (\vec p + \vecp{p} )\, e^{2iE_p x^0} = 0,
\end{eqnarray}
\textit{cf.}\ equation~(\ref{=0}). Therefore, according to 
equation~(\ref{uub-vvd}) the relation 
\begin{equation}\label{bd=02}
\{ b(p,s), d(p',s') \} = 0
\end{equation}
ensues. We can take the hermitian conjugate of the anticommutators of 
equations~(\ref{bd=01}) and~(\ref{bd=02}) which produces another four 
anticommutators that are zero.

It remains to consider 
$\{ B(p,s), B^\dagger(p',s') \}$ and 
$\{ D(p,s), D^\dagger(p',s') \}$.
We concentrate on the first anticommutator, the second one is treated
analogously. In order to obtain $B^\dagger(p',s')$, we have to proceed in 
the same way as in equation~(\ref{auxiliary}). Then we compute
\begin{eqnarray}
\lefteqn{ \{ B(p,s), B^\dagger(p',s') \} = 
\frac{1}{(2\pi)^3 \sqrt{2E_p\, 2E_{p'}}} }
\nonumber \\ && \times 
\int \dd^3x \left( \bar u(p,s) \gamma ^0 \right)_a e^{i p \cdot x} 
\int \dd^3y\, u_b(p',s') e^{-i p' \cdot y}
\{ \psi_a(x), \left( \bar\psi(y)\gamma ^0 \right)_b \}_{x^0 = y^0}
\nonumber \\ && =
\frac{1}{\sqrt{2E_p\, 2E_{p'}}}\,\bar u(p,s) \gamma ^0 u(p',s') 
\delta ( \vec p - \vecp{p} ) 
\nonumber \\ && = 
\frac{1}{2E_p} \times \frac{E_p}{m} \bar u(p,s) u(p,\epsilon s)\, 
\delta ( \vec p - \vecp{p} ).
\label{BB}
\end{eqnarray}
In the last step we have applied the first relation in equation~(\ref{L0LL}).
Moreover, we have taken into account that $s' = \epsilon s$ for $p' = p$. 
Finally, equation~(\ref{uub}) allows us relate the anticommutator 
$\{ b(p,s), b^\dagger(p',s') \}$ to equation~(\ref{BB}):
\begin{equation}\label{bb}
\frac{1}{2m}\, \bar u(p,s) u(p,\epsilon s)\, \delta ( \vec p - \vecp{p} ) =
\frac{1}{4m^2} \bar u(p,s) u(p,s)\,\bar u(p',s') u(p',s')\,
\{ b(p,s), b^\dagger(p',s') \}.
\end{equation}
This equation shows that $\{ b(p,s), b^\dagger(p',s') \}$ must be 
proportional to the delta function and to $\delta_{1\epsilon}$. 
The same applies to 
$\{ d(p,s), d^\dagger(p',s') \}$. Using the simplest normalization 
conditions for these anticommutators, \textit{i.e.}\
\begin{equation}\label{bbdd}
\{ b(p,s), b^\dagger(p',s') \} = \{ d(p,s), d^\dagger(p',s') \} = 
\delta_{ss'} \delta ( \vec p - \vecp{p} ),
\end{equation}
the normalization conditions for the spinors $u$ and $v$ have to be
\begin{equation}\label{normalization}
\bar u(p,s) u(p,s) = 2m \quad \mbox{and} \quad \bar v(p,s) v(p,s) = -2m,
\end{equation}
respectively, in order to satisfy equation~(\ref{bb}).
Note that the minus sign in the normalization of $v$ follows from the minus 
in the second relation of equation~(\ref{L0LL}) because this leads to a 
minus sign in the analogue of equation~(\ref{BB}):
\begin{equation}
\{ D(p,s), D^\dagger(p',s') \} = -\frac{1}{2m}
\bar v(p,s) v(p,\epsilon s)\, \delta ( \vec p - \vecp{p} ).
\end{equation}
We have thus derived all possible anticommutator relations invoking 
the annihilation and creation operators and fixed their normalization. The 
latter step fixes at the same time 
the normalization of the spinors $u$ and $v$.

Note that in equation~(\ref{bbdd}) we have written the Kronecker symbol 
$\delta_{ss'}$ referring the spin vectors, because this is commonly used 
in text books. Actually, conforming to the notation above, 
we should have replaced $s'$ by $\epsilon s$ in the argument 
of the operators and $\delta_{ss'}$ by $\delta_{1\epsilon}$. 
However, this would be a highly unusual notation.

\section{Time reversal}
There are several arguments leading to the conclusion that time reversal 
is realized as an antiunitary operator. Here we argue from the point of 
view of the Schr\"odinger equation.

Time reversal $\mt$ is a norm-preserving transformation acting on the 
Hilbert space of states such that the time evolution is reversed. 
Denoting a state by $| \alpha, t \rangle$, 
where $\alpha$ is a collection of quantum numbers 
fully specifying the state, then
\begin{equation}
\mt | \alpha, t \rangle = | \alpha_T, -t \rangle,
\end{equation}
where $\alpha_T$ describes the time-reversed state. 
For instance, a one-particle state with a definite momentum $\vec p$ 
has $\vec p \in \alpha$ and $-\vec p \in \alpha_T$ because time 
reversal effects $\vec p \to -\vec p$.
Time evolution is governed by the Schr\"odinger equation
\begin{equation}\label{SE0}
i \frac{\partial}{\partial t} | \alpha, t \rangle = 
H | \alpha, t \rangle.
\end{equation}
Rewriting the Schr\"odinger equation in terms of the time-reversed state 
$| \alpha_T, -t \rangle$, we obtain
\begin{equation}
\mt i \mt^{-1} \frac{\partial}{\partial t} | \alpha_T, -t \rangle = 
\mt H \mt^{-1} | \alpha_T, -t \rangle
\end{equation}
or 
\begin{equation}\label{SE1}
-\mt i \mt^{-1} \frac{\partial}{\partial t'} | \alpha_T, t' \rangle = 
\mt H \mt^{-1} | \alpha_T, t' \rangle
\quad \mbox{with} \quad t' = -t.
\end{equation}
The requirement that a theory is invariant under time reversal means that 
$| \alpha_T, -t \rangle$ obeys the Schr\"odinger equation if 
$| \alpha, t \rangle$ does so. One possibility of equation~(\ref{SE1}) to 
concur with the Schr\"odinger equation is obtained by the requirement
$\mt i \mt^{-1} = i$ and $\mt H \mt^{-1} = -H$. 
However, this is to be discarded because it would mean that for every 
state with definite energy $E$ there would be time-reversed state with 
energy $-E$. Evidently this is wrong because, for instance, a free 
one-particle state always has a positive energy without a counterpart 
with negative energy. Therefore, we are lead to
\begin{equation}
\mt i \mt^{-1} = -i \quad \mbox{and} \quad \mt H \mt^{-1} = H.
\end{equation}
This means that $\mt$ is an antiunitary operator. 

Let us now consider the example of plane waves in 
quantum mechanics. According to the discussion above, 
time reversal has to be realized as the mapping~\cite{bailin}
\begin{equation}\label{phit}
\varphi(t,\vec x) = e^{-i(Et - \vec p \cdot \vec x)} \mapsto
\varphi_T(t,\vec x) = \varphi^*(-t,\vec x) = 
e^{-i(Et + \vec p \cdot \vec x)}. 
\end{equation}
Indeed, this makes sense because the state with momentum $\vec p$ is 
transformed into a state with momentum $-\vec p$, which is just 
what one expects from time reversal. Moreover, in equation~(\ref{phit}) 
we can replace plane waves by general normalizable wave functions $\varphi$. 
Then $\varphi(t,\vec x) \mapsto \varphi^*(-t,\vec x)$ is a mapping that 
reverses time und preserves probabilities, \textit{i.e.}\ it has the 
desired properties.

In the following it is useful to distinguish between Dirac wave functions, 
denoted by $\Psi$, and Dirac fields, denoted by $\psi$. The discussions 
above suggest that, if $\Psi(x)$ is a solution of the Dirac equation, 
then the time-reversed solution of the Dirac equation 
has the form 
\begin{equation}\label{TPsi}
\Psi_T(x) = T \Psi^*(x')
\quad \mbox{with} \quad 
x' = \left( \begin{array}{c} -x^0 \\ \vec x \end{array} \right)
\end{equation}
and a $4 \times 4$ matrix $T$ yet to be determined.
The requirement 
\begin{equation}
\left( i \gamma^\mu \partial_\mu - m \right) \Psi_T(x) = 0
\end{equation}
leads, after complex conjugation, to
\begin{equation}
\left( -i \left( \gamma^\mu \right)^* \partial_\mu - m \right) 
T^* \Psi(x') = 
T^* \left( i\epsilon(\mu) \left( T^{-1} \gamma^\mu\, T \right)^* \partial'_\mu - 
m \right) \Psi(x') = 0
\end{equation}
with
\begin{equation}
\partial'_\mu = \frac{\partial}{\partial {x'}^\mu}.
\end{equation}
To recover the Dirac equation, we have to request
\begin{equation}\label{Tg}
T^{-1} \gamma^\mu T = \epsilon(\mu) \left( \gamma^\mu \right)^*,
\end{equation}
where $\epsilon(\mu)$ is the sign function defined in equation~(\ref{sign}).
The existence of $T$ is guaranteed by Theorem~\ref{similarity}. Indeed, 
with equation~(\ref{Cbg}) we readily find 
\begin{equation}
T = \left( C \beta^T \right) \left( \gamma^0 \gamma_5 \right)^*.
\end{equation}
Since we have already fixed the normalization of $C \beta^T$ in 
equation~(\ref{cbcb}), the only freedom in $T$ is a phase factor that 
is, however, irrelevant in the following.

The matrix $T$ has the interesting property~\cite{jauch}
\begin{equation}\label{TT*}
T T^* = T^* T = -\bone_4.
\end{equation}
\textbf{Proof:} 
We rewrite $TT^*$ as 
\begin{eqnarray*}
T T^* &=& 
\left( C \beta^T \right) \left( \gamma^0 \gamma_5 \right)^* 
\left( C \beta^T \right)^* \left( \gamma^0 \gamma_5 \right) 
\\ &=& 
\left( C \beta^T \right) \left( C \beta^T \right)^* 
\left( \left( C \beta^T \right)^{-1} \gamma^0 \gamma_5 \left( C \beta^T \right) 
\right)^* \gamma^0 \gamma_5.
\end{eqnarray*}
Taking into account equation~(\ref{cbcb}), $TT^*$ simplifies to 
\[
TT^* = 
\left( \left( C \beta^T \right)^{-1} \gamma^0 \gamma_5 \left( C \beta^T \right) 
\right)^* \gamma^0 \gamma_5 = 
\gamma^0 \gamma_5 \gamma^0 \gamma_5 = -\bone_4.
\]
In the second to last step we have applied equation~(\ref{Cbg}).
Q.E.D.

Equation~(\ref{TT*}) tells us that, when applying time reversal two times 
to a Dirac wave function, we obtain
\begin{equation}\label{TTP}
\Psi(x) \to T \Psi^*(x') \to TT^* \Psi(x) = -\Psi(x).
\end{equation}

Now we move on to Dirac fields. Consistency with equation~(\ref{TPsi})
demands that 
\begin{equation}\label{Tpsi}
\mathcal{T}^{-1} \psi(x) \mathcal{T} = T^* \psi(x').
\end{equation}
Let us dwell a little on this point. We may rewrite equation~(\ref{psi}) as 
\begin{equation}
\psi(x) = \int \dd^3 p \sum_{\epsilon = \pm 1} \left(
\Psi_u(p,\epsilon s;x)\, b(p,\epsilon s) + 
\Psi_v(p,\epsilon s;x)\, d^\dagger(p,\epsilon s) \right).
\end{equation}
The functions $\Psi_u(p,\epsilon s;x)$ and $\Psi_v(p,\epsilon s;x)$ are 
the plane-wave solutions of the Dirac equation ocurring in $\psi(x)$. 
In order to shorten the notation, we define
\begin{equation}
\mathcal{D} = i \gamma^\mu \partial_\mu - m. 
\end{equation}
Therefore,
\begin{subequations}
\begin{eqnarray}
\mathcal{D} \Psi_u(p,\epsilon s;x) = 0 & \Rightarrow &
\mathcal{D} T \Psi^*_u(p,\epsilon s;x') = 0, \\
\mathcal{D} \Psi_v(p,\epsilon s;x) = 0 & \Rightarrow &
\mathcal{D} T \Psi^*_v(p,\epsilon s;x') = 0.
\end{eqnarray}
\end{subequations}
Applying equation~(\ref{Tpsi}) to $\mathcal{D} \psi(x) = 0$, 
we are lead to
\begin{eqnarray}
0 &=& 
\mt^{-1} \mathcal{D} \psi(x) \mt \nonumber \\
&=& \mathcal{D}^* T^* \psi(x') \nonumber \\
&=&
\int \dd^3 p \sum_{\epsilon = \pm 1} \left(
\left( \mathcal{D} T \Psi^*_u(p,\epsilon s;x') \right)^* b(p,\epsilon s) + 
\left( \mathcal{D} T \Psi^*_v(p,\epsilon s;x') \right)^* 
d^\dagger(p,\epsilon s) \right).
\end{eqnarray}
This demonstrates consistency of equation~(\ref{Tpsi}) 
with equation~(\ref{TPsi}).
Note that, applying $\mt$ two times on the quantized field, 
results in~\cite{jauch}
\begin{equation}
\mathcal{T}^{-2} \psi(x) \mathcal{T}^2 = T^*T \psi(x) = -\psi(x),
\end{equation}
just as in the case of Dirac wave functions---see equation~(\ref{TTP}).

As the last topic of this section we consider the transformation property of 
the current 
\begin{equation}
j^\mu(x) = \bar\psi(x) \gamma^\mu \psi(x) 
\end{equation}
under time reversal. For this purpose we need to know how the 
conjugate Dirac field transforms. This derives from equation~(\ref{Tpsi}):
\begin{equation}
\mathcal{T}^{-1} \bar\psi(x) \mathcal{T} = 
\left( \mathcal{T}^{-1} \psi(x) \mathcal{T} \right)^\dagger \beta^* = 
\bar\psi(x') \beta^{-1} T^T \beta^*.
\end{equation}
To proceed further we prove the relation
\begin{equation}\label{bTbT}
\beta^{-1} T^T \beta^* T^* = \bone_4.
\end{equation}
\textbf{Proof:} With the help of equation~(\ref{Cb1}) we compute 
\begin{eqnarray*}
T^\dagger \beta T &=& 
\left( \gamma^0 \gamma_5 \right)^T \left( C \beta^T \right)^\dagger 
\beta \left( C \beta^T \right) \left( \gamma^0 \gamma_5 \right)^* 
\\ &=&
\left( \gamma^0 \gamma_5 \right)^T 
\left( \left( C \beta^T \right)^\dagger \beta C \right) \beta^T 
\left( \gamma^0 \gamma_5 \right)^* 
\\ &=&
- \left( \gamma_5^\dagger \left( \gamma^0 \right)^\dagger 
\beta\, \gamma^0 \gamma_5 \right)^*
\\ &=&
\left( \beta \gamma_5 \gamma^0 \gamma^0 \gamma_5 \right)^* \;=\; \beta^*.
\end{eqnarray*}
(Note that we stick to the phase convention $\beta^\dagger = \beta$.)
This result is equivalent to 
$\left( \beta^{-1} T^T \beta^* T^* \right) ^* = \bone_4$. 
Therefore, equation~(\ref{bTbT}) is correct. Q.E.D.
\\[2mm]
Now the treatment of the current $j^\mu$ is straightforward:
\begin{eqnarray}
\mt^{-1} j^\mu(x) \mt &=&  
\left( \mt^{-1} \bar\psi(x) \mt \right) 
\left( \mt^{-1} \gamma^\mu \mt \right) 
\left( \mt^{-1} \psi(x) \mt \right) 
\nonumber \\
&=&
\bar\psi(x') \beta^{-1} T^T \beta^* \left( \gamma^\mu \right)^* T^* \psi(x')
\nonumber \\
&=&
\bar\psi(x') \beta^{-1} T^T \beta^* T^* 
\left( T^{-1} \gamma^\mu T \right)^* \psi(x').
\end{eqnarray}
Finally, with equations~(\ref{Tg}) and~(\ref{bTbT}) we arrive at the 
result
\begin{equation}
\mt^{-1} j^\mu(x) \mt = \epsilon(\mu) j^\mu(x').
\end{equation}
This is in accordance with the transformation property 
\begin{equation}
\mt^{-1} A^\mu(x) \mt = \epsilon(\mu) A^\mu(x')
\end{equation}
of the electromagnetic vector potential that can be gathered from the fact 
that the magnetic field changes sign under time reversal wheras the 
electric field does not. Conceiving $j^\mu$ as an electric current density, 
this accordance means 
\begin{equation}
\mt^{-1} \left( j^\mu(x) A_\mu(x) \right) \mt = j^\mu(x') A_\mu(x').
\end{equation}
As a consequence, QED is invariant under time reversal.

\section{Expectation value of the spin operator}
Before we come to the actual topic of this section, we derive 
the angular-momentum tensor density for the Lagrangian of 
equation~(\ref{lagrangian}), because from this density we can read off the spin 
operator. 

The general relationship between a symmetry of the action 
$\int \dd^4x\, \mathcal{L}$ and 
a conservation law is provided by Noether's Theorem. Suppose the action is 
invariant under 
\begin{equation}\label{dpsi}
\delta \psi_a = \lambda\, \Delta \psi_a,
\end{equation}
where $\lambda$ is an infinitesimal parameter and $\Delta \psi_a$ 
is some deformation of the field.\footnote{In this notation we specialize 
already to a single fermion field, having in mind the application of 
Noether's Theorem to the Lagrangian of equation~(\ref{lagrangian}).} 
Invariance of the action is equivalent to 
the Lagrangian transforming into a 4-divergence, \textit{i.e.}\
\begin{equation}\label{J}
\delta \mathcal{L} = \lambda\, \partial_\mu \mathcal{J}^\mu
\end{equation}
under the transformation of equation~(\ref{dpsi}).
Then Noether's Theorem furnishes us with the conserved current\footnote{The 
second term in the conserved current will be zero in our case because  
equation~(\ref{lagrangian}) does not contain derivatives of 
$\bar\psi_a$.}
\begin{equation}\label{conscurr}
j^\mu = \frac{\partial \mathcal{L}}{\partial(\partial_\mu \psi_a)} 
\Delta \psi_a + 
\frac{\partial \mathcal{L}}{\partial(\partial_\mu \bar\psi_a)} 
\Delta \bar\psi_a - \mathcal{J}^\mu.
\end{equation}

Now we apply Noether's Theorem to the Lagrangian of 
equation~(\ref{lagrangian}) which is invariant under 
Lorentz transformations of the fermion field, equation~(\ref{Spsi}). 
The infinitesimal parameter is assumed to be
$\lambda = \omega_{\alpha\beta} = -\omega_{\beta\alpha}$, 
with fixed indices $\alpha$ and $\beta$ ($\alpha \neq \beta$).
Equations~(\ref{Spsi}), (\ref{Linfinitesimal}) and~(\ref{Sexp}) readily 
lead to
\begin{equation}
\Delta \psi = \left( -\frac{i}{2}\, \sigma^{\alpha\beta} + 
x^\alpha \partial^\beta - x^\beta \partial^\alpha \right) \psi 
\end{equation}
and the corresponding $\Delta \bar\psi$. Plugging these two expressions 
into the Lagrangian of equation~(\ref{lagrangian}) and 
applying equation~(\ref{sigmacond}), 
the current of equation~(\ref{J}) is computed as
\begin{equation}
\mathcal{J}^\mu = 
\left( x^\alpha g^{\mu\beta} - x^\beta g^{\mu\alpha} \right) \mathcal{L}.
\end{equation}
Therefore, with equation~(\ref{conscurr}), 
we obtain for every pair of indices $\alpha \neq \beta$ 
the conserved current~\cite{bjorken}
\begin{equation}\label{Mmab}
M^{\mu\alpha\beta} = \bar\psi \gamma^\mu 
\left( \frac{1}{2} \sigma^{\alpha\beta} + i \left( 
x^\alpha \partial^\beta - x^\beta \partial^\alpha \right) \right) \psi - 
\left( x^\alpha g^{\mu\beta} - x^\beta g^{\mu\alpha} \right) \mathcal{L}.
\end{equation}
Since $M^{\mu\alpha\beta} = -M^{\mu\beta\alpha}$, there are six independent 
conserved currents, with associated conservation laws formulated as
\begin{equation}
\partial_\mu M^{\mu\alpha\beta} = 0 
\quad \Rightarrow \quad \frac{\dd}{\dd t} \int \dd^3 x\, M^{0\alpha\beta} = 0
\;\;\forall\, \alpha \neq \beta.
\end{equation}
With $\mu = 0$ and specializing to spatial indices $\alpha = k$, 
$\beta = l$, the second term in equation~(\ref{Mmab}) 
drops out and we read off the angular momentum 
\begin{equation}\label{J_n}
J_n = \frac{1}{2}\sum_{k,l=1}^3 \varepsilon_{nkl} \int \dd^3x\,M^{0kl} = 
\frac{1}{2}\sum_{k,l=1}^3 \varepsilon_{nkl} \int \dd^3x\, \bar\psi \gamma^0 
\left( \frac{1}{2} \sigma^{kl} + 
i\left( x^k \partial^l - x^l \partial^k \right)\right) \psi.
\end{equation}

The angular momentum is the sum of spin and orbital angular momentum.
Obviously, 
\begin{equation}
S_j = \frac{i}{4} \sum_{k,l=1}^3 \varepsilon_{jkl} \int \dd^3 x\, :\bar\psi(x) 
\gamma^0 \gamma^k \gamma^l \psi(x):
\end{equation}
is the spin operator. To arrive at this form, we have used 
$\sigma^{kl} = i\gamma^k \gamma^l$ ($k \neq l$). Moreover, to emphasize that 
$S_j$ is now considered as an operator, we have indicated normal ordering 
by putting the integrand between colons.
The position of the index $j$ (up or down) 
is irrelevant since $\vec S$ is a vector in space only.
For the computation of the 
expectation value of $\vec S$ the following relation is essential:
\begin{equation}\label{sigma-gamma}
\frac{i}{2} \sum_{k,l=1}^3 \varepsilon_{jkl}\, \Sigma_+(s) 
\gamma^0 \gamma^k \gamma^l \Sigma_+(s)  =  s^j\, \Sigma_+(s).
\end{equation}
\textbf{Proof:} 
We shift the left spin projector through the three Dirac matrices:
\[
\frac{i}{2} \sum_{k,l=1}^3 \varepsilon_{jkl}\, 
\Sigma_+(s) \gamma^0 \gamma^k \gamma^l \Sigma_+(s)  = 
\frac{i}{2} \sum_{k,l=1}^3 \varepsilon_{jkl}\, 
\gamma^0 \gamma^k \gamma^l \left( \Sigma_-(s) - \gamma_5 s^j \gamma^j 
\right) \Sigma_+(s).
\]
There is no summation over $j$ on the right-hand side because $j$ is 
a fixed index.
The justification for this intermediate result is that $\gamma_5 \gamma^n$ 
anticommutes with $\gamma^0 \gamma^k \gamma^l$ for $n=0,k,l$, 
while it commutes for $n=j$. Since $\Sigma_-(s)\Sigma_+(s) = 0$, we obtain
\[
\frac{i}{2} \sum_{k,l=1}^3 
\varepsilon_{jkl}\, \Sigma_+(s) \gamma^0 \gamma^k \gamma^l\, \Sigma_+(s)  = 
\frac{i}{2} \sum_{k,l=1}^3
\varepsilon_{jkl}\, \gamma^0 \gamma^k \gamma^l \gamma^j \gamma_5\, 
s^j\, \Sigma_+(s).
\]
Finally, we observe that
\[
\frac{i}{2} \sum_{k,l=1}^3
\varepsilon_{jkl}\, \gamma^0 \gamma^k \gamma^l \gamma^j = \gamma_5.
\]
Because of $\gamma_5^2 = \bone_4$, we arrive at the announced result.
Q.E.D.

One-particle and one-antiparticle states are created by 
\begin{equation}
| p,s \rangle_P = b^\dagger (p,s) |0\rangle 
\quad \mbox{and} \quad
| p,s \rangle_A = d^\dagger (p,s) |0\rangle,
\end{equation}
respectively. Before we compute the expectation values of $\vec S$ in 
these one-particle states, a word is in order concerning the spin vector $s$.
It is evident that, for a given 4-momentum $p$, the 4-vector 
\begin{equation}\label{s}
s \equiv s(p,\hat s) = \frac{1}{\sqrt{1 - (\hat s \cdot \vec p/E_p)^2}}
\left( \begin{array}{c} \hat s \cdot \vec p/E_p \\ \hat s
\end{array} \right) 
\quad \mbox{with} \quad \left( \hat s  \right)^2 = 1,
\end{equation}
fulfills the defining relations of $s$ given by equation~(\ref{ps}).  
In this equation, $\hat s$ is an arbitrary unit vector in three dimensions 
whose direction is independent of $\vec p$.
An alternative but useful form of the normalization factor 
in equation~(\ref{s}) is 
\begin{equation}\label{alternative}
\sqrt{1 - (\hat s \cdot \vec p/E_p)^2} = 
\frac{\sqrt{m^2 + {\vec p_\bot}^{\,2}}}{E_p}
\quad \mbox{with} \quad 
\vec p_\bot = \vec p - (\vec p \cdot \hat s) \hat s.
\end{equation}
In the following 
we will tacitly assume that $s$ and $s'$ have the same $\hat s$.
In this way we guarantee that $s' \to s$ for $\vecp{p} \to \vec p$. 
With equation~(\ref{sigma-gamma}) it is straightforward to compute 
the expectation value of $\vec S$. For definiteness we display the 
computation for the particle:
\begin{eqnarray*}
_P\langle p,s|S_j\,|p',s' \rangle_P &=& 
\frac{i}{4} \sum_{k,l=1}^3
\varepsilon_{jkl}\, \bar u(p,s) 
\gamma^0 \gamma^k \gamma^l u(p,s) \times \frac{1}{2E_p} \times 
\delta(\vec p - \vecp{p}) \\
&=&
\frac{i}{4} \sum_{k,l=1}^3 \varepsilon_{jkl}\, \bar u(p,s) \Sigma_+(s) 
\gamma^0 \gamma^k \gamma^l \Sigma_+(s) u(p,s) \times \frac{1}{2E_p} \times 
\delta(\vec p - \vecp{p}) \\
&=& 
\frac{1}{2}\, s^j  \bar u(p,s) u(p,s) \times \frac{1}{2E_p} \times 
\delta(\vec p - \vecp{p}).
\end{eqnarray*}
The computation for the antiparticle proceeds analogously, we only have 
to take into consideration that because of the normal ordering we have 
$-\bar v(p,s) v(p,s)$ instead of $\bar u(p,s) u(p,s)$. 
With the normalization of the spinors, equation~(\ref{normalization}), and 
the form of $s$, equation~(\ref{s}), we obtain the final result
\begin{subequations}\label{<s>}
\begin{eqnarray}
_P\langle p,s|\vec S\,|p',s' \rangle_P &=& \frac{1}{2}\,
\delta(\vec p - \vecp{p})\, \frac{\hat s m}{\sqrt{m^2 + {\vec p_\bot}^{\,2}}},
\label{sP}
\\
_A\langle p,s|\vec S\,|p',s' \rangle_A &=& \frac{1}{2}\,
\delta(\vec p - \vecp{p})\, \frac{\hat s m}{\sqrt{m^2 + {\vec p_\bot}^{\,2}}},
\label{sA}
\end{eqnarray}
\end{subequations}
with
\begin{equation}\label{with}
s = s(p,\hat s),\;\; s' = s'(p',\hat s).
\end{equation}
Note that, in the case of the antiparticle, the minus sign from the normal 
ordering is cancelled by the minus sign of the normalization of the 
$v$-spinor. Therefore, both expectation values, equations~(\ref{sP}) 
and~(\ref{sA}), have the same sign.

We stress that $\hat s$ is 
in general different from the unit vector $\hat s_r$ associated with the 
spin vector $s_r$ in the rest frame of the particle or antiparticle.
In the following we will elaborate on the relation between $\hat s$ and 
$\hat s_r$. In the rest frame the energy-momentum vector $p$ and the 
spin vector $s$ have the form
\begin{equation}\label{rest}
p_r = \left( \begin{array}{c} m \\ \vec 0 \end{array} \right)
\quad \mbox{and} \quad 
s_r = \left( \begin{array}{c} 0 \\ \hat s_r \end{array} \right)
\quad \mbox{with} \quad \left( \hat s_r \right)^2 = 1,
\end{equation}
respectively. It suggests itself to use the results of appendix~\ref{sl2c} 
and apply the Lorentz boost $L(\vec v\,)$, equation~(\ref{L(v)}), 
to $p_r$ and $s_r$. With 
\begin{equation}\label{pv}
\hat p \equiv \frac{\vec v}{|\vec v\,|}, \; \frac{E_p}{m} = \gamma ,\; 
\frac{\vec p}{m} = \gamma \vec v,
\end{equation}
the corresponding $L(\vec v\,)$ obviously effects  
$L(\vec v\,) p_r = p$---\textit{cf.}\ equation~(\ref{pr->p}). 
Therefore, we have to apply this Lorentz boost to $s_r$.
In this way we obtain 
\begin{equation}\label{ssr}
s = L(\vec v\,) s_r = \left( 
\begin{array}{c}
\frac{\displaystyle\hat s_r \cdot \vec p}{\displaystyle m} \\[1mm]
\hat s_r + 
\left( \frac{\displaystyle E_p}{\displaystyle m} - 1 \right) 
\left( \hat s_r \cdot \hat p\right) \hat p
\end{array} \right).
\end{equation}
This spin vector does not yet have the form of equation~(\ref{s}),
however, equation~(\ref{ssr}) suggests that $\hat s $ is given by
\begin{equation}\label{N}
N \hat s = \hat s_r + 
\left( \frac{\displaystyle E_p}{\displaystyle m} - 1 \right) 
\left( \hat s_r \cdot \hat p\right) \hat p
\quad \mbox{with} \quad \left( \hat s \right)^2 = 1 
\quad \mbox{and} \quad
N = \sqrt{1 + \left( \frac{\hat s_r \cdot \vec p}{m} \right)^2}.
\end{equation}
Indeed, taking the scalar product of $N \hat s$ with $\vec p/E_p$, 
we obtain 
\begin{equation}\label{Nsp}
N\,
\frac{\displaystyle\hat s \cdot \vec p}{\displaystyle E_p} =  
\frac{\displaystyle\hat s_r \cdot \vec p}{\displaystyle m}, 
\end{equation}
which allows us to reformulate equation~(\ref{ssr}) as  
\begin{equation}
s = N \left( \begin{array}{c} \hat s \cdot \vec p/E_p \\ \hat s
\end{array} \right).
\end{equation}
Therefore, consistency with equation~(\ref{s}) is obvious. 
Moreover, taking the square of equation~(\ref{Nsp}), we can express 
$\hat s_r \cdot \vec p/m$ by 
$\hat s \cdot \vec p/E_p$ and obtain in this way 
the alternative form 
\begin{equation}\label{N1}
N = \frac{1}{\sqrt{1 - \left( 
\frac{\displaystyle\hat s \cdot \vec p}{\displaystyle E_p} \right)^2}}
\end{equation}
of the normalization factor $N$.
We have thus full agreement of equation~(\ref{ssr}) with equation~(\ref{s}).
This confirms that equation~(\ref{N}) constitutes the relation between 
$\hat s$ and $\hat s_r$ we have been looking for.
It is now straightforward to rewrite equation~(\ref{<s>}) 
in terms of $\hat s_r$. Reading off 
\begin{equation}
\frac{1}{\sqrt{m^2 + {\vec p_\bot}^{\,2}}} = \frac{N}{E_p}
\end{equation}
from equation~(\ref{alternative}) und $N \hat s$ from equation~(\ref{N}),
we obtain
\begin{equation}
_P\langle p,s|\vec S\,|p',s' \rangle_P = \,
_A\langle p,s|\vec S\,|p',s' \rangle_A = 
\frac{1}{2}\, \delta(\vec p - \vecp{p})
\left( \hat s_{r \|} + \frac{m}{E_p}\,\hat s_{r \bot} \right)
\end{equation}
where we have defined
\begin{equation}
\hat s_{r \|} = \left( \hat s_r \cdot \hat p \right) \hat p 
\quad \mbox{and} \quad
\hat s_{r \bot} = \hat s_r - \left( \hat s_r \cdot \hat p \right) \hat p.
\end{equation}

\section{Ultrarelativistic particles and helicity}
\label{ultrarelativistic}
We have seen in equation~(\ref{<s>}) that the expectation value 
of the spin operator $\vec S$ goes to zero for 
$\left| \vec p_\bot \right| \to \infty$, while it is $\pm 1/2$ for 
$\vec p_\bot = \vec 0$, irrespective of the magnitude of $|\vec p\,|$.
In the latter case, $\vec p$ and $\hat s$ have the same or the opposite 
direction, \textit{i.e.}\ $\hat p = \pm \hat s$.
This consideration suggests to use the helicity operator
\begin{equation}\label{helicity}
h = \frac{\vec S \cdot \vec P}{\big| \vec P \big|} 
\end{equation}
for ultrarelativistic particles instead of the spin operator $\vec S$. 
In equation~(\ref{helicity}), $\vec P$ denotes the momentum operator in space.
The eigenvalues of $h$ are $h_\pm = \pm 1/2$, as can be read off from 
equation~(\ref{<s>}) for $\vec p_\bot = \vec 0$. Inspection of 
equations~(\ref{s}) and~(\ref{alternative}) 
leads to the following characterization 
of the helicity states:
\begin{subequations}
\begin{eqnarray}
h_+ = +\frac{1}{2}: && \hat s = \hat p, \;\;
s = \frac{1}{m} \left( \begin{array}{c} |\vec p\,| \\[1mm] E_p\, \hat p
\end{array} \right), 
\label{h+s}
\\
h_- = -\frac{1}{2}: && \hat s = -\hat p, \;\;
s = -\frac{1}{m} \left( \begin{array}{c} |\vec p\,| \\[1mm] E_p\, \hat p
\end{array} \right).
\end{eqnarray}
\end{subequations}
In the remainder of this section we will denote the vector $s$ appearing in 
equation~(\ref{h+s}) by $s_+$. For the further discussion it is useful 
to rewrite $s_+$ as
\begin{equation}\label{s+}
s_+ = \frac{1}{m} \left( \begin{array}{c} |\vec p\,| \\[1mm] E_p\, \hat p
\end{array} \right) =
\frac{p}{m} + \frac{mz}{E_p + |\vec p\,|}
\quad \mbox{with} \quad 
z = \left( \begin{array}{c} -1 \\ \hat p 
\end{array} \right).
\end{equation}
We denote by $u(p,+)$ a spinor corresponding to positive helicity and so on.

In the Standard Model, before spontaneous gauge-symmetry breaking, 
the fields participating in electroweak interactions are 
chiral fields, \textit{i.e.}\ eigenvectors of the chiral projectors
of equation~(\ref{chiral projectors}). Neutrinos are a special case because, 
to our present knowledge (leaving out gravity), 
neutrinos participate only in weak interactions so that the neutrino fields 
invariably appear as left-handed fields, \textit{i.e.}\ they 
are multiplied by the chiral projector $\gamma_-$.
Since neutrino masses are below 1\,eV and neutrino detection requires 
energies above 100\,keV, detectable neutrinos are always ultrarelativistic.
The following consideration can, however, be used for any ultrarelativistic 
fermion produced in weak interactions with a $W^\pm$ boson because its 
interaction with fermions always has an inherent chiral projector $\gamma_-$.

It is straightforward to understand the effect of the chiral projectors on 
spinors $u(p,\pm)$ and $v(p,\pm)$. Taking 
for definiteness $\gamma_-$ because of its importance for weak 
interactions and using
\begin{equation}\label{aux}
\gamma_- \gamma_5 = -\gamma_- \quad \mbox{and} \quad
\frac{\slashed{p}}{m}  u(p,+) = u(p,+),
\end{equation}
we obtain
\begin{equation}
\gamma_- u(p,+) = \gamma_- \Sigma_+(s_+) u(p,+) =
\frac{1}{2} \gamma_- \left( \bone_4 - \slashed{s}_+ \right) u(p,+) = 
-\frac{m}{2(E_p + |\vec p\,|)} \gamma_- \slashed{z} u(p,+).
\end{equation}
This suggests that, in the ultrarelativistic limit of a weak process, amplitudes 
which have a particle with positive helicity on an external line vanish.
However, arguing more to the point, we better take into account the 
$p$-dependence inherent in the spinor as well and reproduce what is really 
done in the treatment of an external line when computing a cross section.
Since such a computation leads to the expressions $u \bar u$ and $v \bar v$,
we need the relations
\begin{equation}\label{uuvvP}
u(p,\pm) \bar u(p,\pm) = 2m \Lambda_+(p) \Sigma_\pm(s_+),
\quad
v(p,\pm) \bar v(p,\pm) = -2m \Lambda_+(p) \Sigma_\pm(s_+),
\end{equation}
which follow from the properties of the spinors and projectors,
and the normalization conditions of equation~(\ref{normalization}).
Now we revert to the consideration of $\gamma_- u(p,+)$ and compute 
\begin{subequations}\label{u+u+}
\begin{eqnarray}
\lefteqn{
\gamma_- u(p,+) \left( \gamma_- u(p,+) \right)^\dagger \beta = }
\nonumber \\ &&
\gamma_- u(p,+) \bar u(p,+) \gamma_+ =
2m \gamma_- \Lambda_+(p) \Sigma_+(s_+) \gamma_+ =
\label{l1} \\ &&
\frac{1}{2} \gamma_- \left( \slashed{p} + m \right) 
\left( \bone_4 + \gamma_5 \slashed{s}_+ \right) \gamma_+ = 
\frac{1}{2} \gamma_- \left( \slashed{p} + 
m \gamma_5 \slashed{s}_+ \right) \gamma_+ = 
\label{l2} \\ &&
\frac{1}{2} \gamma_- \left( \slashed{p} - 
m \slashed{s}_+ \right) \gamma_+ = 
-\frac{m^2 \slashed{z}}{2(E_p + |\vec p\,|)}\,\gamma_+.
\label{l3}
\end{eqnarray}
\end{subequations}
In equation~(\ref{l2}) we have taken into account that an expression with 
an even number of Dirac matrices sandwiched between $\gamma_-$ and $\gamma_+$ 
vanishes. The last line, equation~(\ref{l3}), has been achieved by 
taking advantage of the first relation in equation~(\ref{aux}) and 
using the form of $s_+$, equation~(\ref{s+}).
With the procedure introduced in equation~(\ref{u+u+}), we compute the 
remaining expressions and arrive at the following list:
\begin{subequations}\label{uuvv}
\begin{eqnarray}
\gamma_- u(p,+) \left( \gamma_- u(p,+) \right)^\dagger \beta 
&=& -\left( 0 + 
\frac{m^2 \slashed{z}}{2(E_p + |\vec p\,|)} \right) \gamma_+,
\label{la} \\
\gamma_- u(p,-) \left( \gamma_- u(p,-) \right)^\dagger \beta 
&=& + 
\left( \slashed{p} + \frac{m^2 \slashed{z}}{2(E_p + |\vec p\,|)} 
\right) \gamma_+,
\label{lb} \\
\gamma_- v(p,+) \left( \gamma_- v(p,+) \right)^\dagger \beta 
&=& 
+\left( \slashed{p} + \frac{m^2 \slashed{z}}{2(E_p + |\vec p\,|)} 
\right) \gamma_+,
\label{lc} \\
\gamma_- v(p,-) \left( \gamma_- v(p,-) \right)^\dagger \beta 
&=& - \left( 0 + 
\frac{m^2 \slashed{z}}{2(E_p + |\vec p\,|)} \right) \gamma_+.
\label{ld} 
\end{eqnarray}
\end{subequations}
We have kept the zero matrix 
on the right-hand side of equations~(\ref{la}) and~(\ref{ld}), 
to order to highlight the contrast to the occurrence of   
$\slashed{p}$ in the case of the other two equations. Note that 
we have put $\gamma_+$ on the right-hand side of $\slashed{p}$ and 
$\slashed{z}$, but we 
could as well put $\gamma_-$ on the left-hand side, or sandwich 
$\slashed{z}$ between these two chiral projectors. 

In computing 
cross sections, the right-hand sides of equation~(\ref{uuvv}) appear 
in traces. It is thus plausible that cross sections with an 
ultrarelativistic particle with positive helicity on an external line are 
suppressed compared to those with a negative helicity, 
and vice versa for an external antiparticle line.

For Dirac neutrinos this means in practice that, for 
a given momentum $\vec p$, out of the four possible states (neutrino vs.\ 
antineutrino, positive vs.\ negative helicity) only two can be 
observed, namely neutrinos with 
negative helicity and antineutrinos with positive helicity. 
This is the reason that it is very difficult to 
distinguish between Dirac and Majorana neutrinos because 
the two observable Dirac neutrino states could be just the 
two helicity states of a Majorana neutrino.

\section{Weyl basis}
\label{weyl basis}
\subsection{Dirac matrices in the Weyl basis}
\label{dirac matrices}
The Weyl basis is defined by
\begin{equation}\label{weyl}
\gamma^\mu = \left( \begin{array}{cc} 
0 & \sigma^\mu \\ {\bar \sigma}^\mu & 0
\end{array} \right)
\quad \mbox{with} \quad 
\left( \sigma^\mu \right) = \left( \bone, \sigma^1, \sigma^2, \sigma^3 \right),
\quad
\left( {\bar\sigma}^\mu \right) = 
\left( \bone, -\sigma^1, -\sigma^2, -\sigma^3 \right),
\end{equation}
where the $\sigma^j$ ($j=1,2,3$) denote the usual Pauli matrices---see also
section~\textit{Notation}. Note that the Dirac matrices
of equation~(\ref{weyl}) obey equation~(\ref{special}), therefore, 
$\beta \equiv \gamma^0$. 
Moreover,
\begin{equation}\label{gammaT}
\left( \gamma^0 \right)^T = \gamma^0, \;\;
\left( \gamma^1 \right)^T = -\gamma^1, \;\;
\left( \gamma^2 \right)^T = \gamma^2, \;\;
\left( \gamma^3 \right)^T = -\gamma^3,
\end{equation}
which leads to a charge conjugation matrix $C \propto \gamma^2 \gamma^0$.
A phase convention for $C$ commonly used in the literature is 
\begin{equation}
C = i \gamma^2 \gamma^0 = 
\left( \begin{array}{cc} i\sigma^2 & 0 \\ 0 & -i\sigma^2
\end{array} \right).
\end{equation}
Finally, in the Weyl basis $\gamma_5$ is 
given by
\begin{equation}
\gamma_5 = \left( \begin{array}{rr} -\bone & 0 \\ 0 & \bone
\end{array} \right)
\end{equation}
and the chiral projectors have the exceedingly simple form
\begin{equation}\label{gamma+-}
\gamma_- = \left( \begin{array}{cc} \bone & 0 \\ 0 & 0
\end{array} \right)
\quad \mbox{and} \quad
\gamma_+ = \left( \begin{array}{cc} 0 & 0 \\ 0 & \bone
\end{array} \right).
\end{equation}

\subsection{Weyl basis and Lorentz invariance}
\label{wbli}
In this section we compute, in the Weyl basis, the explicit form of 
$\mathcal{S}(\omega)$ introduced in section~\ref{deli}.
For this purpose we parameterize the antisymmetric coefficient matrix 
$\left( \omega_{\alpha\beta} \right)$ by~\cite{sexl}
\begin{equation}
\left( \omega_{\alpha\beta} \right) = 
\left( \begin{array}{cccc}
0 & w_1 & w_2 & w_3 \\
-w_1 & 0 & \alpha_3 & -\alpha_2 \\
-w_2 & -\alpha_3 & 0 & \alpha_1 \\
-w_3 & \alpha_2 & -\alpha_1 & 0
\end{array} \right).
\end{equation}
Note that $\omega_{jk} = \varepsilon_{jkl} \alpha_l$ for spatial indices 
$j,k,l$ and summation over $l$. In the next step we have to compute
\begin{equation}
\frac{1}{2} \omega_{\alpha\beta}\, \sigma^{\alpha\beta} = 
\omega_{0j}\, \sigma^{0j} + \sum_{j < k} \omega_{jk}\, \sigma^{jk}.
\end{equation}
This is easily done with
\begin{equation}
\sigma^{0j} = i \left( 
\begin{array}{cc} -\sigma^j & 0 \\ 0 & \sigma^j
\end{array} \right),
\quad
\sigma^{jk} = \varepsilon_{jkl} \left( 
\begin{array}{cc} \sigma^l & 0 \\ 0 & \sigma^l
\end{array} \right).
\end{equation}
The result is 
\begin{equation}
\frac{1}{2} \omega_{\alpha\beta}\, \sigma^{\alpha\beta} = 
\left( \begin{array}{cc}
\left( \vec \alpha - i \vec w\, \right) \cdot \vec \sigma & 0 \\
0 & \left( \vec \alpha + i \vec w\, \right) \cdot \vec \sigma
\end{array} \right).
\end{equation}
Therefore, 
\begin{equation}\label{SA}
\mathcal{S}(\omega) = \left(
\begin{array}{cc}
A & 0 \\ 0 & \left( A^{-1} \right)^\dagger
\end{array} \right)
\quad \mbox{with} \quad 
A = \exp \left( -\frac{i}{2} 
\left( \vec \alpha - i \vec w\, \right) \cdot \vec \sigma \right).
\end{equation}
We see that  
in the Weyl basis it is evident that the matrices $\mathcal{S}(\omega)$ 
generate a reducible representation of $SL(2,\mathbbm{C})$. 
Moreover, with the chiral projectors, equation~(\ref{gamma+-}), we 
obtain an explicit version of equation~(\ref{S+S-}):
\begin{equation}
\mathcal{S}_-(\omega) = 
\left( \begin{array}{cc} A & 0 \\ 0 & \bone 
\end{array} \right),
\quad
\mathcal{S}_+(\omega) = 
\left( \begin{array}{cc} \bone & 0 \\ 0 & \left( A^{-1} \right)^\dagger
\end{array} \right). 
\end{equation}
This result for $\mathcal{S}_\pm(\omega)$, obtained in the Weyl basis, 
agrees exactly with the result displayed in equation~(\ref{S+-z}), 
obtained in a basis-independent way, 
due to the fact that in the Weyl basis the matrices $T_j$ are given by
\begin{equation}
T_j = \left( \begin{array}{cc} \sigma^j & 0 \\ 0 & \sigma^j
\end{array} \right).
\end{equation}

In equation~(\ref{SA}), the matrix $A \in SL(2,\mathbbm{C})$ is represented 
as an exponential---see footnote~\ref{side note} for remarks on the exponential 
function in this group. If $\vec w = \vec 0$ in $A$, then $A \in SU(2)$ 
and the associated Lorentz transformation is a rotation in space. If 
$\vec\alpha = \vec 0$, then $A = A^\dagger$ corresponds to a Lorentz boost 
on 4-vectors. This is demonstrated in appendix~\ref{sl2c} where more details 
on $SL(2,\mathbbm{C})$ and Lorentz transformations and supplementary 
information can be found.

\subsection{Plane-wave solutions in the Weyl basis}
\label{plane-wave weyl basis}
Plane-wave solutions of the Dirac equation, associated with the 4-momentum 
vector $p$ ($p^2 = m^2$), lead to the equations 
\begin{equation}\label{DE}
\left( \slashed{p} - m \right) u(p,\xi) = 0,
\quad
\left( \slashed{p} + m \right) v(p,\eta) = 0
\end{equation}
for the Dirac spinors $u$ for positive energies and $v$ for negative energies. 
Explicitly, in the Weyl basis the spinors $u$, $v$ are given by
\begin{equation}\label{uv}
u(p,\xi) = \left(
\begin{array}{c}
\sqrt{p \cdot \sigma}\, \xi \\ \sqrt{p \cdot \bar\sigma}\, \xi
\end{array} \right),
\quad
v(p,\eta) = \left(
\begin{array}{r}
\sqrt{p \cdot \sigma}\, \eta \\ -\sqrt{p \cdot \bar\sigma}\, \eta
\end{array} \right),
\end{equation}
respectively. Note that $p \cdot \sigma = p_\mu \sigma^\mu$, 
$p \cdot \bar\sigma = p_\mu {\bar\sigma}^\mu$. The
$\xi,\, \eta \in \mathbbm{C}^2$ are in principle arbitrary, however, 
henceforth we will assume that they are unit vectors. We will see below 
that for unit vectors the normalization conditions of 
equation~(\ref{normalization}) are fulfilled.

A word is in order concerning the meaning of the square root in 
equation~(\ref{uv}). Any positive matrix $A$, which is by definition 
hermitian, has eigenvalues $\lambda_j \geq 0$, and the square root 
$\sqrt{A}$ is uniquely defined as the matrix 
with the same eigenvectors as $A$ but with eigenvalues 
$\sqrt{\lambda_j} \geq 0$. 
Let us consider the hermitian $2 \times 2$ matrices $p \cdot \sigma$ and 
$p \cdot \bar\sigma$ of equation~(\ref{uv}). The part with the Pauli matrices 
is given by 
$\vec p \cdot \vec\sigma \equiv \sum_{j=1}^3 p^j \sigma^j$.
Because of $\left( \vec p \cdot \vec\sigma \right)^2 = {\vec p}^{\,2} \bone$,
the eigenvalues of $\vec p \cdot \vec\sigma$ are $\pm |\vec p\,|$. 
Therefore, there are eigenvectors $\zeta_\pm$ such that
\begin{equation}\label{zeta+-}
\left( \vec p \cdot \vec\sigma \right) \zeta_\pm = \pm |\vec p\,|\, \zeta_\pm
\quad \mbox{with} \quad
\zeta_+^\dagger \zeta_+ = \zeta_-^\dagger \zeta_- = 1, \;\;
\zeta_+^\dagger \zeta_- = \zeta_-^\dagger \zeta_+ = 0.
\end{equation}
Consequently, since
\begin{equation}
E_p \equiv p^0 = \sqrt{{\vec p}^{\,2} + m^2}, 
\end{equation}
the matrices $p \cdot \sigma$ and $p \cdot \bar\sigma$
have positive eigenvalues 
\begin{equation}
E_p \pm |\vec p\,| \geq 0. 
\end{equation}
Therefore, the square roots in equation~(\ref{uv}) are well-defined. 
Formally, these square roots are thus given by
\begin{subequations}\label{rootsps}
\begin{eqnarray}
\sqrt{p \cdot \sigma} &=& 
\sqrt{E_p - |\vec p\,|}\, \zeta_+ \zeta_+^\dagger + 
\sqrt{E_p + |\vec p\,|}\, \zeta_- \zeta_-^\dagger,
\label{root-sigma}
\\
\sqrt{p \cdot \bar\sigma} &=& 
\sqrt{E_p + |\vec p\,|}\, \zeta_+ \zeta_+^\dagger + 
\sqrt{E_p - |\vec p\,|}\, \zeta_- \zeta_-^\dagger.
\label{root-sigmabar}
\end{eqnarray}
\end{subequations}
A simple application of equation~(\ref{rootsps}) is
\begin{equation}\label{ppm}
\sqrt{p \cdot \sigma} \sqrt{p \cdot \bar\sigma} = 
\sqrt{p \cdot \bar\sigma} \sqrt{p \cdot \sigma} = 
m \left( \zeta_+ \zeta_+^\dagger + \zeta_- \zeta_-^\dagger \right) =
m \bone.
\end{equation}
However, this result could have been obtained in a simpler way by using 
the obvious relation $\sqrt{A} \sqrt{B} = \sqrt{AB}$ 
for $A \geq 0$, $B \geq 0$ and $[A,B] = 0$ and 
$\left( p \cdot \sigma \right) \left( p \cdot \bar\sigma \right) = m^2 \bone$.

With equation~(\ref{ppm}) and taking into account that $\beta = \gamma^0$, 
the normalization conditions of equation~(\ref{normalization}) are 
obviously satisfied.

Now we turn to the poof of equation~(\ref{DE}), which is easy to do 
by taking advantage of equation~(\ref{ppm}). Let us 
consider the first relation:
\begin{eqnarray}
\left( \slashed{p} - m \right) u(p,\xi) &=& 
\left( 
\begin{array}{cc}
-m \bone & p \cdot \sigma \\ p \cdot \bar\sigma & -m \bone
\end{array} \right) \left(
\begin{array}{c}
\sqrt{p \cdot \sigma}\, \xi \\ \sqrt{p \cdot \bar\sigma}\, \xi
\end{array} \right) \nonumber \\ &=&
\left( \begin{array}{c} 
-m \sqrt{p \cdot \sigma} + \sqrt{p \cdot \sigma}\, m \\[1mm]
+ \sqrt{p \cdot \bar\sigma}\, m - m \sqrt{p \cdot \bar\sigma}
\end{array} \right)
\left( \begin{array}{c} \xi \\ \xi 
\end{array} \right) = 0.
\end{eqnarray}
The proof for the second relation proceeds in the same way.

Equation~(\ref{rootsps}) can be expressed more directly, without resorting 
to the vectors $\zeta_\pm$, by defining the projectors 
\begin{equation}\label{P+-}
\mathcal{P}_\pm = \frac{1}{2} 
\left( \bone \pm \hat p \cdot \vec \sigma \right)
\quad \mbox{with} \quad \hat p = \frac{\vec p}{|\vec p\,|}.
\end{equation}
Then the identification
\begin{equation}\label{Pzeta}
\mathcal{P}_+ = \zeta_+ \zeta_+^\dagger, \quad
\mathcal{P}_- = \zeta_- \zeta_-^\dagger
\end{equation}
is trivial and an explicit version of the roots is given by
\begin{subequations}\label{rootsP}
\begin{eqnarray}
\sqrt{p \cdot \sigma} &=& 
\sqrt{E_p - |\vec p\,|}\, \mathcal{P}_+ +
\sqrt{E_p + |\vec p\,|}\, \mathcal{P}_-
\\
\sqrt{p \cdot \bar\sigma} &=& 
\sqrt{E_p + |\vec p\,|}\, \mathcal{P}_+ +
\sqrt{E_p - |\vec p\,|}\, \mathcal{P}_-.
\end{eqnarray}
\end{subequations}

\subsection{The spin vector}
\label{spin vector}
On the one hand, 
in section~\ref{plane-wave} we have characterized the plane-wave 
solutions of the Dirac equation by projectors, without using 
any special basis of the Dirac matrices. Such solutions were determined 
by just two vectors, the 4-momentum $p$ and the spin vector $s$.
On the other hand, in section~\ref{plane-wave weyl basis}, 
where we performed computations in the Weyl basis, instead of $s$ we had 
complex two-component vectors $\xi, \, \eta \in \mathbbm{C}^2$.
So the question which imposes on oneself is 
which spin vectors are associated with 
the solutions $u(p,\xi)$ and $v(p,\eta)$ of section~\ref{plane-wave weyl basis}.
The strategy that we will pursue is to determine first the spin vector 
$s_r$ in the rest frame of the particle and thereafter make a boost into 
the frame where the particle has the 4-momentum $p$.
For the vectors $p_r$ and $s_r$ we refer the reader to equation~(\ref{rest}).

In the rest frame, the spin projector is given by
\begin{equation}
\Sigma_+(s_r) = \frac{1}{2} \left( 
\begin{array}{cc}
\bone & \hat s_r \cdot \vec \sigma \\
\hat s_r \cdot \vec \sigma & \bone 
\end{array} \right),
\end{equation}
which can easily be obtained with the formulas of section~\ref{dirac matrices}.
Moreover, setting $p = p_r$ in equation~(\ref{uv}) leads to 
\begin{equation}\label{uvr}
u(p_r, \xi) = \sqrt{m} \left( \begin{array}{c} \xi \\ \xi 
\end{array} \right),
\quad
v(p_r, \eta) = \sqrt{m} \left( \begin{array}{c} \eta \\ -\eta 
\end{array} \right).
\end{equation}
Let us postulate that we have the same vector 
$\hat s_r$ for both types of solutions $u(p_r, \xi)$ and $v(p_r, \eta)$,
\textit{i.e.}
\begin{equation}
\Sigma_+(s_r) u(p_r, \xi) = u(p_r, \xi) 
\quad \mbox{and} \quad
\Sigma_+(s_r) v(p_r, \eta) = v(p_r, \eta). 
\end{equation}
It is easy to see that this leads to the equations 
\begin{equation}\label{hat s_r}
\left( \hat s_r \cdot \vec \sigma \right) \xi = \xi
\quad \mbox{and} \quad
\left( \hat s_r \cdot \vec \sigma \right) \eta = -\eta.
\end{equation}
Since $\xi$ and $\eta$ are eigenvectors of the hermitian 
$2 \times 2$ matrix $\hat s_r \cdot \vec \sigma$ 
with different eigenvalues, we infer $\xi \bot \eta$.

Consequently, given $\xi$ or $\eta$ with $\xi \bot \eta$,
the spin vector $\hat s_r$ is determined 
by equation~(\ref{hat s_r}). The solution is
\begin{equation}\label{sols}
\hat s_r = \xi^\dagger \vec \sigma \xi = -
\eta^\dagger \vec \sigma \eta.
\end{equation}
\textbf{Proof:} Solving equation~(\ref{hat s_r}) 
for $\hat s_r$ is easily done if one uses the identity
\begin{equation}
\sum_{j=1}^3 \sigma^j_{ab} \sigma^j_{cd} = 
2 \delta_{ad} \delta_{cb} - \delta_{ab} \delta_{cd}
\end{equation}
of the Pauli matrices. It implies
\begin{subequations}
\begin{eqnarray}
&&
\left( \xi^\dagger \vec \sigma \xi \right) \cdot 
\left( \xi^\dagger \vec \sigma \xi \right) =
\left( \eta^\dagger \vec \sigma \eta \right) \cdot 
\left( \eta^\dagger \vec \sigma \eta \right) = 1,
\\
&&
\left( \xi^\dagger \vec \sigma \xi \right) \cdot 
\left( \eta^\dagger \vec \sigma \eta \right) = -1,
\\
&&
\left( \xi^\dagger \vec \sigma \xi \right) \cdot 
\left( \xi^\dagger \vec \sigma \eta \right) =
\left( \xi^\dagger \vec \sigma \xi \right) \cdot 
\left( \eta^\dagger \vec \sigma \xi \right) = 0,
\label{(c)}
\end{eqnarray}
\end{subequations}
provided $\xi^\dagger \xi = \eta^\dagger \eta = 1$ and $\xi^\dagger \eta = 0$. 
Therefore, multiplying the relations in equation~(\ref{hat s_r})
from the left by $\xi^\dagger$ and $\eta^\dagger$, respectively, one obtains 
\begin{equation}\label{solution hat s_r}
\hat s_r \cdot \left( \xi^\dagger \vec \sigma \xi \right) = 
-\hat s_r \cdot \left( \eta^\dagger \vec \sigma \eta \right) = 1.
\end{equation}
Since $\hat s_r$, $\xi^\dagger \vec \sigma \xi$ and 
$\eta^\dagger \vec \sigma \eta$ are unit vectors of $\mathbbm{R}^3$, 
equation~(\ref{sols}) ensues. It remains to check 
\[
\eta^\dagger \left( \hat s_r \cdot \vec \sigma \right) \xi = 
\xi^\dagger \left( \hat s_r \cdot \vec \sigma \right) \eta = 0
\]
for consistency, but this is obviously fulfilled according to 
equation~(\ref{(c)}). Q.E.D.

The determination of $\hat s_r$ through equation~(\ref{hat s_r}) 
suggests that the spin vector of $u(p,\xi)$ and $v(p,\eta)$ 
is given by equation~(\ref{ssr}), \textit{i.e.}\ 
by $s = L(\vec v\,) s_r$ where $L(\vec v\,)$ is the Lorentz boost that 
performs $p_r \to p$ with $\vec v$ determined  
from $p$ via the prescription of equation~(\ref{pv}). Now we want to show from 
the perspective of $SL(2,\mathbbm{C})$ and the Weyl basis that this 
is indeed the case. The key to do so is the observation that 
\begin{equation}\label{A(p)}
A(p) \equiv \sqrt{\frac{p \cdot \sigma}{m}} \in SL(2,\mathbbm{C})
\end{equation}
is exactly the matrix that generates the Lorentz transformation 
$L(\vec v\,)$~\cite{peskin}.
This is proven in general in appendix~\ref{sl2c}---\textit{cf.}\ 
equation~(\ref{H(v)}) with $u = p/m$. However, skipping all intricacies 
of the general proof, with equation~(\ref{basic-sl2c}) one can 
trivially see that $A(p)$ induces $p_r \to p$:
\begin{equation}\label{AprA}
A(p) \left( p_r \cdot \sigma \right) A^\dagger(p) = 
\sqrt{\frac{p \cdot \sigma}{m}} \left( m \bone \right) 
\sqrt{\frac{p \cdot \sigma}{m}} = p \cdot \sigma.
\end{equation}
Moreover, equation~(\ref{ppm}) tells us that 
\begin{equation}
A^{-1}(p) = \sqrt{\frac{p \cdot \bar\sigma}{m}}.
\end{equation}
Therefore, using equation~(\ref{SA}), $A(p)$
allows us to define a transformation 
\begin{equation}
\mathcal{S}(p) = \left( \begin{array}{cc}
A(p) & 0 \\ 0 & A^{-1}(p) 
\end{array} \right)
\end{equation}
on the Dirac spinors~\cite{peskin}. Notice that in this equation 
we have taken into account that $A(p)$ is hermitian. It is then obvious that 
\begin{equation}
\mathcal{S}(p) u(p_r, \xi) = u(p,\xi), \quad 
\mathcal{S}(p) v(p_r, \eta) = v(p,\eta),
\end{equation}
\textit{cf.}\ equations~(\ref{uv}) and~(\ref{uvr}). 
This is also consistent with our discussion of $A(p)$ above; 
while $A(p)$ induces the Lorentz boost $L(\vec v\,)$ on 4-vectors 
such that $p = L(\vec v\,) p_r$, $\mathcal{S}(p)$ performs 
the corresponding boost on Dirac spinors. Eventually, 
we come to our last issue, 
that is the announced consistency check that $s = L(\vec v\,) s_r$ is indeed 
the spin vector associated with $u(p,\xi)$ and $v(p,\eta)$. 
The argument goes as follows:
\begin{eqnarray}
\lefteqn{ 
u(p,\xi) = 
\mathcal{S}(p) u(p_r,\xi) =} \nonumber \\ && 
\mathcal{S}(p) \Sigma_+(s_r) u(p_r,\xi) =
\frac{1}{2} \left( \bone_4 + \gamma_5 
\mathcal{S}(p) \slashed{s}_r \mathcal{S}^{-1}(p) \right) 
\mathcal{S}(p) u(p_r,\xi) =
\nonumber \\
&&
\frac{1}{2} \left( \bone_4 + \gamma_5 
\gamma_\lambda L^\lambda_{\hphantom{\lambda}\mu}(\vec v\,) 
s_r^\mu \right) u(p,\xi) = \Sigma_+(s) u(p,\xi).
\label{s->sr}
\end{eqnarray}
In the step from the second to the third line we have taken advantage of 
equation~(\ref{Sg1}). We have also used that $\mathcal{S}(p)$ commutes 
with $\gamma_5$. For $v$-spinors one simply has to make 
the replacements $u \to v$ and $\xi \to \eta$ in 
equation~(\ref{s->sr}). We finish this section with the following 
important conclusion:
\begin{quote}
If we have two solutions $u(p,\xi)$ and $v(p,\eta)$ that are supposed 
to correspond to the \emph{same} spin vector $s$, then 
necessarily $\eta \bot \xi$.
\end{quote}

\subsection{Charge-conjugation and plane-wave solutions}
\label{cc and pws}
In the Weyl basis, the matrix relevant for charge conjugation is given by
\begin{equation}
C \beta^T = i \gamma^2 = 
\left( \begin{array}{cc} 0 & i\sigma^2 \\ -i\sigma^2 & 0
\end{array} \right) 
\quad \Rightarrow \quad 
\left( C \beta^T \right) \left( C \beta^T \right)^* = 
\left( i\gamma^2 \right)^2 = \bone_4,
\end{equation}
\textit{cf.}\ section~\ref{dirac matrices}. The latter result 
agrees with equation~(\ref{cbcb}) and demonstrates once more that the constant 
$d$ of equation~(\ref{cbd}) is positive. Note that 
$i \sigma^2 \equiv \varepsilon$ with $\varepsilon$ 
defined in equation~(\ref{varepsilon}).

In order to investigate the behaviour of the spinors $u$ and $v$ of 
equation~(\ref{uv}) under charge conjugation, we need the relations
\begin{equation}\label{ccc}
\varepsilon \left( \sqrt{p \cdot \sigma} \right)^* \varepsilon^{-1} =
\sqrt{p \cdot \bar\sigma}, \quad
\varepsilon \left( \sqrt{p \cdot \bar\sigma} \right)^* \varepsilon^{-1} =
\sqrt{p \cdot \sigma}.
\end{equation}
\textbf{Proof:} We use equation~(\ref{rootsps}) to accomplish the proof.
Thus we have to consider 
\begin{equation}
\varepsilon \left( \zeta_+ \zeta_+^\dagger \right)^* \varepsilon^{-1} = 
\left( \varepsilon \zeta_+^* \right) 
\left( \varepsilon \zeta_+^* \right)^\dagger =
\zeta_- \zeta_-^\dagger.
\end{equation}
The justification for the last step is given by the fact that 
$\varepsilon \zeta_+^* \bot \zeta_+$. Therefore, $\varepsilon \zeta_+^*$ 
agrees with $\zeta_-$ apart from a phase factor which is, however, 
irrelevant in $\zeta_- \zeta_-^\dagger$. Obviously, 
\begin{equation}
\varepsilon \left( \zeta_- \zeta_-^\dagger \right)^* \varepsilon^{-1} = 
\zeta_+ \zeta_+^\dagger
\end{equation}
holds as well. Therefore, the operation 
$\sqrt{p \cdot \sigma} \to
\varepsilon \left( \sqrt{p \cdot \sigma} \right)^* \varepsilon^{-1}$
corresponds to the exchange
$\zeta_+ \zeta_+^\dagger \leftrightarrow \zeta_- \zeta_-^\dagger$
in $\sqrt{p \cdot \sigma}$ and the same is true for 
$\sqrt{p \cdot \bar\sigma}$. As we read off from 
equation~(\ref{rootsps}) this proves equation~(\ref{ccc}). Q.E.D.

Let us apply charge conjugation to $u(p,\xi)$ of equation~(\ref{uv}):
\begin{equation}
u^c(p,\xi) = 
\left( \begin{array}{cc} 0 & \varepsilon \\ -\varepsilon & 0 
\end{array} \right) 
\left(
\begin{array}{c}
\sqrt{p \cdot \sigma}\, \xi \\ \sqrt{p \cdot \bar\sigma}\, \xi
\end{array} \right)^* = 
\left(
\begin{array}{r}
\varepsilon \left( \sqrt{p \cdot \bar\sigma} \right)^* \varepsilon^{-1} 
\left( \varepsilon \xi^* \right) \\ 
-\varepsilon \left( \sqrt{p \cdot \sigma}  \right)^* \varepsilon^{-1}
\left( \varepsilon \xi^* \right) 
\end{array} \right) = v(p, \varepsilon \xi^*).
\end{equation}
In the last step we have used equation~(\ref{ccc}). In a similar fashion we 
can treat $v^c(p,\eta)$. Summarizing, our result is
\begin{equation}
u^c(p,\xi) = v(p, \varepsilon \xi^*) \quad \mbox{and} \quad
v^c(p,\eta) = u(p, -\varepsilon \eta^*).
\end{equation}
We have thus worked out how charge conjugation is explicitly realized in 
the Weyl basis.

\subsection{Helicity states and chiral projectors in the Weyl basis}
In section~\ref{ultrarelativistic} we have performed a general discussion of 
helicity. Here we transfer it to the Weyl basis. Firstly, we want 
to investigate which vectors $\xi$ and $\eta$ in 
$u(p,\xi)$ and $v(p,\eta)$, respectively, belong to which helicities.
In order to obtain the answer to this question, we have to set 
$\hat s_r = \hat p$ in equation~(\ref{hat s_r}), in which case the eigenvectors 
are given by the $\zeta_\pm$ of equation~(\ref{zeta+-}). Therefore, 
equation~(\ref{hat s_r}) tells us that 
\begin{equation}
\begin{array}{rl}
\mbox{particle:} & 
\left\{ \begin{array}{ccc}
h_+ = +1/2 & \Rightarrow & \xi = \zeta_+, \\
h_- = -1/2 & \Rightarrow & \xi = \zeta_-, \\
\end{array} \right.
\\[4mm]
\mbox{antiparticle:} & 
\left\{ \begin{array}{ccc}
h_+ = +1/2 & \Rightarrow & \eta = \zeta_-, \\
h_- = -1/2 & \Rightarrow & \eta = \zeta_+. \\
\end{array} \right.
\end{array}
\end{equation}
As for the spinors $u(p,\pm)$ and $v(p,\pm)$ with definite helicities, 
with the formulas in section~\ref{plane-wave weyl basis} we easily find
\begin{equation}\label{uv+-}
u(p,\pm) = 
\left( \begin{array}{c}
\sqrt{E_p \mp |\vec p\,|}\, \zeta_\pm \\[2mm]
\sqrt{E_p \pm |\vec p\,|}\, \zeta_\pm
\end{array} \right),
\quad
v(p,\pm) = 
\left( \begin{array}{r}
 \sqrt{E_p \pm |\vec p\,|}\, \zeta_\mp \\[2mm]
-\sqrt{E_p \mp |\vec p\,|}\, \zeta_\mp
\end{array} \right).
\end{equation}

Using $\sqrt{E_p - |\vec p\,|} = m/\sqrt{E_p + |\vec p\,|}$ and
applying the projector $\gamma_-$ to the spinors in equation~(\ref{uv+-}),
we obtain
\begin{subequations}\label{guv+-}
\begin{eqnarray}
\gamma_- u(p,+) = \frac{m}{\sqrt{E_p + |\vec p\,|}}
\left( \begin{array}{c} \zeta_+ \\ 0 \end{array} \right),
&&
\gamma_- u(p,-) = \sqrt{E_p + |\vec p\,|}
\left( \begin{array}{c} \zeta_- \\ 0 \end{array} \right),
\\
\gamma_- v(p,+) = \sqrt{E_p + |\vec p\,|}
\left( \begin{array}{c} \zeta_- \\ 0 \end{array} \right),
&&
\gamma_- v(p,-) = \frac{m}{\sqrt{E_p + |\vec p\,|}}
\left( \begin{array}{c} \zeta_+ \\ 0 \end{array} \right).
\end{eqnarray}
\end{subequations}
Therefore, the Weyl-basis result shows explicitly that, 
in the ultrarelativistic limit,  
the weak amplitude of particles with positive helicity is suppressed compared 
to the weak amplitude with negative helicity and vice versa for 
antiparticles:
\begin{quote}
$\gamma_- u(p,+)$ suppressed by 
$\frac{\displaystyle m}{\displaystyle E_p +|\vec p\,|}$ 
compared to $\gamma_- u(p,-)$, \\[2mm]
$\gamma_- v(p,-)$ suppressed by 
$\frac{\displaystyle m}{\displaystyle E_p +|\vec p\,|}$ 
compared to $\gamma_- v(p,+)$.
\end{quote}

Secondly, let us check if our Weyl-basis results conform to 
equation~(\ref{uuvv}). For this purpose we reformulate 
equation~(\ref{guv+-}) as
\begin{subequations}\label{weyl-uu}
\begin{eqnarray}
\hspace{-15mm}\label{uuaa}
&&
\gamma_- u(p,+) \left( \gamma_- u(p,+) \right)^\dagger \beta = 
\gamma_- v(p,-) \left( \gamma_- v(p,-) \right)^\dagger \beta = 
\left( E_p - |\vec p\,| \right)
\left( \begin{array}{cc} 0 & \zeta_+ \zeta_+^\dagger  \\ 0 & 0 
\end{array} \right),
\\
\hspace{-15mm}\label{uubb}
&&
\gamma_- u(p,-) \left( \gamma_- u(p,-) \right)^\dagger \beta = 
\gamma_- v(p,+) \left( \gamma_- v(p,+) \right)^\dagger \beta = 
\left( E_p + |\vec p\,| \right)
\left( \begin{array}{cc} 0 & \zeta_- \zeta_-^\dagger  \\ 0 & 0 
\end{array} \right).
\end{eqnarray}
\end{subequations}
For a comparison of this with equation~(\ref{uuvv}), we evaluate the 
right-hand sides of the latter in the Weyl basis. The key 
to this evaluation is given by equations~(\ref{P+-}) and~(\ref{Pzeta}),
which readily lead to
\begin{equation}
\slashed{z} = -\gamma^0 - \hat p \cdot \vec \gamma = 
\left( \begin{array}{cc} 
0 & -\bone - \hat p \cdot \vec \sigma \\ 
-\bone + \hat p \cdot \vec \sigma & 0 
\end{array} \right) = -2 
\left( \begin{array}{cc}
0 & \zeta_+ \zeta_+^\dagger \\ \zeta_- \zeta_-^\dagger & 0
\end{array} \right)
\end{equation}
or 
\begin{equation}
-\frac{m^2 \slashed{z}}{2(E_p + |\vec p\,|)} \gamma_+ = 
\left( E_p - |\vec p\,| \right)
\left( \begin{array}{cc} 0 & \zeta_+ \zeta_+^\dagger  \\ 0 & 0 
\end{array} \right).
\end{equation}
This proves agreement with equation~(\ref{uuaa}).
For the computation of $\slashed{p}$ in the Weyl basis we use 
\begin{equation}
\zeta_+ \zeta_+^\dagger + \zeta_- \zeta_-^\dagger = \bone, \quad
\zeta_+ \zeta_+^\dagger - \zeta_- \zeta_-^\dagger = 
\hat p \cdot \vec \sigma 
\end{equation}
and obtain 
\begin{equation}
\slashed{p} = \left(
\begin{array}{cc}
0 &
\left( E_p - |\vec p\,| \right) \zeta_+ \zeta_+^\dagger +
\left( E_p + |\vec p\,| \right) \zeta_- \zeta_-^\dagger 
\\
\left( E_p + |\vec p\,| \right) \zeta_+ \zeta_+^\dagger +
\left( E_p - |\vec p\,| \right) \zeta_- \zeta_-^\dagger 
& 0
\end{array} \right)
\end{equation}
or
\begin{equation} 
\left( \slashed{p} + \frac{m^2 \slashed{z}}{2(E_p + |\vec p\,|)} 
\right) \gamma_+ = 
\left( E_p + |\vec p\,| \right) 
\left( \begin{array}{cc} 0 & \zeta_- \zeta_-^\dagger  \\ 0 & 0 
\end{array} \right).
\end{equation}
This proves agreement with equation~(\ref{uubb}).

\section{Conclusions}
%
In these notes we have provided methods that allow us to perform computations 
in the Dirac theory without ever resorting to a special basis of the gamma 
matrices. We believe that this approach is elegant and insightful at the same 
time. Let us highlight the gist of the present notes:
\begin{itemize}
\item
\textit{Pauli's Theorem:} It states that the Dirac matrices are unique up to 
similarity transformations. With surprising effortlessness the proof of 
it can be performed for a general number of space-time dimensions.
\item
\textit{Uniqueness of similarity transformations connecting 
two sets of Dirac matrices:} Such uniqueness up to a multiplicative 
constant guarantees the existence of the charge conjugation matrix $C$ and 
the matrix $\beta$. It also ensures Lorentz invariance of the Dirac equation;  
moreover, assuming that the transformation group of the Dirac spinors 
is a Lie group with traceless generators,
the freedom of the multiplicative constant is removed---apart from a 
sign---and, in this way, $SL(2,\mathbbm{C})$ is obtained.
\item
\textit{Exclusive usage of the anticommutation relations:} 
These are sufficient to treat all topics presented in these notes. 
In a similar vein, characterization of plane-wave solutions of the 
Dirac equation by projectors is adequate for all purposes. 
\item
\textit{Keeping $\beta$ and $\gamma^0$ apart:}
We have emphasized the totally different physics and mathematics background 
of these matrices. Only if one assumes hermiticity properties of the Dirac 
matrices they can be identified. However, in a general basis, 
assuming the phase convention $\beta^\dagger = \beta$, one can only state  
that the product $\beta \gamma^0$ is a definite matrix. This has 
some bearing on the definition of charge conjugation as a selfinverse 
operation on Dirac spinors.
\item
\textit{Charge conjugation:}
We have shown in a completely basis-independent way that charge conjugation 
can be defined as a selfinverse operation on Dirac spinors.
\end{itemize}

\section*{Acknowledgments} 
The author thanks L.~Lavoura and H.~Neufeld for helpful discussions.

\newpage
\appendix

\setcounter{equation}{0}
\renewcommand{\theequation}{A.\arabic{equation}}

\section{A group-theoretical proof of Pauli's Theorem}
\label{a-proof}
\paragraph{Definition of the group:}
We conceive the Dirac matrices $\gamma^\mu$ $(\mu = 0,\ldots,N-1)$ 
as abstract objects and assume that they generate 
a group denoted by $G_D$. 
In order to incorporate the anticommutation relations of 
equation~(\ref{anticom_N}), we have to introduce, apart from the unit element 
$e$, an additional element $a$ which represents the minus sign. 
We thus have $N+1$ group generators. They fulfill the relations 
\begin{equation}\label{a}
a^2 = e, \quad
\left( \gamma^0 \right)^2 = e, \quad
\left( \gamma^j \right)^2 = a \;\;(j=1,\ldots,N-1), \quad
a \gamma^\mu = \gamma^\mu a
\end{equation}
and
\begin{equation}\label{anticom_G}
\gamma^\mu \gamma^\nu = a\, \gamma^\nu \gamma^\mu \quad (\mu \neq \nu).
\end{equation}
The latter relation replaces equation~(\ref{anticom_N}) 
in the case of $\mu \neq \nu$.
The set of $N+1$ generators together with the relations of 
equations~(\ref{a}) and~(\ref{anticom_G}) constitute a presentation of 
the group $G_D$. This group consists of the $2^{N+1}$ elements 
$G^r$ and $a G^r$ ($r = 1,\ldots,2^N$). For the definition of $G^r$ see 
equation~(\ref{G}).\footnote{Obviously, $\bone_d$ is replaced by $e$.} 
For later convenience we stipulate $G^1 = e$ and 
$G^{2^N} \equiv A = \gamma^0 \gamma^1 \cdots \gamma^{N-1}$.

In order to discuss the dimensions of the irreps of $G_D$, 
we use the following two theorems of the theory of finite groups (see 
for instance~\cite{ramond,grimus}):
\begin{enumerate}
\renewcommand{\labelenumi}{(\alph{enumi})}
\item
The number of conjugacy classes equals the number of inequivalent irreps.
\item
The number of group elements, $\mbox{ord}\,G$, can be expressed as  
\begin{equation}\label{od}
\mbox{ord}\,G = \sum_k d_k^2,
\end{equation}
where the sum on the right-hand side runs over 
the dimensions~$d_k$ of \emph{all} inequivalent irreps.
\end{enumerate}

We begin the discussion with the  irreps where $a \mapsto 1$. Such 
irreps are commutative and, therefore, one-dimensional.\footnote{Note that 
$N=1$ is a special case because there 
$G_D \cong \mathbbm{Z}_2 \times \mathbbm{Z}_2$ and thus \emph{all} irreps 
are one-dimensional.}
Moreover, in this case no distinction between $N$ even and odd is necessary. 
In these representations the Dirac matrices are just $\pm 1$ and 
not interesting for the theory of fermions, 
but we have to determine the number of these irreps because we want 
to apply equation~(\ref{od}) for finding 
the dimension(s) of the physically interesting representation(s).

\paragraph{Irreps featuring \boldmath$a \mapsto 1$\unboldmath:}
In this case we necessarily have 
\begin{equation}
a \mapsto 1 \quad \Rightarrow \quad \left( \gamma^\mu \right)^2 \mapsto 1.
\end{equation}
Therefore, 
\begin{equation}
\gamma^\mu \mapsto s(\mu) \quad \mbox{with} \quad 
\left( s(\mu) \right)^2 = 1 \; \forall \mu.
\end{equation}
Consequently, any sequence of $N$ signs denoted by 
\begin{equation}
\langle s \rangle \equiv \langle s(0), s(1), \ldots, s(N-1) \rangle
\end{equation}
defines a one-dimensional (irreducible) representation
$D^{(\langle s \rangle)}$. There are $2^N$ such representations. 
The sequence $\langle 1,1,\ldots,1 \rangle$ refers to the trivial
representation.

\paragraph{\boldmath$N$\unboldmath\ even:}
First we determine the classes of $G_D$. Just as in the beginning 
of the proof of Theorem~\ref{Neven}, we argue that for $p$ even 
we have 
\begin{equation}\label{mu1}
\gamma^{\mu_1} 
\left( \gamma^{\mu_1} \gamma^{\mu_2} \cdots \gamma^{\mu_p} \right) 
\left( \gamma^{\mu_1} \right)^{-1} = 
a \left( \gamma^{\mu_1} \gamma^{\mu_2} \cdots \gamma^{\mu_p} \right).
\end{equation}
If $p$ is odd and taking into account that $N$ is even, 
there is always an index $\nu$ different from all 
$\mu_1, \mu_2, \ldots, \mu_p$. Then 
\begin{equation}\label{nu}
\gamma^\nu 
\left( \gamma^{\mu_1} \gamma^{\mu_2} \cdots \gamma^{\mu_p} \right) 
\left( \gamma^\nu \right)^{-1} = 
a \left( \gamma^{\mu_1} \gamma^{\mu_2} \cdots \gamma^{\mu_p} \right).
\end{equation}
Therefore, including the trivial classes,
the conjugacy classes of $G_D$ are given by 
\begin{equation}
C_1 = \{ e \}, \quad C_2 = \{ a \}, \quad 
C_{r+1} = \{ G^r, \, a G^r \} \;\;(r = 2, \ldots, 2^N).
\end{equation}
There are thus $2^N + 1$ inequivalent irreps of $G_D$,
if $N$ is even.
We know already $2^N$ one-dimensional representations. The dimension $d$ 
of the remaining irrep can thus be determined by 
the formula of equation~(\ref{od}):
\begin{equation}
\mbox{ord}\,G = 2^{N+1} = 2^N \times 1^2 + d^2 
\quad \Rightarrow \quad
d = \sqrt{2^{N+1} - 2^N} = 2^{N/2}, 
\end{equation}
in agreement 
with the result of Theorem~\ref{irrep}.
We have thus found a group-theoretical proof in the case of $N$ even 
that there is a unique dimension $d = 2^{N/2}$ where the 
Dirac \emph{algebra} can be realized as $d \times d$ matrices. 
We denote this irrep by $D^{(\gamma)}$.

\paragraph{\boldmath$N$\unboldmath\ odd:}
In this case the element $A \equiv \gamma^0 \gamma^1 \cdots \gamma^{N-1}$ 
commutes with all $\gamma^\mu$ and, therefore, with all elements of $G_D$. 
There are thus $2^N + 2$ conjugacy classes given by
\begin{equation}\label{classes-odd}
\begin{array}{l}
C_1 = \{ e \}, \quad C_2 = \{ a \}, \quad 
C_{r+1} = \{ G^r, \, a G^r \} \;\;(r = 2, \ldots, 2^N-1), \\[2mm]
C_{2^N + 1} = \{ A \}, \quad 
C_{2^N + 2} = \{ a A \}.
\end{array}
\end{equation}
Since we have already determined all irreps featuring $a \mapsto 1$,
\textit{i.e.}\ the $2^N$ one-dimensional ones, the remaining two irreps 
necessarily have 
\begin{equation}
a \mapsto -\bone_d 
\quad \mbox{and} \quad 
A \mapsto \omega \bone_d.
\end{equation}
The mapping of $A$ is dictated by Schur's Lemma and $\omega$ is some 
constant. Now we follow the arguments presented in section~\ref{caseNodd}. 
We claim that the missing two inequivalent irreps must have the same 
dimension $d$ and differ in the sign of the $\gamma^\mu$. To corroborate 
this claim, let us assume that $D^{(+)}$ is one of the missing irreps and 
let us consider an irrep $D^{(-)}$ obtained from $D^{(+)}$ by changing 
the signs of the representation matrices of $\gamma^\mu$, \textit{i.e.}
\begin{equation}
D^{(-)}(\gamma^\mu) = -D^{(+)}(\gamma^\mu).
\end{equation}
Then these two irreps are necessarily inequivalent, because 
$A$ is a product of an odd number of Dirac matrices and,
if $A \mapsto \omega \bone_d$ for $D^{(+)}$, then 
$A \mapsto -\omega \bone_d$ for $D^{(-)}$; 
however, $-\omega \bone_d$ can never be obtained from $+\omega \bone_d$ 
by a similarity transformation, which concludes the chain of arguments. 
Knowing this, it is easy to determine $d$ by using equation~(\ref{od}):
\begin{equation}
\mbox{ord}\,G = 2^{N+1} = 2^N \times 1^2 + 2 \times d^2 
\quad \Rightarrow \quad
d = \sqrt{\left( 2^{N+1} - 2^N \right)/2} = 2^{(N-1)/2}.
\end{equation}
Again we confirm Theorem~\ref{irrep} by group-theoretical methods.

Note that 
\begin{equation}
A^2 = a^{N(N-1)/2} \left( \gamma^0 \right)^2 \left( \gamma^1 \right)^2 
\left( \gamma^{N-1} \right)^2 = 
a^{N(N-1)/2 + (N-1)}
\end{equation}
due to equations~(\ref{a}) and~(\ref{anticom_G}), which is the analogue 
of equation~(\ref{A2}) leading to the same results for $\omega$ 
as in equation~(\ref{omega}).

\paragraph{Checks:}
We want to conclude this paragraph by checking the correctness of the 
above considerations with the help of group characters. Numbering the 
inequivalent irreps of a group $G$ by Greek letters, 
denoting the values of the characters $\chi^{(\alpha)}$ and $\chi^{(\beta)}$ on 
class $C_i$ by $\chi^{(\alpha)}_i$ and $\chi^{(\beta)}_i$, respectively, 
and the number of elements in class $C_i$ by $c_i$, the orthogonality 
relation for characters reads~\cite{ramond,grimus} 
\begin{equation}\label{sp}
( \chi^{(\alpha)} | \chi^{(\beta)} ) \equiv 
\sum_i c_i \left( \chi^{(\alpha)}_i \right)^* \chi^{(\beta)}_i = 
\mbox{ord}\, G\, \delta_{\alpha\beta}.
\end{equation}
In this sense, characters of inequivalent irreps are 
orthogonal to each other. 

We first check the scalar products of the one-dimensional representations.
In this case we obtain\footnote{For $p=1$ the term in the sum is simply~1.} 
\begin{eqnarray}
( \chi^{(\langle s \rangle)} | \chi^{(\langle s' \rangle)} ) &=& 
2\, \sum_{p=0}^N\;\; \sum_{0 \leq \mu_1 < \mu_2 < \cdots < \mu_p \leq N-1}
s(\mu_1)s'(\mu_1)\,s(\mu_2)s'(\mu_2) \cdots s(\mu_p)s'(\mu_p)
\nonumber \\ &=& 2 \prod_{\mu=0}^{N-1}
\Big( 1 + s(\mu)s'(\mu) \Big).
\end{eqnarray}
Note that this formula holds for both $N$ even and odd; though the 
conjugacy class $C_{2^N + 1}$ with two elements for $N$ even is replaced, 
for $N$ odd, by two conjugacy classes $C_{2^N + 1}$ and $C_{2^N + 2}$ 
with one element each, this is irrelevant for the characters 
$\chi^{(\langle s \rangle)}$ due to $a \mapsto 1$. Obviously,
\begin{equation}
( \chi^{(\langle s \rangle)} | \chi^{(\langle s' \rangle)} ) = 0
\quad \mbox{for} \quad \langle s \rangle \neq \langle s' \rangle
\quad \mbox{and} \quad 
( \chi^{(\langle s \rangle)} | \chi^{(\langle s \rangle)} ) = 2^{N+1}.
\end{equation}

Next we specify to $N$ even and consider the character of 
$D^{(\gamma)}$. Due to $a \mapsto -1$, equations~(\ref{mu1}) and~(\ref{nu}) 
lead to
\begin{equation}
\chi^{(\gamma)} = ( d, -d, 
\underbrace{0, \ldots, 0}_{N-1\;\mathrm{times}} )
\quad \mbox{with} \quad d = 2^{N/2}.
\end{equation}
Since the first two entries of the character of $D^{(\langle s \rangle)}$ are 
just~1, \textit{i.e.}
\begin{equation}
\chi^{(\langle s \rangle)} = ( 1, 1, \ldots ),
\end{equation}
we find 
\begin{equation}
( \chi^{(\gamma)} | \chi^{(\langle s \rangle)} ) = 0 
\quad \mbox{and} \quad
( \chi^{(\gamma)} | \chi^{(\gamma)} ) = 2\,d^2 = 2^{N+1}.
\end{equation}

It remains to consider $N$ odd and the characters of $D^{(+)}$ and $D^{(-)}$ 
given by
\begin{subequations}
\begin{equation}
\chi^{(+)} = (\, d, -d, 
\underbrace{0, \ldots, 0}_{N-2\:\mathrm{times}}, \omega d, -\omega d\, )
\end{equation}
and
\begin{equation}
\chi^{(-)} = (\, d, -d, 
\underbrace{0, \ldots, 0}_{N-2\;\mathrm{times}}, -\omega d, \omega d\, )
\end{equation}
\end{subequations}
with $d = 2^{(N-1)/2}$.
Obviously, 
\begin{equation}
( \chi^{(+)} | \chi^{(+)} ) = ( \chi^{(-)} | \chi^{(-)} ) = 4\,d^2 = 2^{N+1}
\quad \mbox{and} \quad
( \chi^{(+)} | \chi^{(-)} ) = 0.
\end{equation}
But 
\begin{equation}
( \chi^{(\pm)} | \chi^{(\langle s \rangle)} ) = 0
\end{equation}
holds too, because not only the first but also the final two entries in 
$\chi^{(\langle s \rangle)}$ are the same due $a \mapsto 1$.

\paragraph{Concluding remarks:}
It is instructive to compare the derivation of 
Pauli's Theorem here in the appendix with that of section~\ref{pauli}. 
The main point is that in section~\ref{pauli} 
we have considered the Dirac \emph{algebra}, \textit{i.e.}\ 
the minus sign in the anticommutation relation, equation~(\ref{anticom_N}), 
was automatically represented by $-\bone_d$. This is evident in the proof of 
Theorem~\ref{Neven} ($N$ even) when we derive that $\tr\,G^r = 0$ for all $r$ 
except for $G^r = \bone_d$. 
In this way we have singled out the irrep whose character is 
zero on all classes except $C_1$ and $C_2$ and we have, therefore,  
not encountered the irreps with $a \mapsto 1$.
For $N$ odd we have proceeded similarly, \textit{cf.} Theorem~\ref{Nodd} 
($N$ odd). In the group-theoretical proof, however, the number of 
one-dimensional irreps, \textit{i.e.}\ those with $a \mapsto 1$, 
was essential in order to determine, by using 
equation~(\ref{od}), the dimension(s) of the physically useful irrep(s).

\setcounter{equation}{0}
\renewcommand{\theequation}{B.\arabic{equation}}

\section{Selfinverse matrices}
\label{selfinverse}
A selfinverse $n \times n$ matrix $A$ is defined via $A^2 = \bone_n$.
It has the following properties.
\begin{lemma}\label{lemma}
Let $A$ be a selfinverse $n \times n$ matrix on the vector space  
$\mathcal{V} = \mathbbm{R}^n$ or $\mathbbm{C}^n$. 
Then 
\begin{enumerate}
\renewcommand{\labelenumi}{(\alph{enumi})}
\item
the matrices
$\mathcal{P}_\pm = \left( \bone_n \pm A \right)/2$ 
have the properties
\[
\mathcal{P}_+^2 = \mathcal{P}_+, 
\quad
\mathcal{P}_-^2 = \mathcal{P}_-,
\quad
\mathcal{P}_+ \mathcal{P}_- = \mathcal{P}_- \mathcal{P}_+ = 0,
\]
\item
the vector space is the direct sum 
\[
\mathcal{V} = \mathcal{V}_+ \oplus \mathcal{V}_- 
\quad \mbox{with} \quad 
\mathcal{P}_\pm \mathcal{V} \equiv \mathcal{V}_\pm,
\]
\item
there exists an invertible $n \times n$ matrix $S$ such that 
\[
S^{-1} A S = \mathrm{diag} \left( \bone_{n_+},\, -\bone_{n_-}  \right)
\quad \mbox{where} \quad 
n_\pm = \dim \mathcal{V}_\pm.
\]
\end{enumerate}
\end{lemma}
\textbf{Proof:}
Exploiting $A^2 = \bone_n$, item~(a) is easy to check.
Consequently, the $\mathcal{P}_\pm$ are projectors but without the 
hermiticity property that is usually required. Because of 
$x = \mathcal{P}_+ x + \mathcal{P}_- x \;\; \forall\, x \in \mathcal{V}$,
we have $\mathcal{V} = \mathcal{V}_+ \cup \mathcal{V}_-$. Moreover, 
if $y \in \mathcal{V}_+ \cap \mathcal{V}_-$, then 
$\mathcal{P}_+ y = \mathcal{P}_- y$ or $Ay = 0$. Since $A$ is invertible,
$y = 0$ follows. Therefore, $\mathcal{V}_+ \cap \mathcal{V}_- = \{ 0 \}$ and 
$\mathcal{V}$ is the direct sum of $\mathcal{V}_+$ and $\mathcal{V}_-$.
Because of
$A \mathcal{P}_\pm = \pm \mathcal{P}_\pm$,
the restrictions of $A$ on the subspaces $\mathcal{V}_\pm$ fulfill
$\left. A \right|_{\mathcal{V}_\pm} = \pm\mbox{id}$.
Then any choice of basis $\{ x_j \,|\, j = 1,\ldots,n_+ \}$ in $\mathcal{V}_+$ 
and $\{ y_k \,|\, k = 1,\ldots,n_- \}$ in $\mathcal{V}_-$ defines a matrix $S$
via 
$S = \left( x_1, \ldots, x_{n_+}, y_1, \ldots, y_{n_-} \right)$.
Q.E.D.

\setcounter{equation}{0}
\renewcommand{\theequation}{C.\arabic{equation}}

\section{Basics of Lorentz transformations and $SL(2,\mathbbm{C})$}
\label{sl2c}
Here we summarize properties of the Lorentz group $\mathbbm{L}$ 
and discuss its relation with $SL(2,\mathbbm{C})$ to the extent that is 
needed in the main part of this manuscript. For details and advanced 
material on this subject we refer the reader to~\cite{sexl}.
\paragraph{The Lorentz group:}
A Lorentz transformation, transforming 4-vectors $x$ from one inertial 
frame to another, is denoted by
\begin{equation}
x' = L x \quad \mbox{or} \quad 
{x'}^\mu = L^\mu_{\hphantom{\mu}\nu} x^\nu.
\end{equation}
Lorentz transformation matrices $L$ are defined via 
\begin{equation}\label{LgLg}
L^T g L = g \quad \mbox{with} \quad g = \diag \left( 1,-1,-1,-1 \right)
\quad \mbox{or} \quad 
L^\mu_{\hphantom{\mu}\lambda}\, g_{\mu\nu} L^\nu_{\hphantom{\nu}\sigma} = 
g_{\lambda\sigma}.
\end{equation}
Note that $\bone_4 \in \mathbbm{L}$ and, 
shifting the Lorentz transformations to the right-hand side 
in equation~(\ref{LgLg}), one obtains $g = \left( L^{-1} \right)^T g L^{-1}$.
In other words, 
$L \in \mathbbm{L} \, \Leftrightarrow \, L^{-1} \in \mathbbm{L}$, which proves 
that equation~(\ref{LgLg}) defines indeed a group.
Moreover, multiplying equation~(\ref{LgLg}) with $(gL)^{-1}$ from 
the right, we find
$L^T = g L^{-1} g$. Therefore, since $g \in \mathbbm{L}$, we obtain 
\begin{equation}
L \in \mathbbm{L} \;\Leftrightarrow\; L^T \in \mathbbm{L}.
\end{equation}
Formulated with indices, the latter statement reads
\begin{equation}\label{LT}
L^\mu_{\hphantom{\mu}\lambda}\, g^{\lambda\sigma} L^\nu_{\hphantom{\nu}\sigma} = 
g^{\mu\nu}.
\end{equation}

Alternatively, a Lorentz transformation $x' = L x$
may be characterized by the requirement ${x'}^2 = x^2$ 
$\forall\, x \in \mathbbm{R}^4$. \\
\textbf{Proof:} If $L$ fulfills equation~(\ref{LgLg}), then 
$x^T L^T g L x = x^T g x$ or ${x'}^2 = x^2$.
The opposite direction of the proof proceeds as follows. For arbitrary 
$x, y \in \mathbbm{R}^4$ we have $(x' + y')^2 = (x + y)^2$ 
and ${x'}^2 = x^2$, ${y'}^2 = y^2$ by assumption.
Therefore, $x' \cdot y' = x \cdot y$ or 
$x^T L^T g L y = x^T g \,y$ $\forall\, x,y \in \mathbbm{R}^4$, whence 
equation~(\ref{LgLg}) follows. Therefore, the two characterizations of 
Lorentz transformations are equivalent. Q.E.D.

From the defining relation of $L$, the properties
\begin{equation}\label{Lprop}
\left( \det L \right)^2 = 1, \quad \left( L^0_{\hphantom{0}0} \right)^2 \geq 1
\end{equation}
ensue. The latter follows from
\begin{equation}\label{L00}
\left( L^T g L \right)_{00} = \left( L^0_{\hphantom{0}0} \right)^2 - 
\sum_{j=1}^3 \left( L^j_{\hphantom{0}0} \right)^2 = 1
\quad \mbox{or} \quad
\left( L g L^T \right)^{00} = \left( L^0_{\hphantom{0}0} \right)^2 - 
\sum_{j=1}^3 \left( L^0_{\hphantom{0}j} \right)^2 = 1.
\end{equation}
\begin{theorem}\label{orthochronous}
The matrices $L$ with $\det L = 1$ and 
$L^0_{\hphantom{0}0} \geq 1$ form a subgroup of $\mathbbm{L}$, the 
proper orthochronous Lorentz group denoted by $\mathbbm{L}^\uparrow_+$.
\end{theorem}
\textbf{Proof:} Let $L$ and $K$ be two Lorentz transformations with the 
properties required in the theorem. Then, 
$\det (LK) = (\det L) (\det K) = 1$ and 
$\det L^{-1} = 1/\det L = 1$. It remains to demonstrate that 
$\left( LK \right)^0_{\hphantom{0}0} \geq 1$ and 
$\left( L^{-1} \right)^0_{\hphantom{0}0} \geq 1$. For this purpose we need 
the inequality
\[
x \geq 1,\; y \geq 1 \quad \Rightarrow \quad
xy - \sqrt{x^2 - 1} \sqrt{y^2 - 1} \geq 1,
\]
which is easy to prove. We tackle the quantity 
$\left( LK \right)^0_{\hphantom{0}0}$ with the Cauchy--Schwarz inequality 
and equation~(\ref{L00}):
\begin{eqnarray*}
\left( LK \right)^0_{\hphantom{0}0} &=& 
L^0_{\hphantom{0}0} K^0_{\hphantom{0}0} + 
\sum_{j=1}^3 L^0_{\hphantom{0}j}  K^j_{\hphantom{0}0} 
\\
& \geq & 
L^0_{\hphantom{0}0} K^0_{\hphantom{0}0} - 
\left( \sum_{j=1}^3 \left( L^0_{\hphantom{0}j} \right)^2 \right)^{1/2}
\left( \sum_{l=1}^3 \left( K^l_{\hphantom{0}0} \right)^2 \right)^{1/2}
\\
&=&
L^0_{\hphantom{0}0} K^0_{\hphantom{0}0} - 
\sqrt{\left( L^0_{\hphantom{0}0} \right)^2 - 1}
\sqrt{\left( K^0_{\hphantom{0}0} \right)^2 - 1}.
\end{eqnarray*}
With $x = L^0_{\hphantom{0}0}$, $y = K^0_{\hphantom{0}0}$ and the above 
inequality we find $\left( LK \right)^0_{\hphantom{0}0} \geq 1$. 
Finally, let us prove $\left( L^{-1} \right)^0_{\hphantom{0}0} \geq 1$
in an indirect way. Thus we assume 
$\left( L^{-1} \right)^0_{\hphantom{0}0} \leq -1$. 
Then $K \equiv -L^{-1}$ has $K^0_{\hphantom{0}0} \geq 1$. Therefore, according 
to what we have just proven, 
$\left( LK \right)^0_{\hphantom{0}0} \geq 1$, but this contradicts 
$LK = -\bone_4$. Q.E.D.

As a consequence of equation~(\ref{Lprop}), 
$\mathbbm{L}$ decays into four disconnected sets that can be characterized by
\begin{equation}
\mathbbm{L}^\uparrow_+, \quad P \mathbbm{L}^\uparrow_+, \quad 
T \mathbbm{L}^\uparrow_+, \quad PT \mathbbm{L}^\uparrow_+
\quad \mbox{with} \quad 
P = \diag\left( 1,-1,-1,-1 \right), \quad T = \diag\left( -1,1,1,1 \right).
\end{equation}
The Lorentz transformations $P$ and $T$ are parity and time reversal, 
respectively. It makes also sense to define
\begin{equation}
\mathbbm{L}^\uparrow \equiv \mathbbm{L}^\uparrow_+ \cup 
P \mathbbm{L}^\uparrow_+
\quad \mbox{and} \quad
\mathbbm{L}^\downarrow \equiv T \mathbbm{L}^\uparrow_+ \cup 
PT \mathbbm{L}^\uparrow_+,
\end{equation}
where $\mathbbm{L}^\uparrow$ is the orthochronous Lorentz group
and $\mathbbm{L}^\downarrow = T \mathbbm{L}^\uparrow$. If
$L \in \mathbbm{L}^\uparrow$ and $K \in \mathbbm{L}^\downarrow$, then both 
$LK$ and $KL$ are elements of $\mathbbm{L}^\downarrow$;
this is obvious from the above proof of 
Theorem~\ref{orthochronous}.

From now on we concentrate on $\mathbbm{L}^\uparrow_+$. 
First we discuss the subgroup of matrices that fulfill 
$L^T = L^{-1}$. For this purpose we reformulate equation~(\ref{LgLg}) as
\begin{equation}\label{LgLga}
\left( L^{-1} \right)^T = g L g.
\end{equation}
Thus the matrices of the subgroup we are searching for have 
$L = g L g$, which leads to 
$L^0_{\hphantom{0}j} = L^j_{\hphantom{0}0} = 0$ ($j=1,2,3$).
We arrive, therefore, at the characterization
\begin{equation}\label{LR}
L^T = L^{-1} \quad \Rightarrow \quad
L = \left( \begin{array}{cl} 1 & {\vec 0}^{\raisebox{3pt}{$\scriptstyle\,T$}} \\ 
\vec 0 & R
\end{array} \right) \equiv L_R
\quad \mbox{with} \quad R^T R = \bone_3
\quad \mbox{and} \quad \det R = 1.
\end{equation}
The subgroup coincides with the group of spatial rotations.
Rotation matrices are parameterized by 
\begin{equation}
R(\alpha,\vec n\,)_{jk} = \cos\alpha\, \delta_{jk} + 
(1 - \cos\alpha )\, n_j n_k + \sin\alpha\, \varepsilon_{jlk} n_l,
\end{equation}
where $\vec n$ is the rotation axis and $\alpha$ the rotation angle and 
$j,k,l = 1,2,3$. Summation over the index $l$ is understood.
This matrix corresponds to an \emph{active} rotation, \textit{i.e.}, looking 
in the direction of $\vec n$, vectors are rotated clockwise about an angle 
$\alpha$.

Next we want to present a full characterization of the elements of 
$\mathbbm{L^\uparrow_+}$~\cite{sexl}.
\begin{theorem}\label{boost}
Every element $L \in \mathbbm{L^\uparrow_+}$ can be written as a product 
$L = L(\vec v\,) L_R$ where 
\begin{equation}\label{L(v)}
L(\vec v\,) = \left( \begin{array}{cc}
\gamma & \gamma {\vec v}^{\,T} \\[2mm]
\gamma \vec v & \bone_3 - 
\frac{\displaystyle \vec v {\vec v}^{\,T}}%
{\rule{0pt}{11pt}\displaystyle v^2} + \gamma
\frac{\displaystyle \vec v {\vec v}^{\,T}}%
{\rule{0pt}{11pt}\displaystyle v^2} \end{array} \right)
\quad \mbox{with} \quad v = |\vec v\,|
\quad \mbox{and} \quad \gamma = \frac{1}{\sqrt{1-v^2}}
\end{equation}
is a Lorentz boost and $L_R$ a rotation.
\end{theorem}
\textbf{Proof:} We make the ansatz
\[
L = \left( \begin{array}{cc}
\gamma & \gamma {\vec w}^T \\ \gamma \vec v & M
\end{array} \right)
\]
with $L^0_{\hphantom{0}0} = \gamma \geq 1$. Evaluation of both 
$L^T g L = g$ and $L g L^T = g$ leads to the conditions 
\begin{eqnarray*}
&&
\gamma^2 \left( 1 - {\vec v}^{\,2} \right) = 
\gamma^2 \left( 1 - {\vec w}^{\,2} \right) = 1,
\\ &&
\gamma \vec w - M^T \vec v = \gamma \vec v - M \vec w = \vec 0,
\\ &&
\gamma^2 \vec w {\vec w}^T - M^T M = 
\gamma^2 \vec v {\vec v}^T - M M^T = -\bone_3.
\end{eqnarray*}
From the first line we infer 
$\gamma = 1/\sqrt{1-v^2}$ and
$|\vec v\,| = |\vec w\,| = v$,
meaning that $\vec v$ and $\vec w$ have the same length $v < 1$.
Defining 
\[
R = M -\frac{\gamma^2}{\gamma + 1}\, \vec v {\vec w}^T,
\]
a straightforward computation using the conditions above gives 
$R^T R = \bone_3$, \textit{i.e.}\ $R$ is an orthogonal matrix~\cite{sexl}. 
In a similar vein we find
\[
R \vec w = \vec v \quad \mbox{or} \quad {\vec w}^T = {\vec v}^T R.
\]
With $R$ we rewrite $L$ as
\[
L = \left( \begin{array}{cc}
\gamma & \gamma {\vec w}^T \\ \gamma \vec v & 
R + \frac{\displaystyle \gamma^2}{\displaystyle \gamma + 1}\, 
\vec v {\vec w}^T
\end{array} \right) = 
\left( \begin{array}{cc}
\gamma & \gamma {\vec v}^T \\ \gamma \vec v & 
\bone_3 + \frac{\displaystyle \gamma^2}{\displaystyle \gamma + 1}\, 
\vec v {\vec v}^T 
\end{array} \right)
\left( \begin{array}{cl} 1 & {\vec 0}^{\raisebox{3pt}{$\scriptstyle\,T$}} \\ 
\vec 0 & R
\end{array} \right).
\]
Finally, we take into account
\[
\frac{\gamma^2}{\gamma + 1} = -\frac{1}{v^2} + \frac{\gamma}{v^2}
\]
and arrive at the form announced in the theorem. 
Since $L(\vec v\,) \to \bone_4$ for $\vec v \to \vec 0$ in a continuous way, 
we infer $\det L(\vec v\,) = \det \bone_4 = 1$. By assumption, 
$\det L = 1$ and, therefore, $\det R = 1$ as well.
Q.E.D.
\\[2mm]
We stress that $L(\vec v\,)$ is an \emph{active} boost. This means that 
$L(\vec v\,)$ transforms the 4-momentum of a particle at rest into 
the 4-momentum of a particle with velocity $\vec v$:
\begin{equation}\label{pr->p}
L(\vec v\,) \left( \begin{array}{c} m \\ \vec 0 \end{array} \right) = 
\left( \begin{array}{c} m \gamma \\ m \gamma \vec v \end{array} \right).
\end{equation}
As a side note, from the forms of 
$L_R$, equation~(\ref{LR}), and $L(\vec v\,)$, equation~(\ref{L(v)}), it 
is obvious that
\begin{equation}
L_R L(\vec v\,) L_R^{-1} = L(R \vec v\,).
\end{equation}

\paragraph{The group \boldmath$SL(2,\mathbbm{C})$:\unboldmath}
This group consists of all complex $2 \times 2$ matrices $A$ with 
$\det A = 1$, \textit{i.e.}
\begin{equation}\label{Aparameterization}
A = \left( \begin{array}{cc} a & b \\ c & d 
\end{array} \right)
\quad \mbox{with} \quad ad-bc = 1.
\end{equation}
\begin{theorem}
The defining representation of $SL(2,\mathbbm{C})$, $A \to A$, is 
equivalent to its contragredient representation 
$A \to \left( A^{-1} \right)^T$.
\end{theorem}
\textbf{Proof:}
We define the matrix
\begin{equation}\label{varepsilon}
\varepsilon = \left( \begin{array}{rr} 0 & 1 \\ -1 & 0 
\end{array} \right),
\end{equation}
which has the properties $\varepsilon^{-1} = \varepsilon^T = -\varepsilon$.
With the parameterization of equation~(\ref{Aparameterization}) we obtain 
\begin{equation}
\left( A^{-1} \right)^T = \left( \begin{array}{rr} d & -c \\ -b & a 
\end{array} \right). 
\end{equation}
The equivalence is demonstrated by 
\begin{equation}\label{eaea}
\varepsilon^{-1} \left( A^{-1} \right)^T \varepsilon = A.
\end{equation}
Q.E.D.

According to the polar decomposition, every $A \in SL(2,\mathbbm{C})$ can 
be written as
\begin{equation}\label{polar}
A =  HU \quad \mbox{with} \quad H^\dagger = H > 0 \;\; \mbox{and} \;\;
U \in SU(2).
\end{equation}
Every $U$ can be parameterized by an angle $\alpha$ and a unit vector 
$\vec n$:
\begin{equation}
U(\alpha,\vec n\,) = \exp \left( 
-i\alpha\, \vec n \cdot \frac{\vec \sigma}{2} \right) = 
\cos \frac{\alpha}{2}\, \bone - i \sin \frac{\alpha}{2}\, 
\vec n \cdot \vec \sigma.
\end{equation}
Every $H$ of equation~(\ref{polar}) can be represented with a 
4-velocity $u$ as
\begin{equation}\label{H(v)}
H(\vec v\,) = \sqrt{u \cdot \sigma}
\quad \mbox{with} \quad 
u = \gamma \left( \begin{array}{c} 1 \\ \vec v 
\end{array} \right), \quad \gamma = \frac{1}{\sqrt{1-v^2}}, \quad
v = |\vec v\,|.
\end{equation}
\textbf{Proof:} Since $H$ is hermitian and $\det H = 1$, it can be written 
as
\begin{equation}\label{Hdiag}
H = \lambda \xi \xi^\dagger + \lambda^{-1} 
\xi_{\raisebox{-2pt}{$\scriptstyle\bot$}} \xi_\bot^\dagger,
\end{equation}
where the vectors $\xi$ and $\xi_\bot$ form an orthonormal basis of 
$\mathbbm{C}^2$. The condition $H > 0$ implies $\lambda > 0$.
In two dimensions, the projectors occurring in 
equation~(\ref{Hdiag}) can be represented with Pauli matrices as
\begin{equation}
\xi \xi^\dagger = 
\frac{1}{2} \left( \bone + \vec m \cdot \vec \sigma \right) 
\equiv \mathcal{Q}_+,
\quad
\xi_{\raisebox{-2pt}{$\scriptstyle\bot$}} \xi_\bot^\dagger = 
\frac{1}{2} \left( \bone - \vec m \cdot \vec \sigma \right)
\equiv \mathcal{Q}_-,
\end{equation}
where the unit vector $\vec m$ is related to $\xi$ via
\begin{equation}
\xi = \left( \begin{array}{c} \xi_1 \\ \xi_2 \end{array} \right)
\quad \Rightarrow \quad
\vec m = \xi^\dagger \vec \sigma \xi = \left( \begin{array}{c}
2 \mbox{Re} \left( \xi_1^* \xi_2 \right) \\
2 \mbox{Im} \left( \xi_1^* \xi_2 \right) \\
|\xi_1|^2 - |\xi_2|^2
\end{array} \right),
\end{equation}
\textit{cf.}\ equation~(\ref{solution hat s_r}).
In the next step, we assume without loss of generality $\lambda \leq 1$ and 
parameterize the eigenvalues as 
\begin{equation}
\lambda = \sqrt{\gamma (1-v)}, \quad 
\lambda^{-1} = \sqrt{\gamma (1+v)}
\quad \mbox{where} \quad \gamma = \frac{1}{\sqrt{1-v^2}}
\quad \mbox{and} \quad 0 \leq v < 1.
\end{equation}
At this stage, we have 
\begin{equation}\label{stage}
H = 
\sqrt{\gamma (1-v)}\,\mathcal{Q}_+ + \sqrt{\gamma (1+v)}\,\mathcal{Q}_-.
\end{equation}
Now we define the matrix
\begin{equation}
H' = u \cdot \sigma 
\quad \mbox{with} \quad 
u = \left( \begin{array}{c} \gamma \\ \gamma v \vec m
\end{array} \right).
\end{equation}
With $\mathcal{Q}_+ + \mathcal{Q}_- = \bone$ and 
$\vec m \cdot \vec \sigma = \mathcal{Q}_+ - \mathcal{Q}_-$, 
$H'$ is written as
\begin{equation}
H' = \gamma (1-v)\,\mathcal{Q}_+ + \gamma (1+v)\,\mathcal{Q}_-.
\end{equation}
Therefore, $H = \sqrt{H'}$. Q.E.D.

\paragraph{The relation between \boldmath$SL(2,\mathbbm{C})$
and $\mathbbm{L}^\uparrow_+$:\unboldmath}
A matrix $A \in SL(2,\mathbbm{C})$ induces a 
Lorentz transformation $L$ on an arbitrary 4-vector $x$ by the prescription 
\begin{equation}\label{basic-sl2c}
A\, \sigma_\mu x^\mu A^\dagger = \sigma_\lambda L^\lambda_{\phantom{\lambda}\nu}
x^\nu.
\end{equation}
\textbf{Proof:} 
The left-hand side of equation~(\ref{basic-sl2c}) 
is hermitian. Every hermitian $2 \times 2$ matrix can be decomposed 
into matrices $\sigma_\lambda$. In other words, 
there is a 4-vector $x'$ such that
\[
A\, \sigma_\mu x^\mu A^\dagger = \sigma_\lambda {x'}^\lambda.
\]
Taking the determinant of this equation, one obtains
\[
\det \left( A\, \sigma_\mu x^\mu A^\dagger \right) = 
\det \left( \sigma_\mu x^\mu \right) = x \cdot x = x' \cdot x'
\]
or $x^2 = {x'}^2$. As a consequence, 
since the prescription $x \to x'$ defines a linear 
transformation, the vectors $x$ and $x'$ are 
related by a Lorentz transformation,
\textit{i.e.}\ there is an $L$ such that $x' = Lx\;\forall x$. Q.E.D.

Note that, as a topological space, $SL(2,\mathbbm{C})$ is connected; 
this means that every $A$ can be reached by a continuous path 
$A(t) \in SL(2,\mathbbm{C})$ ($A(0) = \bone$, 
$A(1) = A$) from the unit matrix $\bone$. \\
\textbf{Proof:} For every non-zero $z \in \mathbbm{C}$ we conveniently define a 
continuous function $\varphi_z(t)$ ($0 \leq t \leq 1$) 
with $\varphi_z(0) = 1$, $\varphi_z(1) = z$ 
such that $\varphi_z(t)$ is non-zero $\forall\, t$.\footnote{If $z$ 
lies on the negative real axis, $\varphi_z(t)$ can for instance  
be chosen as a semicircle. Otherwise, the simplest choice is a straight line.}
We follow the notation of equation~(\ref{Aparameterization}) and make some 
case distinctions. If $a \neq 0$, then $d = (1 + bc)/a$. In this case 
a suitable path with $\det A(t) = 1$ is
\[
A(t) = \left( \begin{array}{cc}
\varphi_a(t) & tb \\ tc & (1 + t^2 bc)/\varphi_a(t)
\end{array} \right).
\]
Obviously, for $d \neq 0$ we can choose 
\[
A(t) = \left( \begin{array}{cc}
(1 + t^2 bc)/\varphi_d(t) & tb \\ tc & \varphi_d(t)
\end{array} \right).
\]
It remains to consider the special case $a = d = 0$, $c = -1/b$.
Then the path 
\[
A(t) = \left( \begin{array}{cc}
\cos \frac{\pi t}{2} & \varphi_b(t) \sin \frac{\pi t}{2} \\ 
-\frac{1}{\varphi_b(t)}\, \sin \frac{\pi t}{2} & \cos \frac{\pi t}{2}
\end{array} \right)
\]
has the desired properties. Q.E.D.

In the following we denote the Lorentz transformation 
occurring in equation~(\ref{basic-sl2c}) by $L_A$, 
in order to stress that it is induced by $A$. As a consequence of 
$SL(2,\mathbbm{C})$ being connected we find that $L_A$ is a 
proper orthochronous Lorentz transformation. The reason is that any continuous 
path $A(t)$ connecting $\bone$ with $A$ induces via 
equation~(\ref{basic-sl2c}) a continuous path $L_{A(t)}$ 
connecting $\bone_4$ with $L_A$. Such a continuous path 
cannot jump from $\mathbbm{L}^\uparrow_+$ to any of the other three 
disconnected components of the Lorentz group. Therefore, we have the mapping 
\begin{equation}\label{mapping}
A \in SL(2,\mathbbm{C}) \to L_A \in \mathbbm{L}^\uparrow_+.
\end{equation}
In addition, the following two theorems 
confirm $L_A \in \mathbbm{L}^\uparrow_+$ in an explicit manner and 
demonstrate, moreover, that this mapping is surjective, \textit{i.e.}\ 
all elements of $\mathbbm{L}^\uparrow_+$ can be obtained from 
$SL(2,\mathbbm{C})$.

The mapping of equation~(\ref{mapping}) has the following properties.
\begin{theorem}\label{U->L_R}
If $A \in SL(2,\mathbbm{C})$ is unitary, then
\[
A = U(\alpha, \vec n\,) \to L_A = 
\left( \begin{array}{cc} 1 & 0 \\ 0 & R(\alpha, \vec n\,)
\end{array} \right).
\]
\end{theorem}
This is a standard topic in quantum mechanics and we do not elaborate on 
it any further. The next one is less standard and we will, therefore, sketch 
a proof.
\begin{theorem}\label{H->L(v)}
If $A \in SL(2,\mathbbm{C})$ is hermitian, then
\[
A = \pm H(\vec v\,) \to L_A = L(\vec v\,).
\]
\end{theorem}
\textbf{Proof:} If $A$ is hermitian, it can be diagonalized as formulated 
in equation~(\ref{Hdiag}), hence $A = \pm H(\vec v\,)$. Since 
$L_A = L_{-A}$, we consider, without loss of generality, $A = H(\vec v\,)$ 
and proceed by direct computation with $H(\vec v\,)$ in the form of 
equation~(\ref{stage}). Since we know at this point that $\vec m$ 
represents the direction of $\vec v$ , we use from now on 
$\hat v \equiv \vec m = \vec v/v$. We insert $H(\vec v\,)$ into 
equation~(\ref{basic-sl2c}):
\begin{eqnarray}
H(\vec v\,) \left( x \cdot \sigma \right) H(\vec v\,) &=& 
x^0 \gamma \left( (1-v) \mathcal{Q}_+ + (1+v) \mathcal{Q}_- \right) \\&&
- \gamma \left( 
\sqrt{1-v}\,\mathcal{Q}_+ + \sqrt{1+v}\,\mathcal{Q}_- \right)
\vec x \cdot \vec \sigma 
\left( \sqrt{1-v}\,\mathcal{Q}_+ + \sqrt{1+v}\,\mathcal{Q}_- \right).
\nonumber
\end{eqnarray}
To proceed further we need the products
\begin{subequations}
\begin{eqnarray}
\mathcal{Q}_+ \left( \vec x \cdot \vec\sigma \right) \mathcal{Q}_+ 
&=&
\hphantom{-} \frac{1}{2} \left( \vec x \cdot \hat v \right)
\left( \bone + \hat v \cdot \vec \sigma \right),
\\
\mathcal{Q}_- \left( \vec x \cdot \vec\sigma \right) \mathcal{Q}_- 
&=&
- \frac{1}{2} \left( \vec x \cdot \hat v \right)
\left( \bone - \hat v \cdot \vec \sigma \right),
\\
\mathcal{Q}_+ \left( \vec x \cdot \vec\sigma \right) \mathcal{Q}_- 
&=&
\hphantom{-} \frac{1}{2} \left( \vec x - 
i \left( \vec x \times \hat v \right) - 
\left( \vec x \cdot \hat v \right) \hat v \right) \cdot \vec \sigma, 
\\
\mathcal{Q}_- \left( \vec x \cdot \vec\sigma \right) \mathcal{Q}_+ 
&=&
\hphantom{-} \frac{1}{2} \left( \vec x + 
i \left( \vec x \times \hat v \right) - 
\left( \vec x \cdot \hat v \right) \hat v \right) \cdot \vec \sigma.
\end{eqnarray}
\end{subequations}
After some algebra we find
\begin{equation}
H(\vec v\,) \left( x \cdot \sigma \right) H(\vec v\,) = 
\gamma \left( x^0 + \vec v\cdot \vec x \right) \bone -
\left[ \gamma \left( \vec v x^0 + \hat v \left( \hat v \cdot \vec x \right) 
\right) + \vec x - \hat v \left( \hat v \cdot \vec x \right) \right] 
\cdot \vec \sigma,
\end{equation}
whence we can read off the induced $L_A$ and check that it agrees 
with $L(\vec v\,)$. Q.E.D. \\[2mm]
The consequence of Theorems~\ref{U->L_R} and~\ref{H->L(v)} is 
that the mapping of equation~(\ref{mapping}) is surjective, 
because every $L \in \mathbbm{L}^\uparrow_+$ can be decomposed as 
$L = L(\vec v\,) L_R$. The mapping is, of course, not injective 
because $A \left( x \cdot \sigma \right) A^\dagger = x \cdot \sigma$ has the 
two solutions $A = \pm \bone$---\textit{cf.}\ the discussion at the end of 
section~\ref{deli}.

As the last topic in appendix~\ref{sl2c} we derive equation~(\ref{Scond}) 
in the Weyl basis, departing from equation~(\ref{basic-sl2c}).
\begin{theorem}\label{AL}
For every $A \in SL(2,\mathbbm{C})$ the following relations hold:
\[
A \sigma_\mu A^\dagger = \sigma_\lambda L^\lambda_{\phantom{\lambda}\mu}, 
\quad
\left( A^{-1} \right)^\dagger \bar\sigma_\mu A^{-1} = 
\bar\sigma_\lambda L^\lambda_{\phantom{\lambda}\mu}.
\]
The second relation follows from the first one and vice versa.
\end{theorem}
\textbf{Proof:}
The first relation is nothing but equation~(\ref{basic-sl2c}) with 
the general 4-vector left out. To derive the second relation, we need 
\[
A^{-1} = \varepsilon^{-1} A^T \varepsilon 
\quad \mbox{and} \quad 
\left( A^{-1} \right)^\dagger = \varepsilon^{-1} A^* \varepsilon 
\]
obtained by rewriting equation~(\ref{eaea}), and in addition
\[
\varepsilon \bar \sigma_\mu \varepsilon^{-1} = \sigma_\mu^*,
\]
which is easy to check. Then we compute
\[ 
\left( A^{-1} \right)^\dagger \bar\sigma_\mu A^{-1} = 
\varepsilon^{-1} A^* \varepsilon\, \bar\sigma_\mu \varepsilon^{-1} A^T 
\varepsilon = 
\varepsilon^{-1} \left( A \sigma_\mu A^\dagger \right)^* \varepsilon = 
\varepsilon^{-1} \sigma_\lambda^* \varepsilon L^\lambda_{\phantom{\lambda}\mu} = 
\bar\sigma_\lambda L^\lambda_{\phantom{\lambda}\mu}.
\]
Q.E.D.

It is a special property of the Weyl basis that the matrix $\mathcal{S}$ 
which performs Lorentz transformations on the 4-spinors is block-diagonal 
and given by
\begin{equation}
\mathcal{S} = \left( \begin{array}{cc} A & 0 \\ 
0 & \left( A^{-1} \right)^\dagger
\end{array} \right)
\quad \mbox{with} \quad A \in SL(2,\mathbbm{C}),
\end{equation}
\textit{cf.}\ section~\ref{wbli}. Therefore, 
\begin{eqnarray}
\lefteqn{\mathcal{S} \gamma_\mu \mathcal{S}^{-1} =} \nonumber \\ &&
\left( \begin{array}{cc} A & 0 \\ 
0 & \left( A^{-1} \right)^\dagger
\end{array} \right)
\left( \begin{array}{cc} 0 & \sigma_\mu \\ \bar\sigma_\mu & 0 
\end{array} \right)
\left( \begin{array}{cc} A^{-1} & 0 \\ 
0 & A^\dagger
\end{array} \right) = 
\left( \begin{array}{cc} 0 & A \sigma_\mu A^\dagger \\ 
\left( A^{-1} \right)^\dagger \bar\sigma_\mu A^{-1} & 0 
\end{array} \right). \hphantom{xxx}
\end{eqnarray}
Using Theorem~\ref{AL}, we obtain 
\begin{equation}\label{Sg1}
\mathcal{S} \gamma_\mu \mathcal{S}^{-1} = \gamma_\lambda 
L^\lambda_{\phantom{\lambda}\mu}.
\end{equation}
Though this computation was performed in the Weyl basis, the result 
is basis-independent. An equivalent relation is 
\begin{equation}\label{Sg2}
\mathcal{S} \gamma^\mu \mathcal{S}^{-1} = 
\left( L^{-1} \right)^\mu_{\phantom{\mu}\lambda} \gamma^\lambda 
\quad \mbox{or} \quad 
\mathcal{S}^{-1} \gamma^\mu \mathcal{S} = 
L^\mu_{\phantom{\mu}\lambda} \gamma^\lambda,   
\end{equation}
which follows from equation~(\ref{Sg1}) by raising the index $\mu$ 
and using the first relation in equation~(\ref{LgLga}).

\end{document}